\documentclass[
reprint,
superscriptaddress,
 amsmath,amssymb,
 aps,
]{revtex4-2}

\usepackage{float}
\makeatletter
\let\newfloat\newfloat@ltx
\makeatother
\usepackage{graphicx}
\usepackage{graphicx}
\usepackage{dcolumn}
\usepackage{bm}
\usepackage{graphicx}
\usepackage{amsmath}
\usepackage{amssymb}
\usepackage{subcaption}
\usepackage{subfig}
\usepackage{qcircuit}
\usepackage{algorithm}
\usepackage{algpseudocode}
\usepackage{cprotect}
\usepackage{lipsum}
\usepackage{booktabs}
\usepackage{array}
\usepackage{comment}
\usepackage{amsmath, amsthm, amssymb}
\usepackage{braket}
\usepackage{bm}
\usepackage{physics}
\usepackage{graphics}
\usepackage[dvipsnames]{xcolor}
\usepackage{booktabs}
\usepackage{mathtools}

\theoremstyle{definition}

\newcommand{\q}{\bm{q}}
\newcommand{\n}{\hat{\bm{n}}}
\newcommand{\z}{\bm{\varsigma}}
\newcommand{\D}{\mathcal{D}}

\DeclareMathOperator*{\argmin}{argmin}

\makeatletter
\renewcommand{\fnum@algorithm}{\fname@algorithm~\thealgorithm}
\makeatother

\makeatletter
\@ifundefined{sf@counterlist}{\newcommand{\sf@counterlist}{}}{}
\makeatother
\usepackage[colorlinks=true, allcolors=blue]{hyperref}

\usepackage{cleveref}

\begin{document}

\title{Gate Freezing Methods for Gradient-Free Variational Quantum Algorithms in Circuit Optimization}

\author{Joona V. Pankkonen}
\email{Corresponding author: joona.pankkonen@aalto.fi}
 \affiliation{%
 Micro and Quantum Systems group, Department of Electronics and Nanoengineering,\\Aalto University, Finland
}%
\author{Lauri Ylinen}
 \affiliation{%
 Department of Mathematics and Statistics,\\University of Jyväskylä, Finland
}%
\author{Matti Raasakka}
 \affiliation{%
 Micro and Quantum Systems group, Department of Electronics and Nanoengineering,\\Aalto University, Finland
}%
\author{Andrea Marchesin}
 \affiliation{%
 Micro and Quantum Systems group, Department of Electronics and Nanoengineering,\\Aalto University, Finland
}%
\author{Ilkka Tittonen}
 \affiliation{%
 Micro and Quantum Systems group, Department of Electronics and Nanoengineering,\\Aalto University, Finland
}%

\date{\today}

\begin{abstract}
Parameterized quantum circuits (PQCs) are pivotal components of variational quantum algorithms (VQAs) that enable flexible encoding of quantum information through tunable quantum gates and have been successfully applied across domains such as quantum chemistry, combinatorial optimization, and quantum machine learning. Despite their potential, PQC performance on real quantum hardware is hindered by noise, decoherence, and the presence of barren plateaus, which can impede optimization. We propose novel methods to improve gradient-free optimizers \verb|Rotosolve|, \verb|Fraxis|, and \verb|FQS|, incorporating information from previous parameter iterations. Our approach conserves computational resources by reallocating the optimization effort toward poorly optimized gates, thereby often improving convergence in the considered regimes. The greatest improvements are observed for \verb|Rotosolve| and \verb|Fraxis|, while the improvements for \verb|FQS| are smaller and problem dependent. These methods are not intended to mitigate the barren plateaus, but rather to improve resource allocation within existing gradient-free optimizers.
\end{abstract}

\maketitle


\section{Introduction}

Parameterized quantum circuits (PQCs) are an essential component of variational quantum algorithms (VQAs), which have emerged as leading candidates to achieve quantum advantage on near-term devices~\cite{cerezo2021variational, peruzzo2014variational}. PQCs provide a flexible and expressive framework for encoding quantum information and solving computational problems in domains such as quantum chemistry~\cite{bauer2020quantum, singh2023benchmarking, delgado2021variational, motta2022emerging, cao2019quantum}, combinatorial optimization~\cite{comb_opt_cite1, comb_opt_cite2, comb_opt_cite3}, and quantum machine learning~\cite{liu2021hybrid, arthur2022hybrid, jerbi2023quantum, shi2022parameterized}. A PQC comprises a sequence of quantum gates with tunable parameters. By optimizing these parameters using a classical feedback loop, the circuit can approximate solutions to complex problems that are classically intractable~\cite{mcclean2016theory, benedetti2019parameterized}. In particular, PQCs have been successfully applied to simulate molecular ground states~\cite{kandala2017hardware, o2016scalable}, and have been applied to approximate solutions to Max-Cut and other NP-hard optimization problems~\cite{perez2024variational, hadfield2019quantum, wang2018quantum, utkarsh2020solving, choi2020quantum, saleem2020max}. Their adaptability and compatibility with noisy intermediate-scale quantum (NISQ) hardware make them a central building block in current hybrid quantum-classical computing paradigms~\cite{cerezo2021variational, bharti_etal}.

Despite the potential of PQCs, their deployment on NISQ devices introduces significant challenges. However, they may still be able to provide quantum advantages~\cite{ mitarai2018quantum}. Current quantum hardware is limited by decoherence, gate infidelity, and readout errors, affecting the precision and convergence of the optimization process~\cite{preskill2018quantum, schuld2020circuit}. 

The presence of barren plateaus can make gradient-based methods inefficient or infeasible in PQC optimization~\cite{mcclean2018barren, cerezo2021cost}. As such, the development of parameter initialization techniques has been proposed, such as Gaussian initialization~\cite{zhang2022escaping}, Gaussian mixture model~\cite{shi2024avoiding}, reduced domain~\cite{wang2024trainability}, as well as the parameter reuse strategy proposed in \cite{Lee2021Parameters} for the Quantum Approximate Optimization Algorithm (QAOA)~\cite{farhi2018quantum}. Sequential and subset-based optimization techniques improve the training efficiency of PQCs by updating only selected parameters or layers in each iteration, rather than the entire parameter set~\cite{skolik2021layerwise, adaptive_pruning}. As the size of the PQC increases, the parameter optimization becomes NP-hard in general~\cite{bittel2021training}. Additionally, the number of measurements required to estimate the observables can grow rapidly as the number of qubits increases. To solve this problem, several strategies have been proposed, such as unitary partitioning~\cite{Zhao2019Measurement, izmaylov2019unitary, ralli2021implementation}, weighted distribution of measurements~\cite{wecker2015progress, rubin2018application, arrasmith2020operator, kubler2020adaptive}, and others~\cite{gokhale2020n, jena2019pauli, hamamura2020efficient, huggins2021efficient}. While barren plateaus are known to hinder both gradient-based and gradient-free optimization in sufficiently deep and random quantum circuits~\cite{barren_plateau_gradientFree_opts, barren_plateau_deep_pqc}, the objective of this work is not to resolve or mitigate barren plateaus but rather to improve the performance of existing sequential single-qubit gate optimizers.

\begin{figure}
    \centering
    \includegraphics[width=0.99\linewidth]{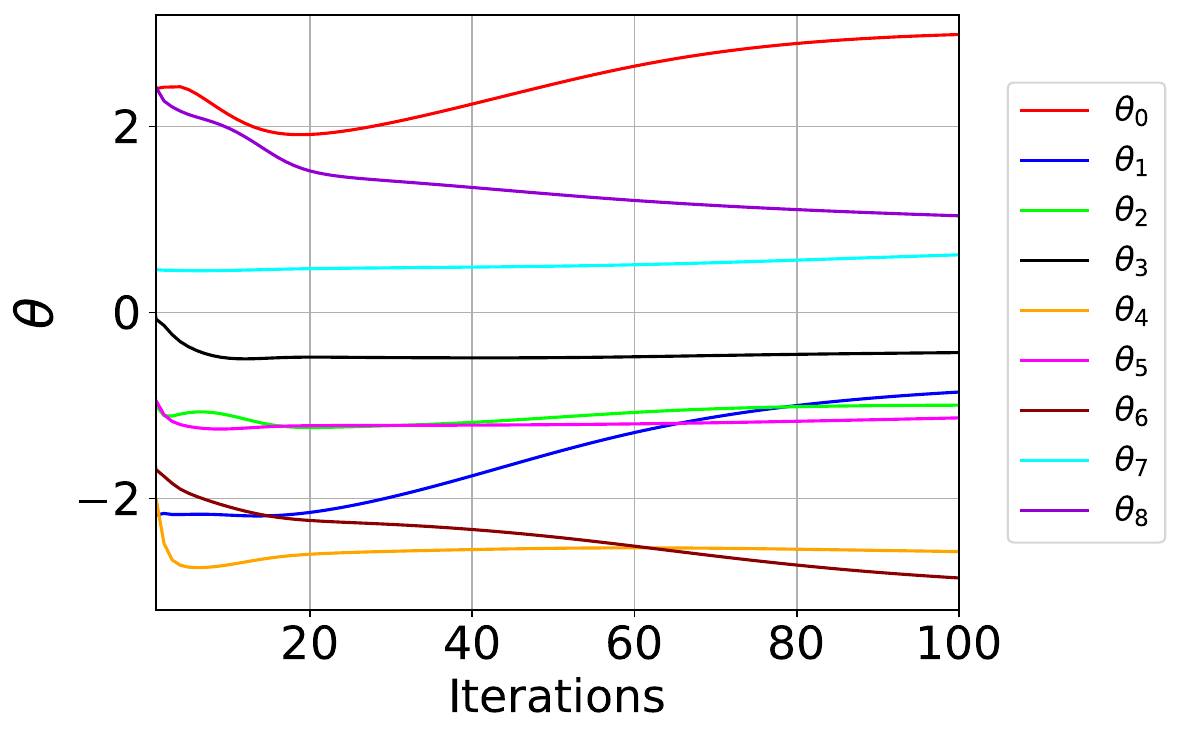}
    \cprotect\caption{A typical parameter evolution over time for randomly chosen angles $\theta_k \in [-\pi, \pi)$ in the PQC using \verb|Rotosolve| optimizer. One-dimensional Heisenberg model with 5 qubits and 3 layers was used with ansatz in Fig.~\ref{roto_gd_ansatz}.}
    \label{params_rotosolve_plot}
\end{figure}

In this work, we develop methods for various sequential single-qubit gate optimizers by incorporating parameter values from previous iterations into the circuit optimization process. The methods are motivated by the observation that typically after only a few iterations in the optimization process, a large subset of the parameters (e.g., angles or unit vectors) in the PQC start to change only very small amounts from iteration to iteration (see Fig.~\ref{params_rotosolve_plot}). We draw a parallel to the layer freezing methods and fine-tuning done in deep learning~\cite{xiao2019fast, goutam2020layerout, wang2023egeria, kumar2019accelerating, li2024smartfrz, yuan2022layer, brock2017freezeout, sorrenti2023selective, wimmer2023dimensionality, liu2021autofreeze}, where deep neural network training is accelerated while having similar or better test accuracy. Thus, by concentrating the optimization steps only on the parameters with significant changes between iterations, we can conserve available computational resources during the optimization process. This often leads to improved convergence across several optimization runs for the optimizers considered in this work. The best improvements were observed for \verb|Rotosolve| and \verb|Fraxis|, while the improvements for \verb|FQS| were smaller and depended on the setting. In this work, we examine the gate freezing methods on deep PQCs, which are analogous to deep neural networks, and demonstrate similar results found in classical machine learning for the layer freezing methods. We also examine the performance in shallow PQCs under finite measurement accuracy, with shot noise used to model the measurement inaccuracy.

This work is structured as follows. First, in Sec.~\ref{VQC_section} we go through the general optimization of PQCs with the gradient-free optimizers \verb|Rotosolve|~\cite{Ostaszewski_2021}, Free-Axis Selection (\verb|Fraxis|)~\cite{fraxis} and Free-Quaternion Selection (\verb|FQS|)~\cite{fqs}. Then, in Sec.~\ref{gate_freeze_section}, we present the proposed gate freezing method that uses gate parameter values and a generalized version that uses matrix norms. In addition, we present the incremental gate freezing method for optimizing PQCs. Next, we demonstrate how to incorporate gate freezing methods into gradient-free optimizers. In Sec.~\ref{Results_section}, we provide results for the one-dimensional Heisenberg model as well as the Fermi-Hubbard lattice model. We also examine the performance of the incremental gate freezing method with \verb|Rotosolve| in various system sizes for the one-dimensional Heisenberg model. Finally, in Sec.~\ref{conclusions_section} we provide conclusions of our work and discuss further research ideas.

\section{Optimization of parameterized quantum circuits} \label{VQC_section}

\subsection{Parameterized Quantum Circuits}
We begin by considering a PQC, which corresponds to a unitary $U(\bm{\theta})$ consisting of parameters $\bm{\theta}$ \cite{cerezo2021variational}. The PQC unitary $U(\bm{\theta})$ usually consists of a layer of parameterized single-qubit gates in each qubit, followed by an entangling layer of two-qubit gates, usually Controlled-Z or CNOT gates. This structure is repeated $L$ times, and the entire PQC unitary $U(\bm{\theta})$ can be written as $U(\bm{\theta}) = U_{L-1}(\bm{\theta}_{L-1}) \cdots U_0(\bm{\theta}_0)$.

Now, if we follow the setting from \cite{pankkonen2025improvingvariationalquantumcircuit} and denote the number of qubits by $n$, the single layer unitary $U_l(\bm{\theta}_l)$ of the PQC can be expressed as follows:
\begin{equation} \label{U_one_layer}
    U_l(\bm{\theta}_l) = W_l \left(\bigotimes_{k=1}^{n} e^{-i\theta_{nl+k} H_{nl+k} / 2} \right),
\end{equation}\\
where $W_l$ is the entangling layer and the index $k$ runs over the individual qubits. The variable $\theta_{nl+k}$ denotes the $k$-th element in the parameter vector $\bm{\theta}_l$ and $H_{nl+k}$ is a Hermitian operator that generates the unitary that acts on the $k$-th qubit in the $l$-th layer. An illustration of a PQC structure is shown in Fig.~\ref{Ansatz_circuit_image}.

\begin{figure}[h]
    \[
    \Qcircuit @C=2em @R=.9em {
    & \mbox{$L$ layers} & & &  \\
    & & & & & &  \\
     \lstick{\ket{0}_1} & \gate{R_{ln + 1}\Bigl(\theta_{ln + 1}\Bigr)}  &  \ctrl{0} & \qw & \qw &  \meter\\
     \lstick{\ket{0}_2}&  \gate{R_{ln + 2}\Bigl(\theta_{ln + 2}\Bigr)} &  \ctrl{-1} & \ctrl{1} & \qw  &   \meter\\
     \lstick{\ket{0}_3}&  \gate{R_{ln + 3}\Bigl(\theta_{ln + 3}\Bigr)}  &  \ctrl{0} & \ctrl{0} & \qw &   \meter\\
     \lstick{\ket{0}_4}&  \gate{R_{ln + 4}\Bigl(\theta_{ln + 4}\Bigr)}  &  \ctrl{-1} & \ctrl{0} & \qw  &  \meter\\
     \lstick{\ket{0}_5}&  \gate{R_{ln + 5}\Bigl(\theta_{ln + 5}\Bigr)} &  \qw & \ctrl{-1} & \qw &  \meter  \gategroup{3}{2}{7}{4}{1.2em}{--} 
    }
    \]
    \cprotect\caption{Ansatz circuit design for PQC optimization for \verb|Fraxis| and \verb|FQS| optimizers.}
    \label{Ansatz_circuit_image}
\end{figure}

To optimize the PQC, we need to define a cost function $C(\bm{\theta})$, which depends on the values of the parameter vector $\bm{\theta}$ \cite{cost_function}. The cost function provides a quantitative measure of how well the PQC is optimized. The cost function $C(\bm{\theta})$ is written as the expectation value of some Hermitian observable $\hat{M}$ as follows
\begin{equation}
    \langle \hat{M} \rangle = \text{Tr}\left(\hat{M}U(\bm{\theta}) \rho_0 U(\bm{\theta})^\dagger \right),
\end{equation}
where $\rho_0$ is the initial state of the PQC that is set to $\rho_0 = |\bm{0}\rangle\langle \bm{0}|$. Here, we denote the $n$-qubit state as $ \vert \bm{0}\rangle \equiv |0\rangle^{\otimes n}$.

Using Eq.~(\ref{U_one_layer}), we can express the cost function in expanded form as
\begin{widetext}
\begin{eqnarray}
    \langle \hat{M} \rangle = \Tr\left(\hat{M}W_{L-1} R_{Ln}'(\theta_{Ln})\cdots W_l R_{(l-1)n}'(\theta_{(l-1)n})\cdots R_1'(\theta_1) \rho_0  R_1'(\theta_1)^\dagger \cdots R_{(l-1)n}'(\theta_{(l-1)n})^\dagger W_l^\dagger \cdots R_{Ln}'(\theta_{Ln})^\dagger W_{L-1}^\dagger \right).  \qquad
\end{eqnarray}
\end{widetext}
Each $R_d'(\theta_d )$ represents a single-qubit gate with a parameter $\theta_d$ that acts on the $k$-th qubit 
\begin{equation}
    R_d'(\theta_d) = I^{\otimes(k-1)} \otimes R_d(\theta_d) \otimes I^{\otimes (n-k)},
\end{equation}
where $R_d(\theta_d)$ is a $2\times 2$ matrix representation of a parameterized single-qubit gate.

We now define the unitaries preceding and following the gate $R_d'$ as $V_1$ and $V_2$, respectively. Here, the unitaries do not depend on the values of $\theta_d$. By doing this, we obtain a more compact form of the cost function as follows
\begin{equation}
     \langle \hat{M} \rangle = \Tr\left( \hat{M} V_1 R_d'(\theta_d) V_2 \rho_0 V_2^\dagger R_d'(\theta_d)^\dagger V_1^\dagger \right).
\end{equation}
By using the cyclic property of the trace operation and defining 
\begin{align}
    M &\equiv V_1^\dagger \hat{M} V_1, \\
    \rho & \equiv V_2 \rho_0 V_2 ^\dagger,
\end{align}
we get the form for the cost function that is commonly used in gradient-free optimizers~\cite{Ostaszewski_2021, fraxis, fqs}
\begin{equation}
    \expval{M} = \Tr \Bigl(MR_d'(\theta_d) \rho R_d'(\theta_d)^\dagger \Bigr).
\end{equation}

The idea behind gradient-free optimizers \verb|Rotosolve|, \verb|Fraxis|, and \verb|FQS| is to choose a single-qubit gate and measure the PQC with different parameter values $\theta_d$ while keeping all other gates fixed. Then, we can compute the optimal parameter value for that gate using the measurements. \verb|Rotosolve| focuses on optimizing the angle of the rotation gate ($R_d \in \{R_X, R_Y, R_Z\}$) using the sinusoidal form of the cost function to its advantage. \verb|Fraxis| and \verb|FQS| are based on matrix factorization, where a symmetric matrix is formed from the measurements. Then its eigenvectors and eigenvalues are computed classically. The optimal axis for \verb|Fraxis| or quaternion for \verb|FQS| corresponds to an eigenvector with the smallest eigenvalue.  For full description of \verb|Rotosolve|, \verb|Fraxis| and \verb|FQS|, see Appendix~\ref{appendix_optimizers}.

Next, we show how to implement our proposed gate freezing method for the \verb|Rotosolve|, \verb|Fraxis|, and \verb|FQS|.

\section{Freezing Threshold for the Parameters} \label{gate_freeze_section}

In this section, we present our proposed gate freezing method for \verb|Rotosolve|, \verb|Fraxis|, and \verb|FQS|. We start by considering a freezing threshold, which determines when the optimizable gates should be ``frozen". That is, we should not optimize the specific gates for $\kappa$ iterations until we begin optimizing them again. To activate gate freezing, we must first define the conditions under which it occurs. We use a freezing threshold $T$, which is a measure of how little the gate parameter $\theta_k$ we allow to change before we freeze the gate. Here, the parameter $\theta_k$ may be multidimensional instead of a single scalar value, such as the parameters for \verb|Fraxis| and \verb|FQS|. We note that the parameters of \verb|Fraxis|, the axis of rotation $\n_d$, can be represented as quaternions by setting the first component of the quaternion to zero as follows $\q_d = (0, n_{x,d}, n_{y,d}, n_{z,d})$. In this section, we use the quaternion representation for both \verb|Fraxis| and \verb|FQS|. The distance between the parameters is denoted by $\D$.


\subsection{Gate Freeze with Parameter Values} \label{param_sec}

We now introduce a freezing method based on the previous parameter values $\theta_k$. The \verb|Rotosolve| optimizer uses only a one-dimensional parameter, an angle $\theta_k$ for the given rotation gate $R_k$, where $R_k \in \{R_X, R_Y, R_Z \}$. When computing the distance between the previous and current parameters, we also need to take into account the cyclic boundary $(-\pi, \pi]$. That is, we define the distance between the parameters as follows
\begin{equation}
\begin{split}
    &\D(\theta_{k,prev}, \theta_{k,new}) \\
    & = \min ( |\theta_{k,prev} - \theta_{k,new}|, 2\pi - |\theta_{k,prev} - \theta_{k,new}|).
\end{split}
\end{equation}

The parameter spaces of the \verb|Fraxis| and \verb|FQS| optimizers can be represented on the three- and four-dimensional unit spheres, respectively. On a spherical surface, the shortest distance is not Euclidean, but rather the distance along the great arc on the sphere $d_g$. The distance $d_g$ for the unit sphere is computed as follows 
\begin{equation}
    d_g(\q_{k,prev}, \q_{k, new}) = \arccos \left(\q_{k,prev} \cdot \q_{k, new} \right). \vspace{0.1cm}
\end{equation}
When evaluating the expectation value of the form $\expval{M} = \Tr(MR_d '\rho R_d'^\dagger)$, we need to take into account that the transformation $\q_k \rightarrow - \q_k$ corresponds to the same orientation and produces the same expectation value $\expval{M}$. We note that this holds for both \verb|FQS| and \verb|Fraxis|. When we compute the distance along the great arch on the sphere between the new and old parameters, we also need to compute the distance from the opposite side of the sphere to the new point. The great circle distance between the opposite axes on the unit sphere is $\pi$, so we need to compute the minimum of these two distances. That is, we wrap the distance $d_{wrap} = \pi \ - \ d_g(\q_{k,prev}, \q_{k, new})$. Then we choose a smaller value of these two as follows
\begin{equation}
    \D(\q_{k,prev}, \q_{k, new}) = \min(d_g, d_{wrap}).
\end{equation}

This works for both \verb|Fraxis| and \verb|FQS| optimizer parameters. Next, we implement gate freezing by evaluating a matrix norm that determines whether a gate should be frozen.

\subsection{Gate Freezing with Matrix Norms} \label{matrix_norm_derivation_sec}

Now we implement a more generalized approach to gate freezing. Each single-qubit gate is a unitary matrix that acts on the given qubit. These unitaries correspond to rotations on a Bloch sphere. Suppose we have unitary matrices $U$ and $V$, representing the gate before and after optimization, respectively, where $U, V \in SU(2)$.
To quantify how much a gate changes in each optimization step, we use the Frobenius norm $\norm{U-V}_F$, which provides a matrix-level distance between the two gates. The Frobenius norm is defined as
\begin{equation}
    \norm{U-V}_F = \sqrt{\text{Tr}\bigl[(U-V)^\dagger(U-V)\bigr]}.
\end{equation}
We remark that in this work, we focus on single-qubit gates, which are $2 \times 2$ unitary matrices. However, this approach can be naturally extended to arbitrary $n$-qubit gates, as it is not limited by specific gate
parameters, unlike the approach discussed in Sec.~\ref{param_sec}. In the context of unitary $2\times2$ matrices, the expression simplifies to
\begin{equation}
    \norm{U - V}_F = \sqrt{ 4 - 2\cdot \Re\bigl(\Tr(U^\dagger V)\bigr)},
\end{equation}
using the fact that $\Tr(U ^\dagger U) = \Tr(V^\dagger V) = \Tr(I) = 2$ for $2\times 2$ unitary matrices $U$ and $V$. Since any global phase factor of $V$ does not influence the physical state, we can multiply $V$ by a complex phase $e^{i\delta}$ without altering observable outcomes. We exploit this degree of freedom to rotate the complex phase of the Frobenius inner product $\Tr(U^\dagger V)$ onto the positive real axis. Specifically, we select $\delta$ such that
\begin{equation}
\Tr(U^\dagger e^{i\delta}V) = e^{i\delta}\Tr(U^\dagger V) = |\Tr(U^\dagger V)|.
\end{equation}
Having forced the overlap to be a non-negative real number, its magnitude is now as large as possible. Thus, we derive the following expression for the distance between unitary matrices by defining
\begin{equation}
    \D(U,V) \equiv \sqrt{ 4 - 2\cdot |\Tr(U^\dagger V)|}.
\end{equation}
Finally, we normalize this matrix norm distance to be strictly in the range $[0,1]$. The expression inside the square root achieves its maximum value when $\Tr(U^\dagger V) = 0$. Therefore, we normalize the derived norm by dividing by a factor of 2, resulting in the normalized distance. Other matrix norms, such as the spectral norm, can also be used to measure the distance between given matrices $U$ and $V$. In this work, we employ the above formulation in our results. We note that the derived expression is related to the process fidelity. The process fidelity $F(U, V)$ between $U$ and $V$ is typically expressed as~\cite{process_fidelity, process_fidelity_ref2} 
\begin{equation}
F(U, V) =  \frac{1}{D^2} |\Tr(U^\dagger V)|^2,    
\end{equation}
where $D$ is the dimension of the Hilbert space. The process infidelity is then 
\begin{equation}
    \text{infidelity} = 1 - \frac{1}{D^2} |\Tr(U^\dagger V)|^2
\end{equation}
Additionally, the derived method can be extended to any $n$-qubit gates, where $n\geq 2$. Then we would need to compute the matrix norms for the $2^n \times 2^n$ matrices. We provide an additional formulation for the $n$-qubit gates in Appendix~\ref{unitary_norms_nqubit_gate_sec}. The gate freezing algorithm is described in Algorithm~\ref{matrix_norm_algorithm}, which can use any distance metric $\D$.

\begin{algorithm}
\vspace{0.1cm}
\caption{Gate freezing method for sequential single-qubit gate optimizers}\label{matrix_norm_algorithm}
\begin{algorithmic}[1]
\State \textbf{Inputs}: A Parameterized Quantum Circuit $U$ with fixed architecture, Hermitian measurement operator as the cost function, a stopping criterion, distance metric $\D$, and an optimizer (e.g. \verb|Rotosolve|).
\State Initialize gate freeze iterations $\kappa \in \mathbb{Z}_+$
\State Initialize the parameter vector $\bm{\theta} = (\theta_1, \theta_2,\ldots,\theta_{Ln})$.
\State  Initialize a fixed freezing threshold $T$.
\Repeat
    \For{$d = 1, \ldots, Ln$}
        \If{$d$-th gate is not frozen}
        \State Fix all gates except the $d$-th one
        \State Optimize the $d$-th gate $R_d$ with given optimizer.
        \State Compute $\Delta R_d \leftarrow \D \left(R_{d,prev} ,R_{d,new} \right)$
        \If {$\Delta R_d < T$ }
        \State Freeze $d$-th gate for $\kappa$ iterations
        \EndIf
        \EndIf
    \EndFor
\Until{stopping criterion is met}
\end{algorithmic}
\end{algorithm}

\subsection{Incremental Increase to Freezing Iterations}\label{gate_freeze_iters_result_sec}

Now that we have created algorithm-specific gate freezing algorithms, we also introduce an algorithm that tunes the freezing iteration hyperparameter $\kappa$ during optimization. Before starting optimization, we initialize a gate freezing iteration vector $\bm{\kappa}$, whose length is equal to the length of the parameter vector $\bm{\theta}$. Each element $\bm{\kappa}_{d}$ is set to 1 for all $d$. During optimization, if the $d$-th gate is frozen, then we increment the element $\bm{\kappa}_d$ by one. This means that the next time the $d$-th gate is frozen, its freezing time increases by one iteration. That is, every time any gate is frozen, its freezing time increases linearly. The heuristic behind this idea is to increasingly penalize the optimization of gates that consistently fall below the freezing threshold, thereby allocating more resources to poorly optimized gates.

We define the incremental increase of the freezing iteration in the Algorithm \ref{incremental_freeze_algo}. The algorithm needs a specific distance metric $\D$ and a threshold $T$ for the given optimizer.

\begin{algorithm}
\caption{Incremental increase to freezing iterations for sequential single-qubit gate optimizers}\label{incremental_freeze_algo}
\begin{algorithmic}[1]
\State \textbf{Inputs}: A Parameterized Quantum Circuit $U$ with fixed architecture, Hermitian measurement operator as the cost function, a stopping criterion, a distance metric $\D$, and an optimizer (e.g. \verb|Rotosolve|).
\State Initialize the parameter vector $\bm{\theta} = (\theta_1, \theta_2,\ldots,\theta_{Ln})$.
\State Initialize gate freeze iterations vector $\bm{\kappa}$, where $\bm{\kappa}_d = 1$ for $d = 1, \ldots, Ln$.
\State  Initialize a fixed freezing threshold $T$.
\Repeat
    \For{$d = 1, \ldots, Ln$}
        \If{$d$-th gate is not frozen}
        \State Fix all gates except the $d$-th one
        \State Optimize the $d$-th gate $R_d$ with given optimizer.
        \State Compute $\Delta R_d \leftarrow \D \left(R_{d,prev} ,R_{d,new} \right)$
        \If {$\Delta R_d < T$ }
        \State Freeze $d$-th gate for $\bm{\kappa}_d$ iterations
        \State $\bm{\kappa}_d \leftarrow \bm{\kappa}_d +  1$
        \EndIf
        \EndIf
    \EndFor
\Until{stopping criterion is met}
\end{algorithmic}
\end{algorithm}

\section{Results} \label{Results_section}
In this section, we examine the results of our proposed method compared to the base versions of the covered optimizers, where gate freezing is not applied at all. First, we go through the results for the 5-qubit Heisenberg Hamiltonian in a one-dimensional lattice with periodic boundary conditions. This was done for all optimizers. Then, we show results with the incremental gate freeze iterations for all optimizers and the individual gate freezing amounts for the whole ansatz circuit. 

\begin{figure}
    \centering
    \[
    \Qcircuit @C=1.5em @R=.9em {
    & &\mbox{$L$ layers} & & & & \\
    & & & & & & & \\
     \lstick{\ket{0}_1} & \gate{R_X\Bigl(\theta_{ln + 1}\Bigr)} & \gate{R_Y\Bigl(\theta_{ln + 6}\Bigr)}  &  \ctrl{0} & \qw & \qw &  \meter\\
     \lstick{\ket{0}_2}&  \gate{R_X\Bigl(\theta_{ln + 2}\Bigr)} & \gate{R_Y\Bigl(\theta_{ln + 7}\Bigr)}  &  \ctrl{-1} & \ctrl{1} & \qw  &   \meter\\
     \lstick{\ket{0}_3}&  \gate{R_X\Bigl(\theta_{ln + 3}\Bigr)} & \gate{R_Y\Bigl(\theta_{ln + 8}\Bigr)}  &  \ctrl{0} & \ctrl{0} & \qw &   \meter\\
     \lstick{\ket{0}_4}&  \gate{R_X\Bigl(\theta_{ln + 4}\Bigr)} & \gate{R_Y\Bigl(\theta_{ln + 9}\Bigr)}  &  \ctrl{-1} & \ctrl{0} & \qw  &  \meter\\
     \lstick{\ket{0}_5}&  \gate{R_X\Bigl(\theta_{ln + 5}\Bigr)} & \gate{R_Y\Bigl(\theta_{ln + 10}\Bigr)}  &  \qw & \ctrl{-1} & \qw &  \meter  \gategroup{3}{2}{7}{6}{1.2em}{--} 
    }
    \]
    \cprotect\caption{Ansatz circuit design for \verb|Rotosolve| optimizer with 5 qubits.}
    \label{roto_gd_ansatz}
\end{figure}

\begin{figure}
    \centering
    \includegraphics[width=0.9\linewidth]{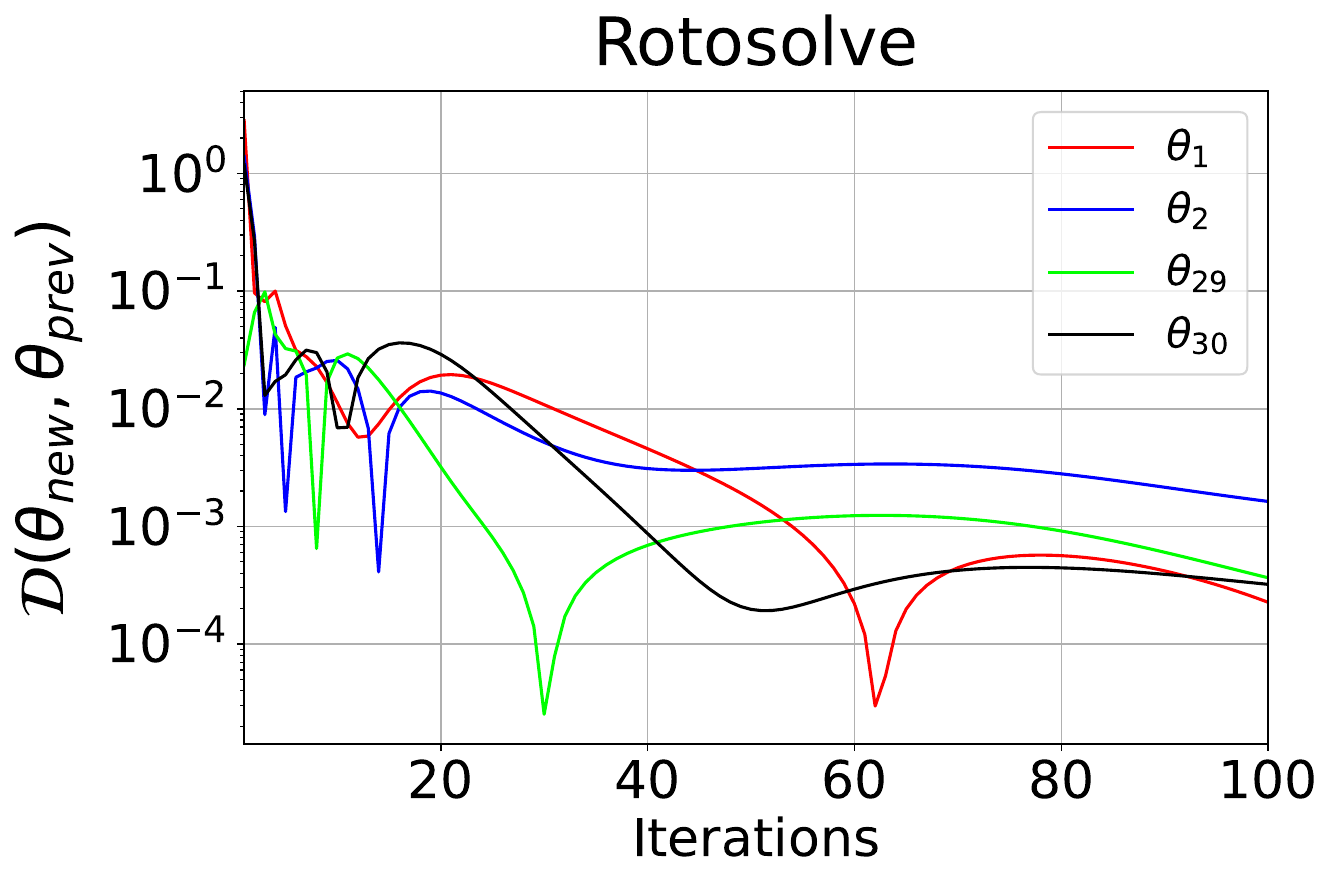}
    \includegraphics[width=0.9\linewidth]{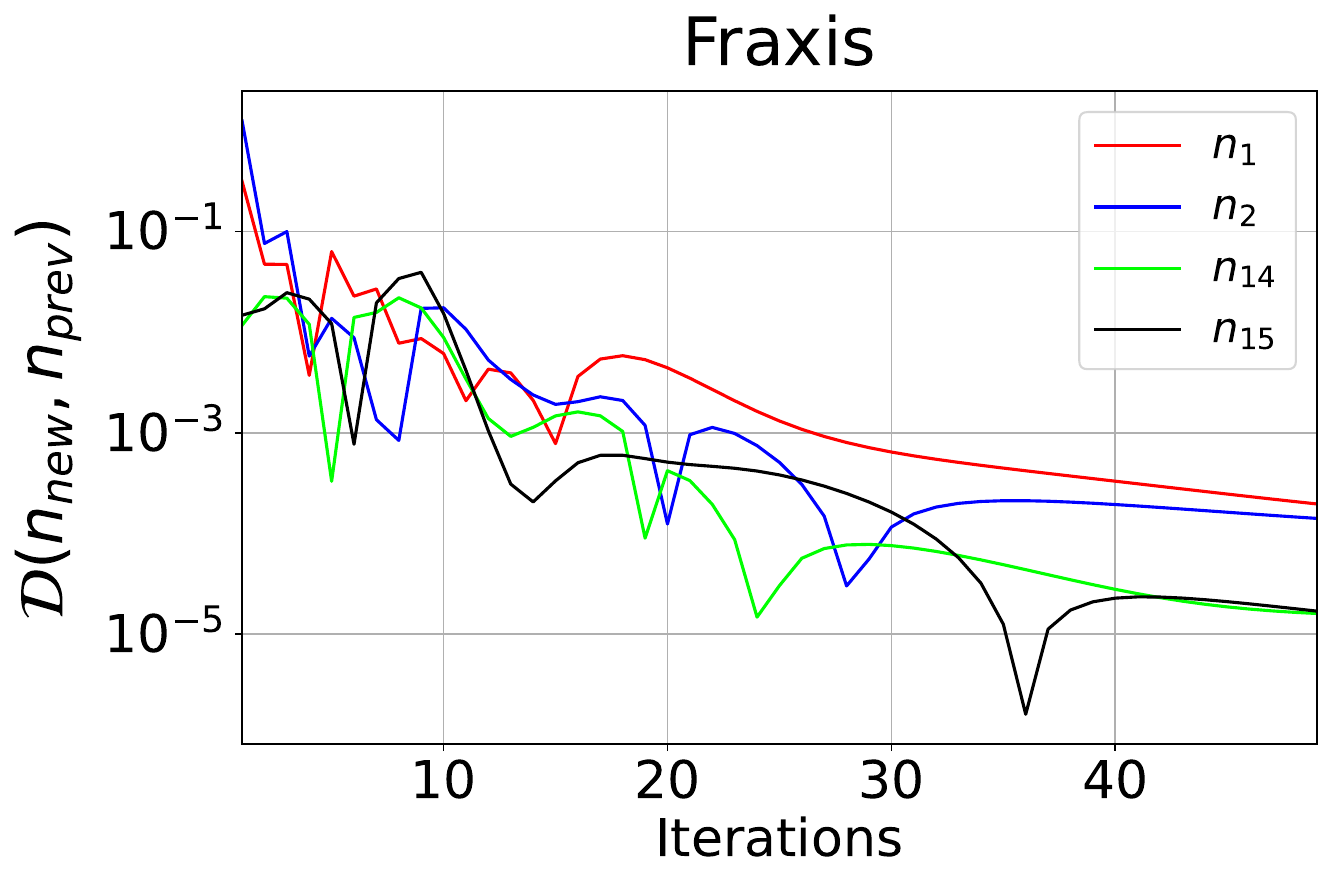}
    \includegraphics[width=0.9\linewidth]{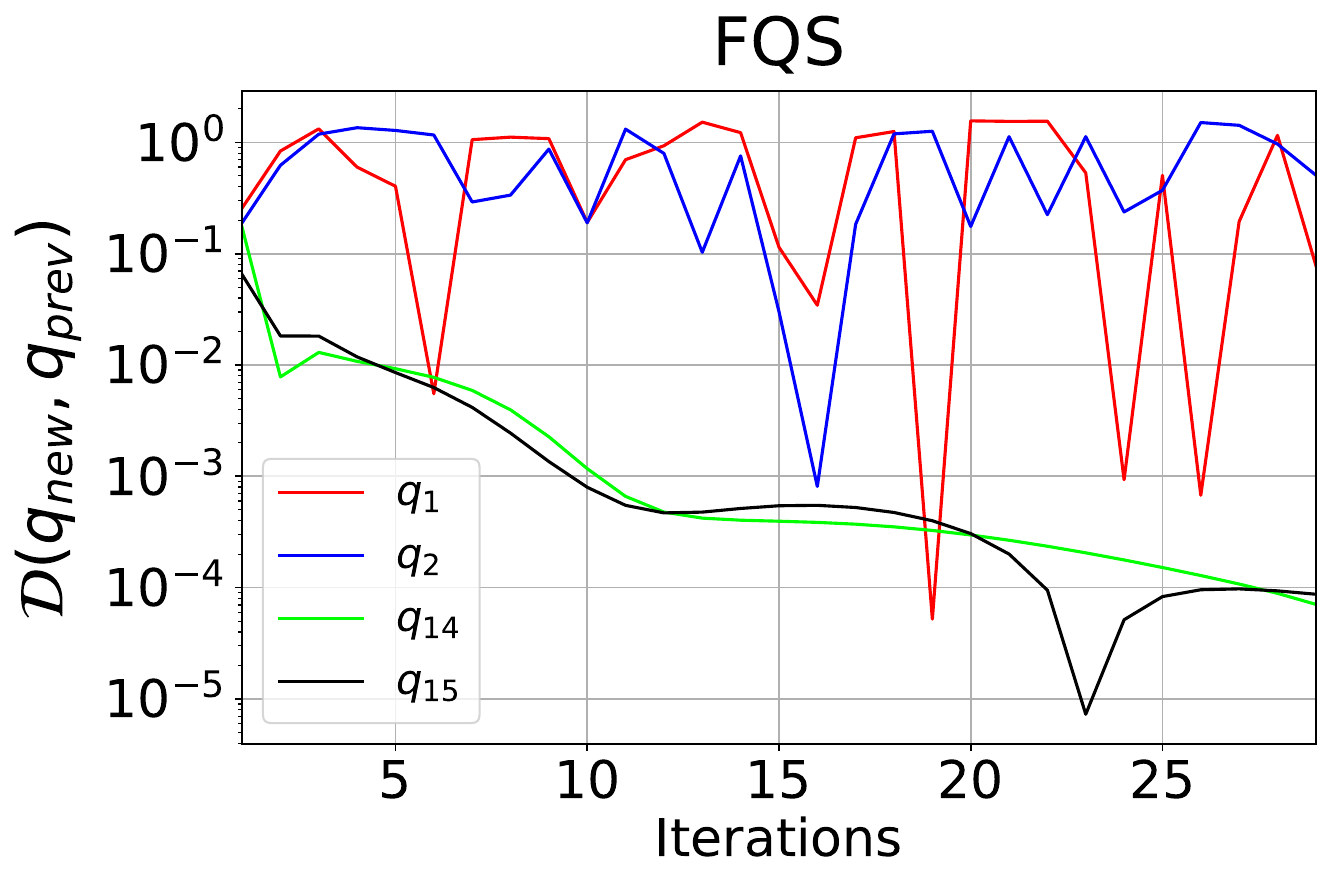}
    \cprotect\caption{Distance between previous and new parameter values over iterations for \verb|Rotosolve|, \verb|Fraxis|, and \verb|FQS| optimizers. The lines correspond to the evolution of the first two (red and blue) and the last two (green and black) gates' parameters in the PQC of one run for a 5-qubit system with 3 layers. The vertical axis is the parameter-based distance between parameter values.}
    \label{Freeze_individual_params_dist}
\end{figure}

\begin{figure*}
    \centering
    \includegraphics[width=0.32\linewidth]{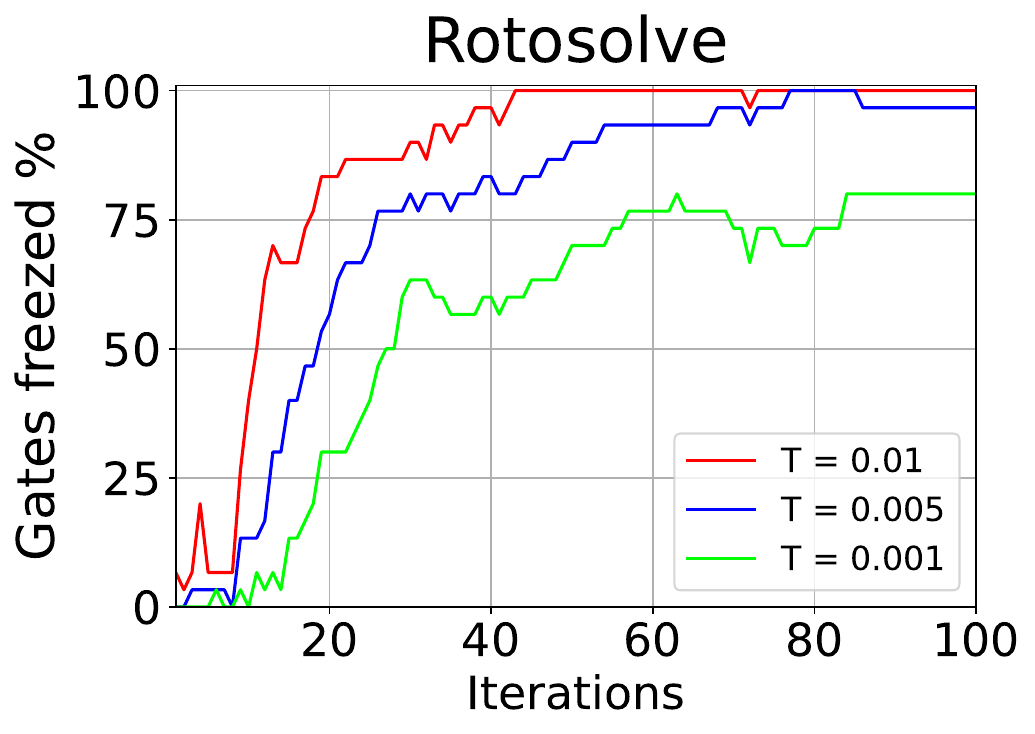}
    \includegraphics[width=0.32\linewidth]{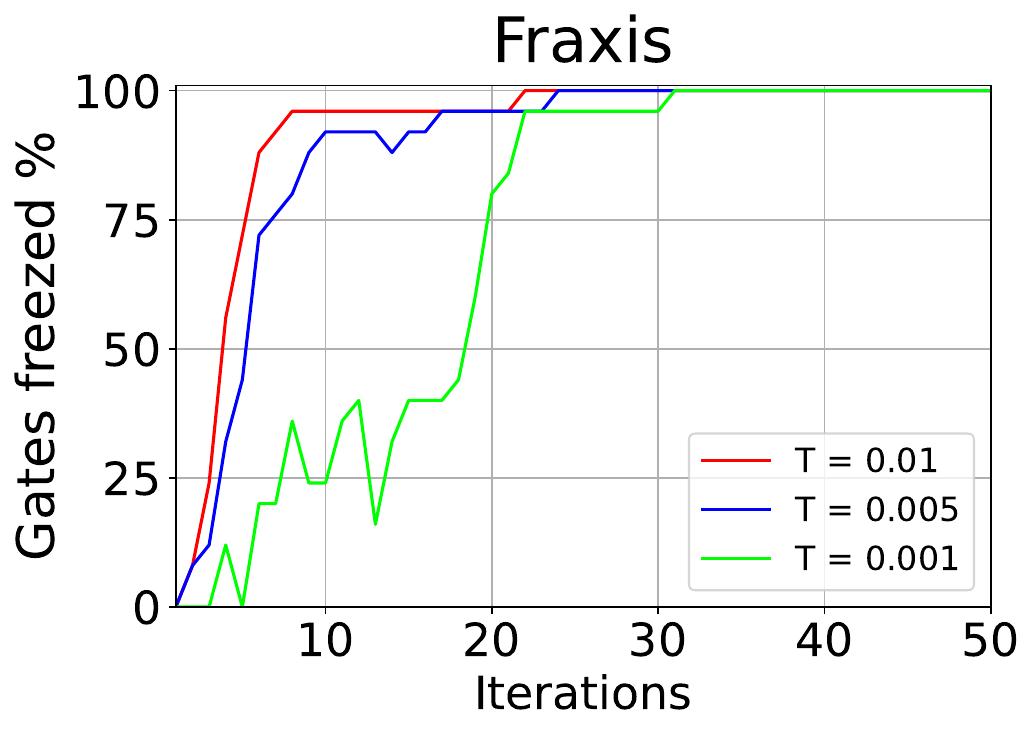}
    \includegraphics[width=0.32\linewidth]{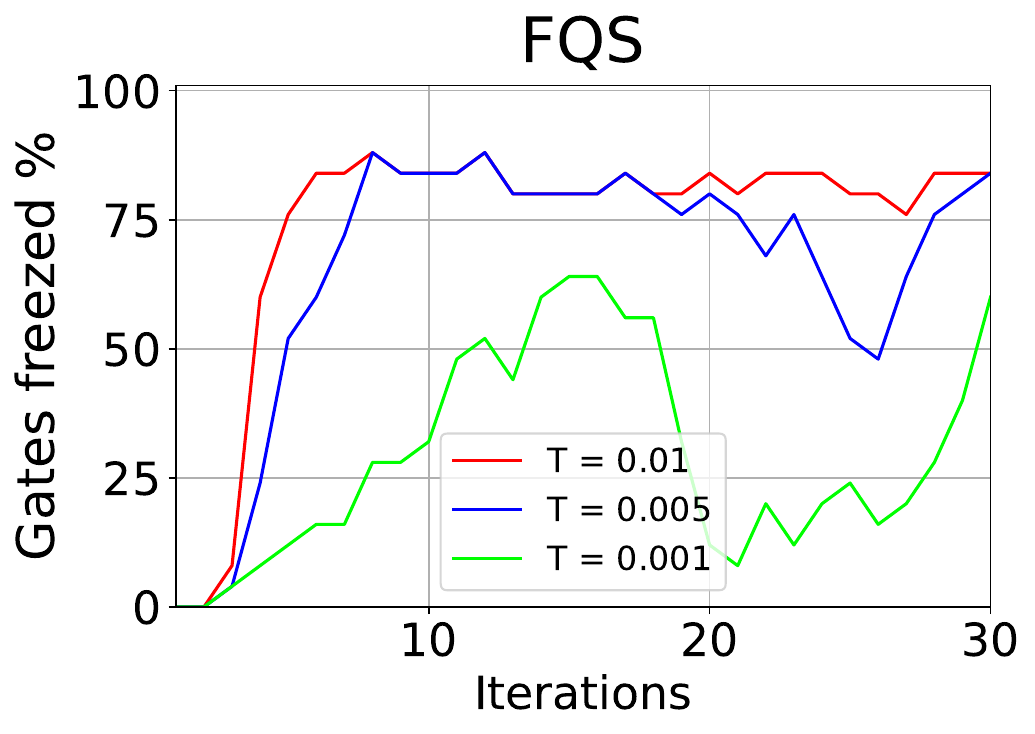}
    \cprotect\caption{The proportion of gates in PQC whose change in parameter values falls below the freezing threshold $T$ over iterations for \verb|Rotosolve|, \verb|Fraxis|, and \verb|FQS| optimizers. Each line corresponds to the percentage (\%) of gates that fall below the threshold $T$ in the PQC of one run. Threshold values $T$ were set to $T=0.01$ (red), $0.005$ (blue), and $0.001$ (green), and a parameter-based metric was used.}
    \label{Freeze_percent_fig}
\end{figure*}

Furthermore, we applied incremental gate freezing for the Fermi-Hubbard model with a 4-qubit system on the $1\times 2$ lattice. 

For the last experiment, we tested the performance of the incremental freezing method with increasing system size for the Heisenberg model using \verb|Rotosolve| optimizer. We used 6 to 9 qubits and set the circuit depth to scale linearly with system size, setting $L = 5n$. Finally, we examined the performance with deep random PQCs with depth of $L=10n$ layers.

In all experiments, we used the ansatz circuit from Fig.~\ref{roto_gd_ansatz} for \verb|Rotosolve| and the ansatz circuit from Fig.~\ref{Ansatz_circuit_image} for \verb|Fraxis| and \verb|FQS| unless otherwise stated. The gates in the circuits were optimized sequentially, layer by layer. We start from the top left of the circuit and then move down the circuit to the next qubit gate after optimizing the first gate. The optimization sequence in one layer is shown in Figs.~\ref{Ansatz_circuit_image} and~\ref{roto_gd_ansatz}. After optimizing all the gates in one layer, we move on to the next layer. This process is continued until all gates in the circuit are optimized once. We define this as one iteration. The base versions of the optimizers' stopping criterion are set to 10 iterations for all of the optimizers. This gives the total number of gate optimizations in one run, which is the number of iterations times the number of parameterized gates in the circuit. The gate freezing method algorithms, on the other hand, optimize the circuit until the set amount of these gate optimizations is reached to get comparable results with the base optimizers. 

The parameters for all optimizers are initialized uniformly. The parameters $\theta_d$ for the \verb|Rotosolve| are sampled from the uniform distribution $(-\pi, \pi]$, the \verb|Fraxis| parameters, unit axis $\n_d$, from the uniform spherical distribution and the parameters of \verb|FQS|, the unit quaternions $\q_d$, from the uniform spherical distribution in four dimensions.

In the optimization, we used the statevector simulator to simulate performance in deep PQCs with an ideal quantum device, free of measurement or hardware noise. Additionally, we examined the performance of the incremental gate freezing method under noisy conditions using shot noise from the measurements. We used 1024 and 4096 shots to approximate each Hamiltonian term, simulating shot noise, and employed shallow PQCs with a few layers. Unless otherwise stated, we use the statevector simulator in experiments without the shot noise. 

We tested the performance of the gate freezing method, where in all experiments the threshold $T$ was heuristically set to one of the following values: $T = 0.01, 0.005, 0.001$ for \verb|Rotosolve| and $T=0.025, 0.01, 0.005$ for \verb|Fraxis| and \verb|FQS| optimizers. We choose these threshold values heuristically by examining how the parameters of individual gates change over time in one run and the proportion of gates in PQC that fall below the freezing threshold $T$. The change in parameter values over the consecutive iterations is illustrated in Fig.~\ref{Freeze_individual_params_dist}. We observed a similar behavior for both parameter-based and matrix norm metrics with \verb|Fraxis| and \verb|FQS| optimizers. Also, the behavior was similar to the Fermi-Hubbard Hamiltonian for all optimizers and metrics. Figure~\ref{Freeze_percent_fig} illustrates the percentage of parameterized gates in the circuit whose parameter distance between consecutive iterations falls below the threshold $T$ as a function of iterations performed. Here, the parameter-based metric was used to compute the distance between consecutive iterations, and a 5-qubit Heisenberg Hamiltonian with 5 layers was employed for all optimizers. With $T=0.01$ and $0.005$, the parameter distance between consecutive iterations for nearly all parameterized gates in PQC falls below the threshold value for both \verb|Rotosolve| and \verb|Fraxis|. For \verb|FQS|, the corresponding percentage plateaus around 80\%. In summary, the parameterized gates of the first layer are rarely frozen when using the \verb|FQS|, which explains the plateauing in Fig.~\ref{Freeze_percent_fig}. For $T=0.001$, \verb|Fraxis| also achieves a nearly fully frozen circuit, but it takes more iterations to reach that point than for the larger threshold values. In contrast, \verb|Rotosolve| and \verb|FQS| do not reach a fully frozen PQC with $T=0.001$.

The values for the freeze iteration parameter were set to $\kappa=2,5$ for \verb|Rotosolve|. For \verb|Fraxis| and \verb|FQS|, we set $\kappa=3$. In all experiments, 20 runs were performed for each choice of gate freezing parameter $\kappa$ and threshold $T$, as well as for the corresponding baseline optimizer without gate freezing. These baseline optimizers serve as a reference for evaluating the effectiveness of gate freezing methods. Each run consisted of 10 iterations for all optimizers. The same experiments were repeated for the incremental gate freezing algorithm using all optimizers. In all the following figures in this section, each line represents a mean of 20 runs, and the shaded area represents a 90\% confidence interval around the mean unless otherwise stated. The box plots summarize the results at the end of 20 runs. The boxes span from the first quartile (Q1) to the third quartile (Q3); the horizontal bar within each box indicates the median, the diamond marker denotes the mean, and the whiskers represent values within 1.5 times the interquartile range (IQR).

In this work, we used the PennyLane package version $0.40$~\cite{pennylane}, and Python version $3.11$ to generate the data for the plots.

\begin{figure*}
    \includegraphics[width=0.85\linewidth]{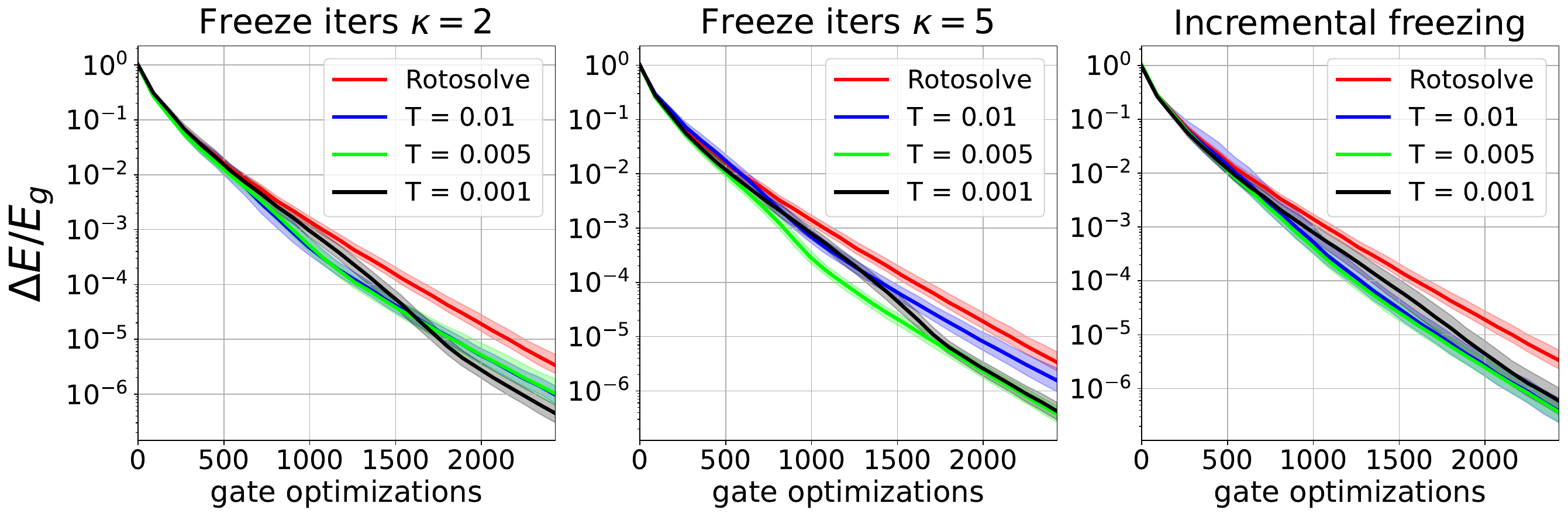}
    \includegraphics[width=0.85\linewidth]{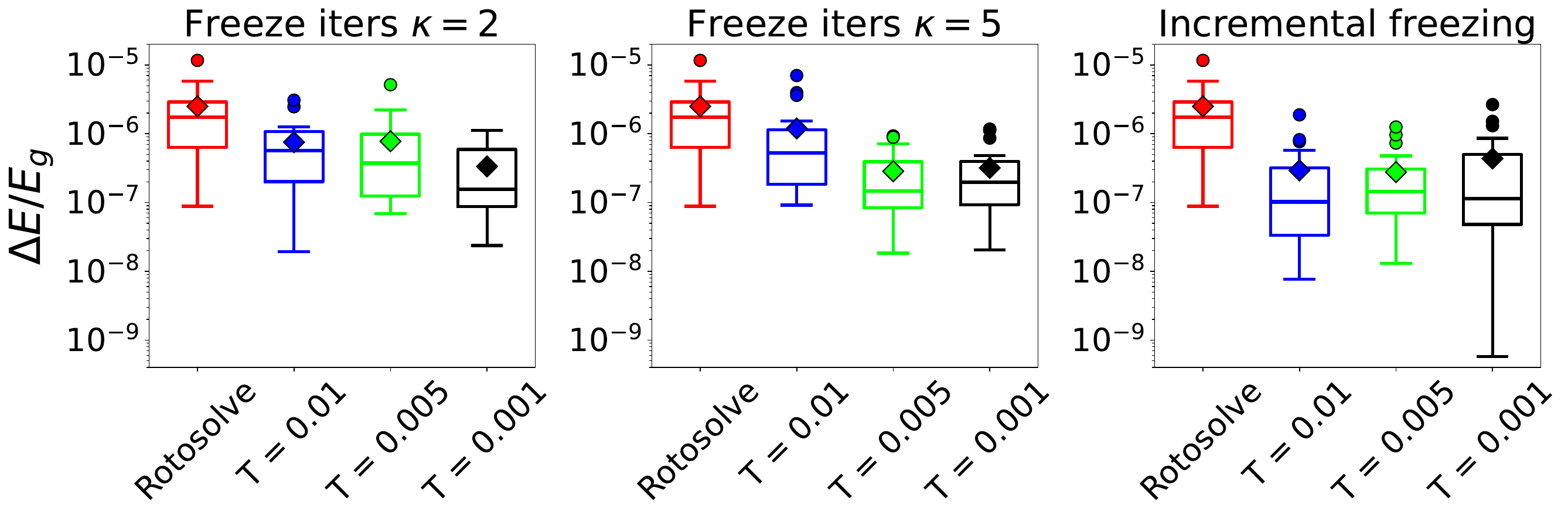}
    \cprotect\caption{Results for 5-qubit Heisenberg model with base \verb|Rotosolve| (red) and gate freezing method with the parameter-based distance metric for $L=5n$ layers. Gate freezing threshold was set to $T = 0.01$ (blue), $0.005$ (green), $0.001$ (black). Freeze iterations $\kappa=2, 5$ and incremental gate freezing were used.}
    \label{rotosolve_heisenberg}
\end{figure*}

\begin{figure}
    \includegraphics[width=0.99\linewidth]{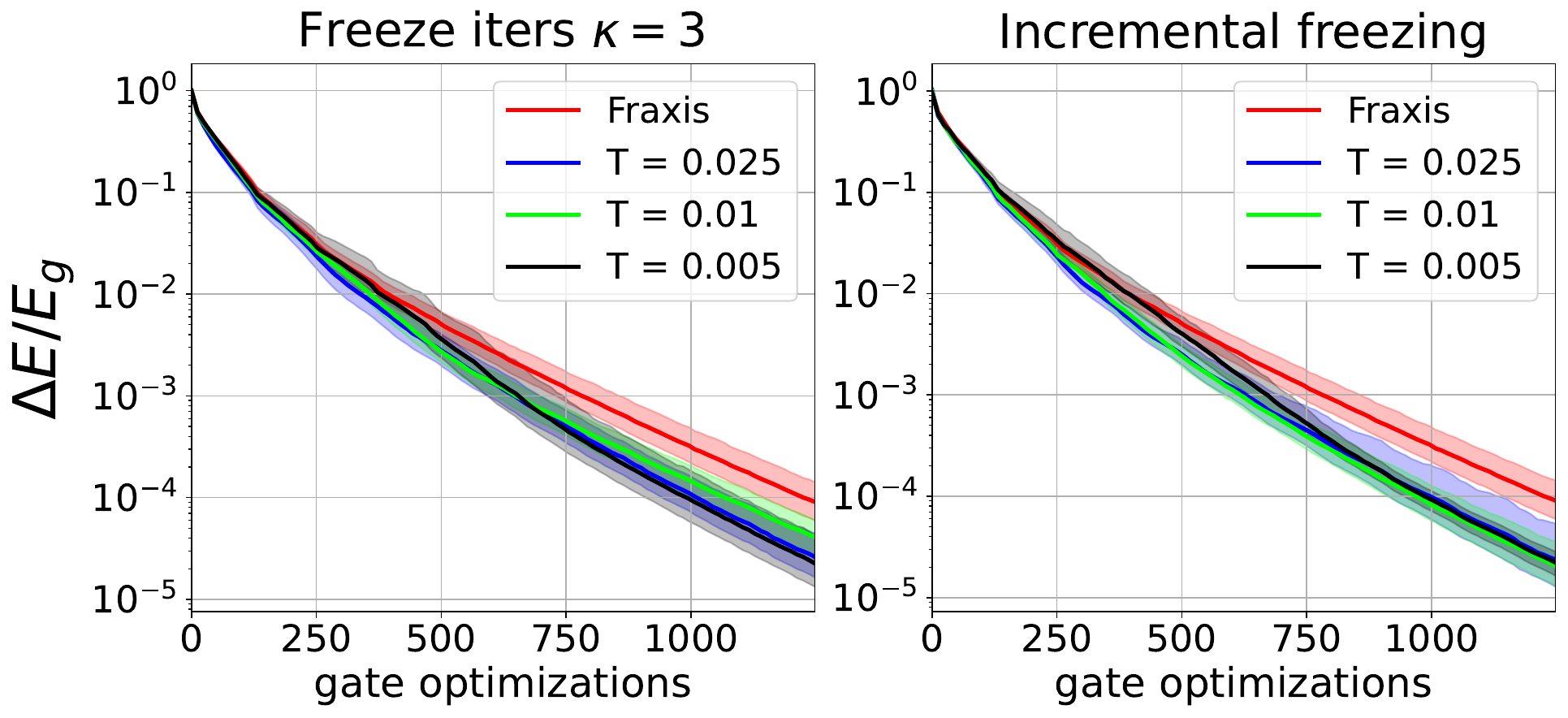}
    \includegraphics[width=0.99\linewidth]{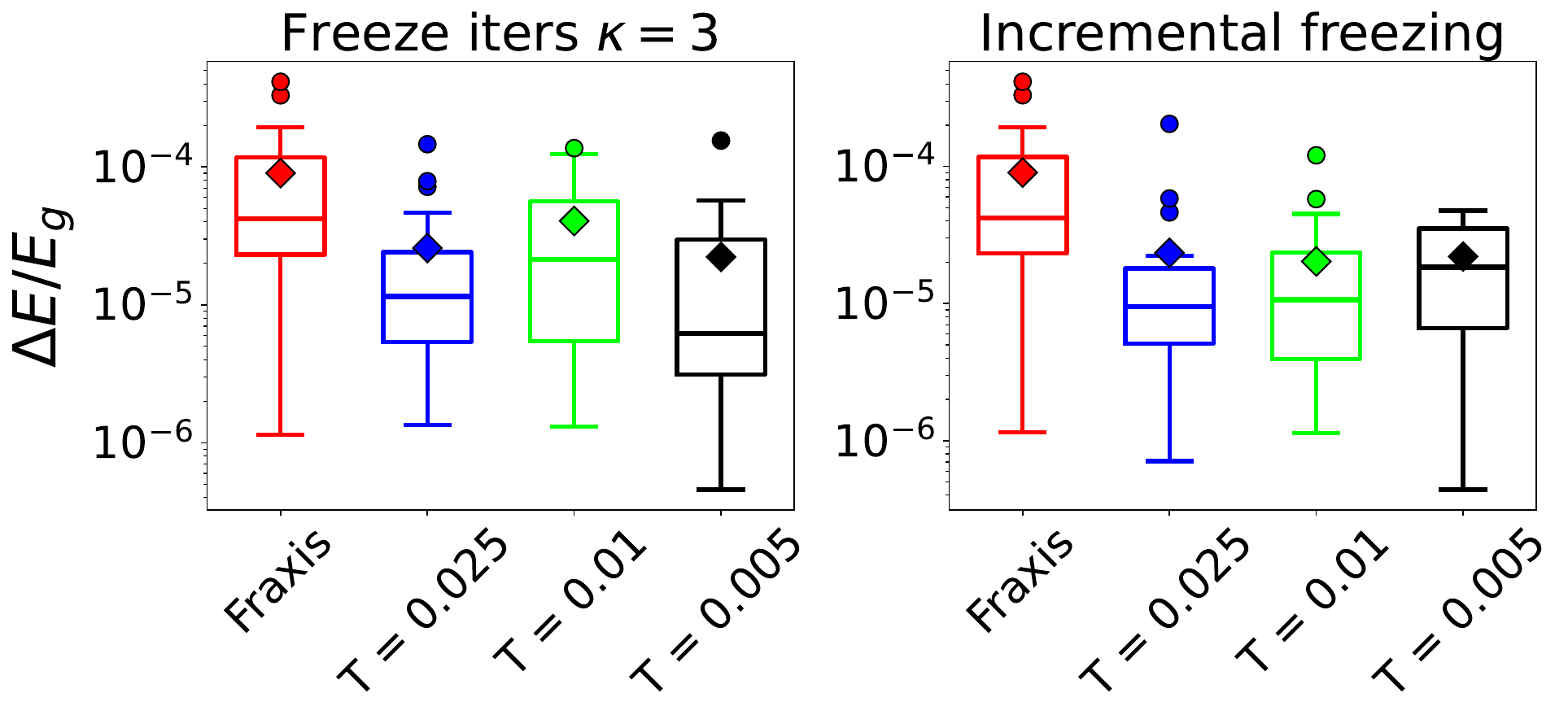}
    \cprotect\caption{Results for 5-qubit Heisenberg model with base \verb|Fraxis| (red) and gate freezing method with the parameter-based distance metric for $L=5n$ layers. Gate freezing threshold was set to $T = 0.025$ (blue), $0.01$ (green), $0.005$ (black). Freeze iterations $\kappa=3$ and incremental gate freezing were used.}
    \label{fraxis_heisenberg}
\end{figure}

\begin{figure}
    \centering
    \includegraphics[width=0.99\linewidth]{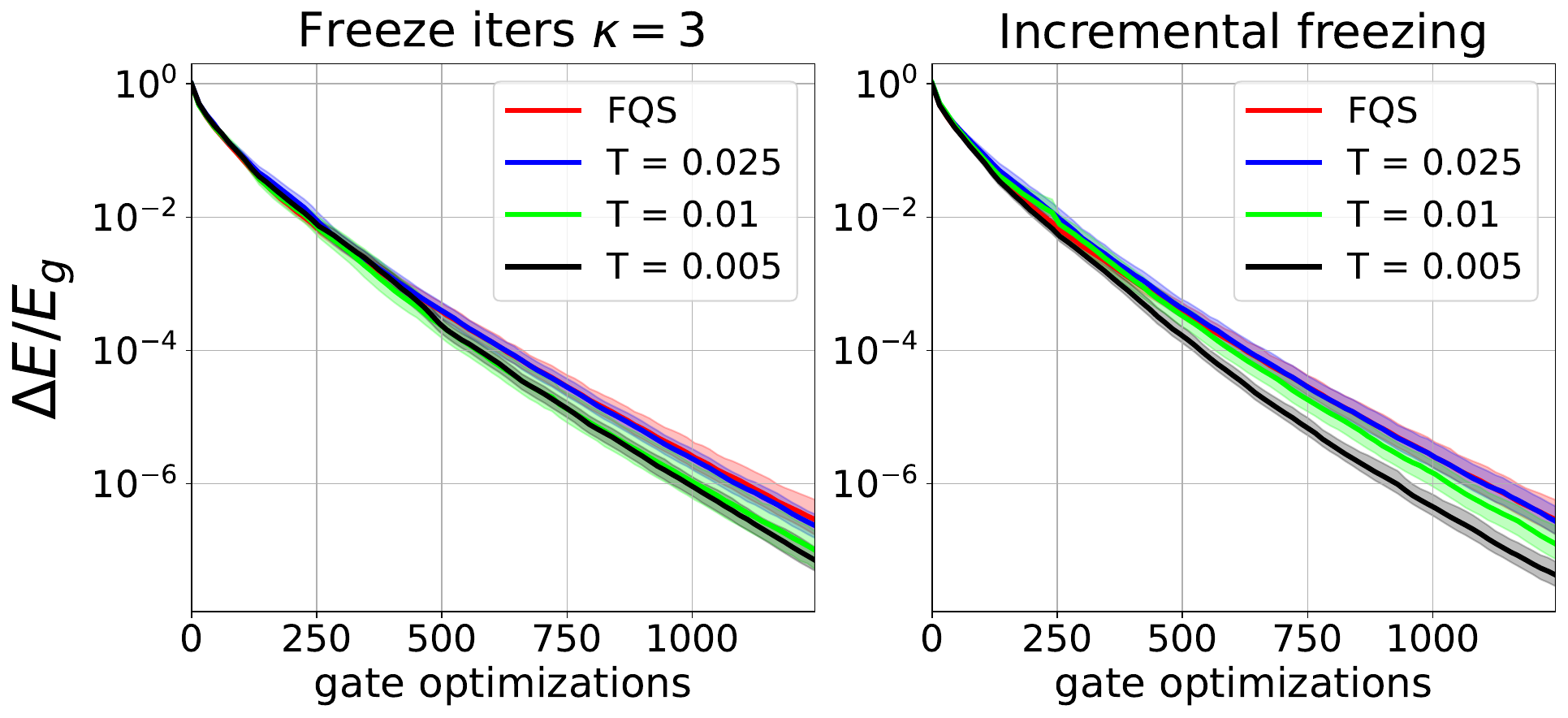}
    \includegraphics[width=0.99\linewidth]{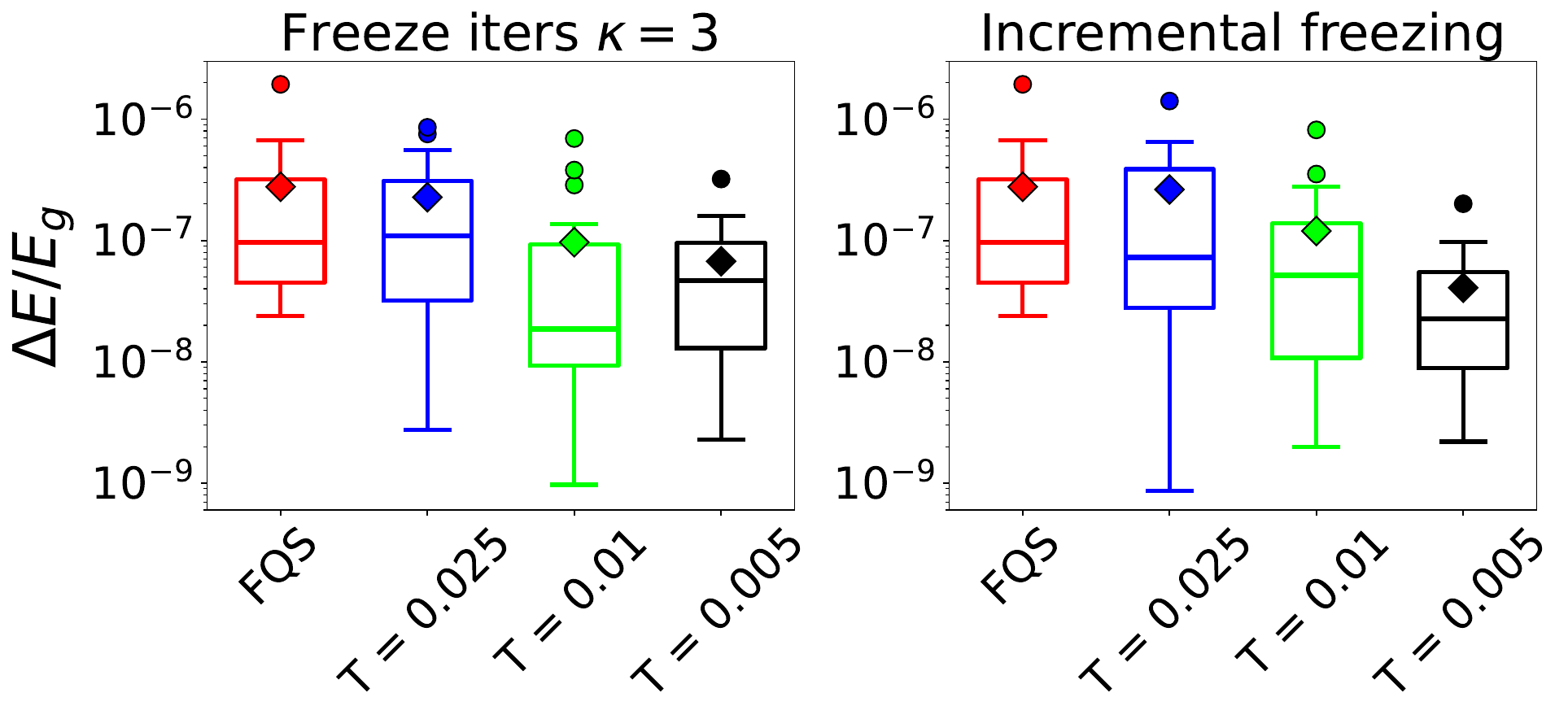}
    \cprotect\caption{Results for 5-qubit Heisenberg model with base \verb|FQS| (red) and gate freezing method with the parameter-based distance metric for $L=5n$ layers. Gate freezing threshold was set to $T = 0.025$ (blue), $0.01$ (green), $0.005$ (black). Freeze iterations $\kappa=3$ and incremental gate freezing were used.}
    \label{fqs_heisenberg}
\end{figure}

\subsection{Heisenberg Model} \label{Heisenberg_section}
Next, we present results of the proposed methods for the 5-qubit cyclic one-dimensional Heisenberg model~\cite{heisenberg_model}. The corresponding Hamiltonian is defined as 
\begin{equation}\label{Heisenberg_hamiltonian}
    H = J   \sum_{i=1}^n \left( X_i X_{i+1} +Y_iY_{i+1} + Z_i Z_{i+1} \right) + h\sum_{i=1}^n Z_i,
\end{equation}
where $J$ is the strength of the spin interaction and $h$ is the magnetic
field strength along the Z-axis. We set $J=h=1$ in this work.

In addition to finding the ground state of the Hamiltonian with the optimizers, we also examine the values the gate freeze iteration vector $\bm{\kappa}$ has after optimizing for each element $\bm{\kappa}_d$ without applying the gate freezing to the gates. The number of layers is set to $L=5n$, which is 25 layers in the 5-qubit system. The ground state of the Hamiltonian is approximately $E_g=-8.47$, and we examined the relative error $\Delta E/E_g$ to the ground state energy, where $\Delta E= \expval{M} - E_g$.

We provide additional results for the incremental gate freezing method in Appendix~\ref{appendix_threshold_counts} using different ansatz circuits from Appendix~\ref{appendix_circuits}.

\subsubsection{Parameter Distances} \label{parameter_results_section}

The results for gate freezing with iterations $\kappa=2,5$ and incremental gate freezing with \verb|Rotosolve| are shown in Fig.~\ref{rotosolve_heisenberg}. In all subplots in Fig.~\ref{rotosolve_heisenberg}, the gate freezing method shows improved convergence compared to base \verb|Rotosolve| in terms of convergence speed and better mean values across the 20 runs. When $\kappa=2$, the $T=0.001$ performs the best and has the best convergence and distribution of runs at the end of the optimization. With the $\kappa = 5$ threshold, $T=0.005$ achieved the same convergence and distribution of runs as $T=0.001$. Subsequently, the next most effective freezing threshold was $T=0.01$ across different values of $\kappa$. In all experiments, the base \verb|Rotosolve| without gate freezing had the worst performance of all. Here, we also emphasize the trend that the gate freezing methods with different threshold values exhibit a faster convergence towards the ground state in the semilog plot in Fig.~\ref{rotosolve_heisenberg}. Additionally, we note that after 10 iterations, the mean is nearly one magnitude better than with the base \verb|Rotosolve|.

For the \verb|Fraxis| optimizer, we obtained similar results as for the \verb|Rotosolve| optimizer. For \verb|Fraxis| we used one optimizable gate for each qubit in a layer instead of the pair gates for the ansatz circuit shown in Fig.~\ref{roto_gd_ansatz}. Otherwise, the structure stays the same. The results are fully shown in Fig.~\ref{fraxis_heisenberg}, where we used $L=5n$ layers with 5 qubits, and the freeze iterations $\kappa$ were set to $\kappa=3$. The results also include the incremental gate freezing method in its subplot. In Fig.~\ref{fraxis_heisenberg}, the gate freezing methods with different threshold values have a similar trend to the results of \verb|Rotosolve|, where gate freezing methods have faster convergence of the mean towards the ground state on the semilog plot. With fixed freezing iterations $\kappa=3$, the smallest threshold value $T=0.005$ has the best mean and boxplot distribution, followed by $T=0.025$ and $T=0.01$, respectively. Incremental freezing exhibits similar performance across threshold values, both in the mean convergence and in the boxplots at the end of the optimization across 20 runs.

For \verb|FQS|, we employed a 5-qubit system and $L=5n$ layers, freezing iterations $\kappa=3$ and incremental gate freezing, as well as the same ansatz circuit used with the \verb|Fraxis| optimizer. The results for \verb|FQS| are shown in Fig.~\ref{fqs_heisenberg}. In contrast to the results for \verb|Rotosolve| and \verb|Fraxis|, the base version of \verb|FQS| performs relatively well compared to the results for \verb|Rotosolve| and \verb|Fraxis|. Regardless, the gate freezing method with $\kappa=3$ obtains slightly better convergence of the mean and better distribution towards the ground state. Here, the best results are obtained with $T=0.01$ or $T=0.005$. The base \verb|FQS| performs similarly to the gate freeze with parameter value $T=0.025$. Incremental freezing has the best results with $T=0.005$, with clearly the best mean and median values, followed by $T=0.01$.

\begin{figure}
    \centering
    \includegraphics[width=0.99\linewidth]{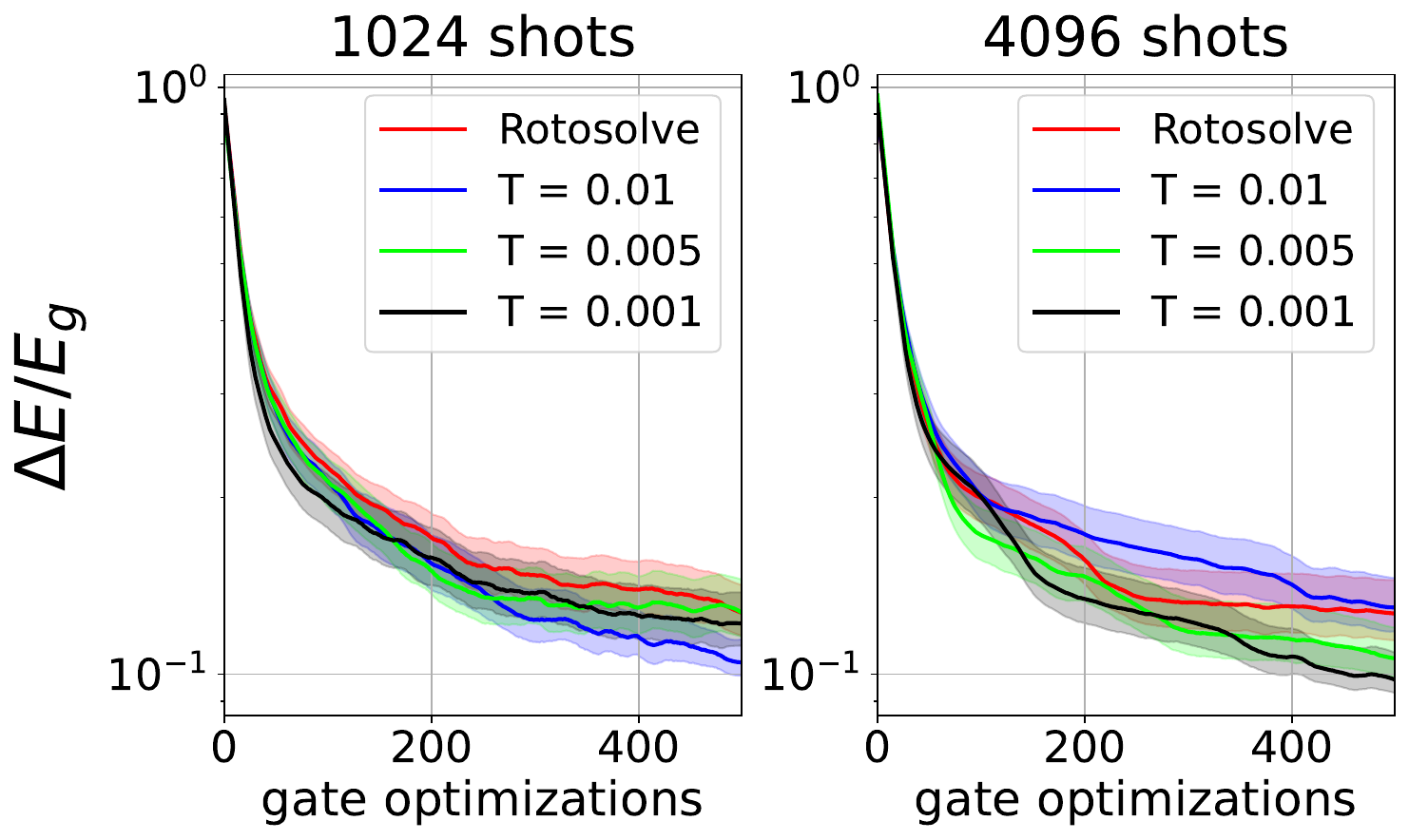}
    \includegraphics[width=0.99\linewidth]{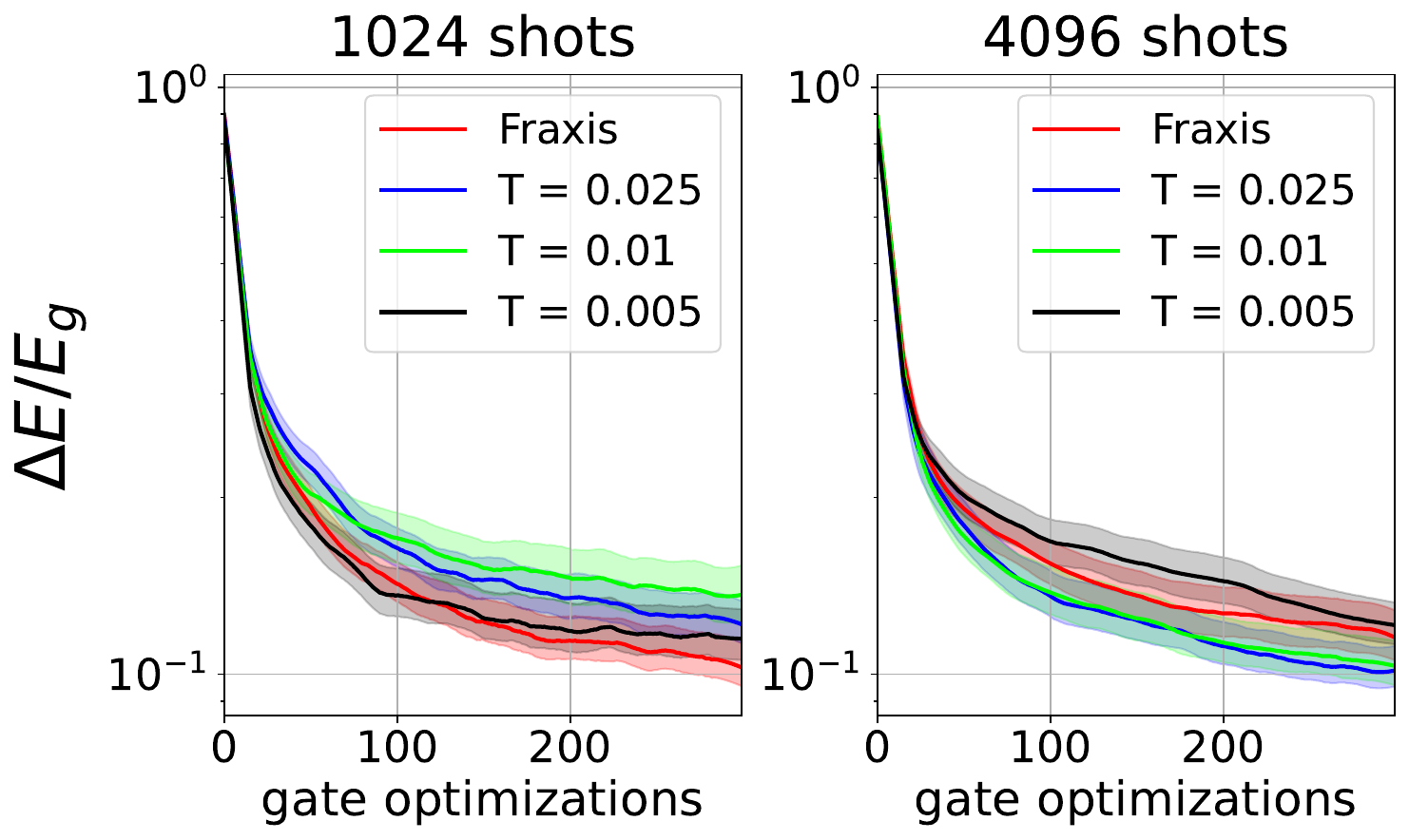}
    \includegraphics[width=0.99\linewidth]{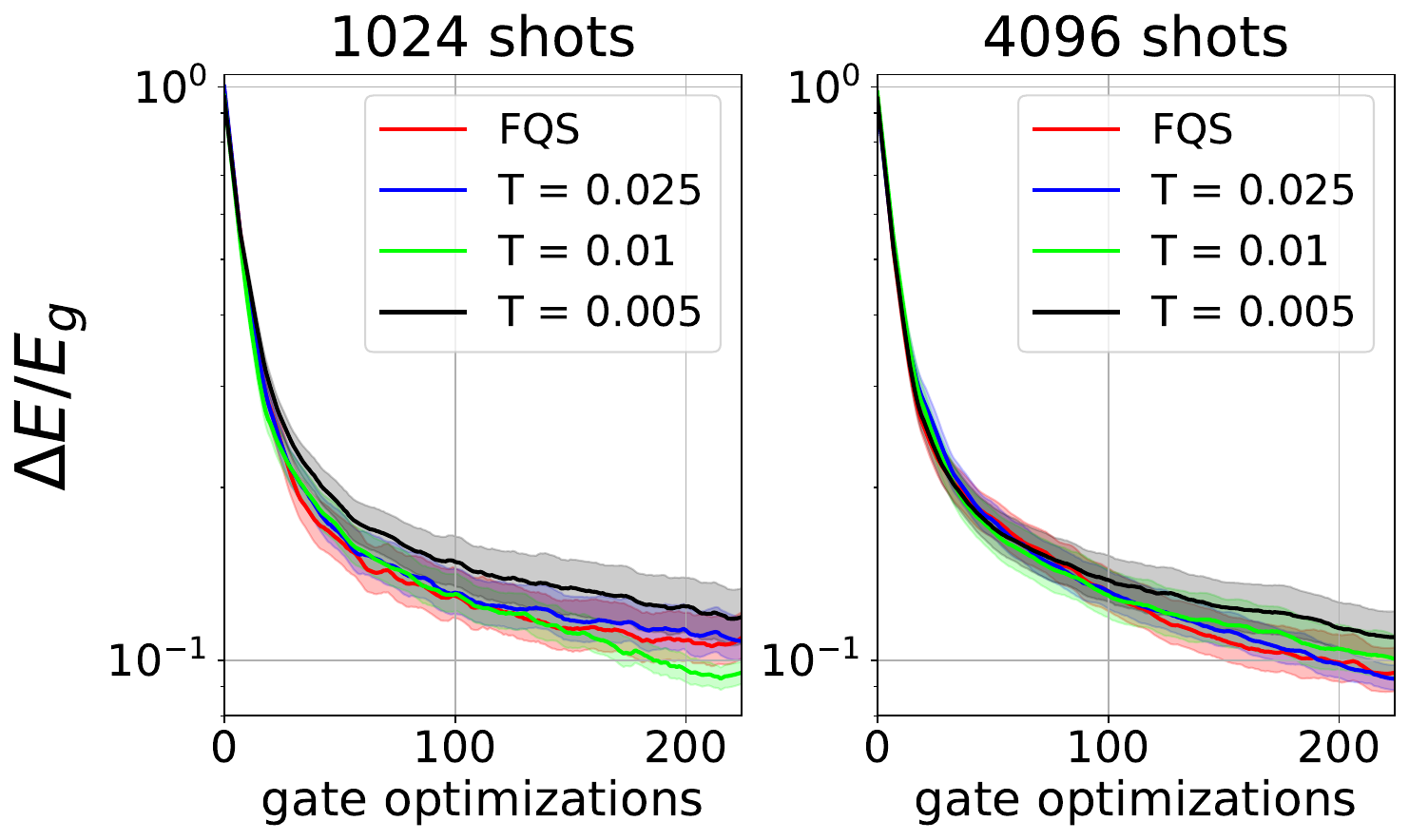}
    \cprotect\caption{Performance of the incremental gate freezing with 5-qubit Heisenberg Hamiltonian for \verb|Rotosolve|, \verb|Fraxis|, and \verb|FQS| optimizers with 2, 3, and 3 layers, respectively. 1024 and 4096 shots were used to approximate each Hamiltonian term to simulate the shot noise. Each line corresponds to a mean of 30 runs, and the shaded area 68\% confidence interval around the mean.}
    \label{1DHeisenberg_noisy}
\end{figure}

Finally, we discuss the results for the incremental gate freezing method with shot noise for the measurements. We conducted the experiments with 30 runs, and the shot counts were set to 1024 and 4096 for \verb|Rotosolve|, \verb|Fraxis|, and \verb|FQS| optimizers. In Fig.~\ref{1DHeisenberg_noisy}, \verb|Rotosolve| exhibits clear improvement with the incremental gate freezing. When using 1024 shots, a larger gate freezing threshold benefits the most, and with 4096 shots, a smaller threshold $T$ is better for the optimization. In both cases, there is an advantage to gate freezing, but it depends on the number of shots used in the measurements and which value of $T$ yields the best results. \verb|Fraxis| optimizer on the other hand only benefits the gate freezing with 4096 shots, but not when using 1024 shots. However, for the \verb|FQS|, we do not see a significant improvement, but rather a small advantage over the base version with 1024 shots and $T=0.01$. In addition, we experimented with deep circuits in the presence of shot noise but did not observe any improvement regardless of the threshold $T$. This behavior is consistent with the optimization difficulty of deep PQCs, which includes the barren plateau-related effects as the number of layers increases~\cite{mcclean2018barren, barren_plateau_deep_pqc}. In deep PQCs, the loss differences can become exponentially small. In the presence of shot noise, the optimization of deep PQC becomes prone to statistical noise and may require an exponential number of shots to resolve meaningful loss differences~\cite{larocca2025barren}. If too few shots are used, the optimization of deep PQCs may result in a random walk in the cost function landscape. As barren plateaus are known to affect gradient-free optimizers~\cite{barren_plateau_gradientFree_opts}, this may explain the observations with deep PQCs when using shot noise in optimization.

Next, we show our results for the matrix norm as a distance measure.

\subsubsection{Matrix Norm Distances} \label{matrix_norm_results}

In this section, we use the matrix norm derived in Sec.~\ref{matrix_norm_derivation_sec} to measure the rotations of the unitaries on a Bloch sphere. Due to the simple nature of \verb|Rotosolve|, we now only consider \verb|Fraxis| and \verb|FQS|. Similarly to the previous section, we use $\kappa=3$ and the incremental freezing method for gate freezing. The threshold values $T$ are set to $0.025, 0.01$, and $0.005$. All runs were performed using 10 iterations per run for both optimizers. The ansatz circuit in Fig.~\ref{roto_gd_ansatz} with $L=5n$ layers and 5 qubits was employed. 

\begin{figure}
    \includegraphics[width=0.99\linewidth]{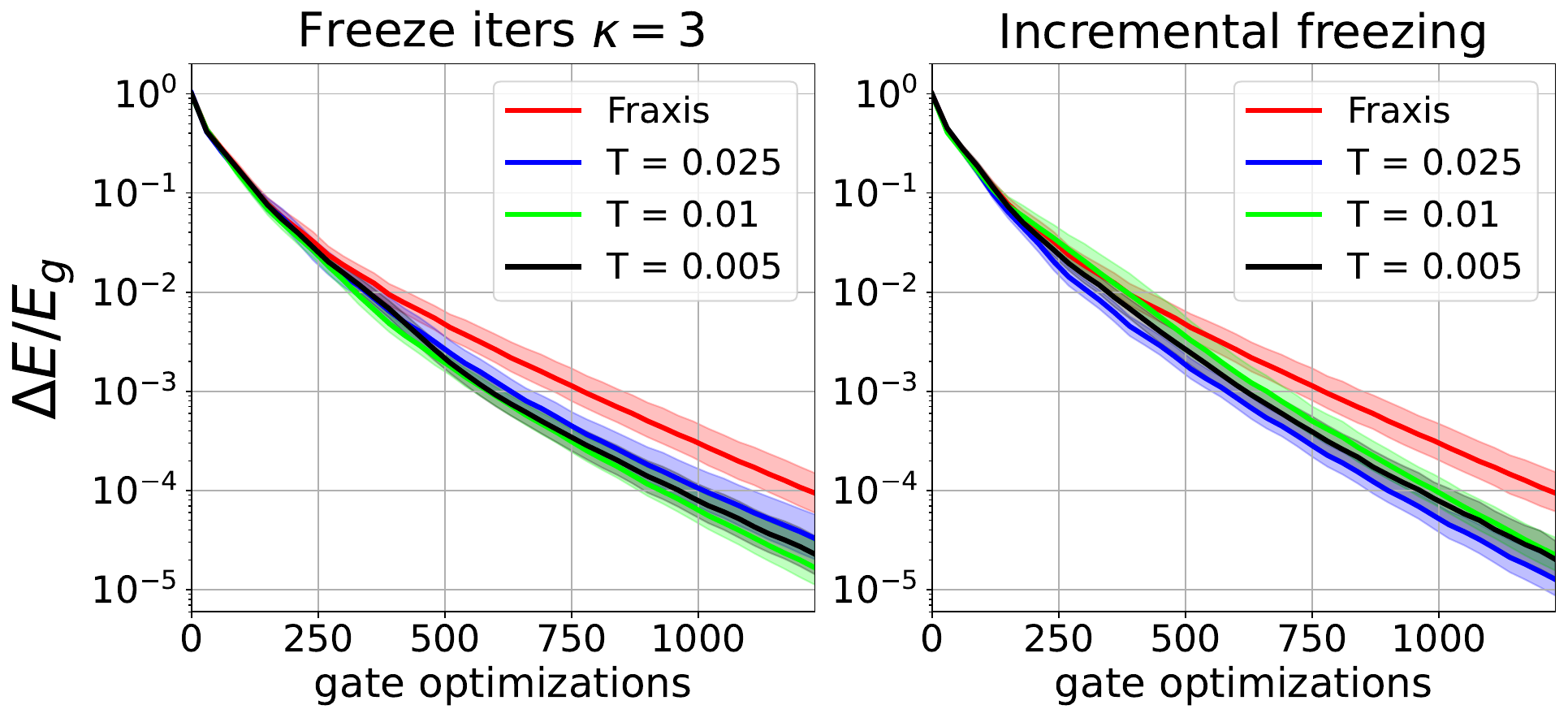}
    \includegraphics[width=0.99\linewidth]{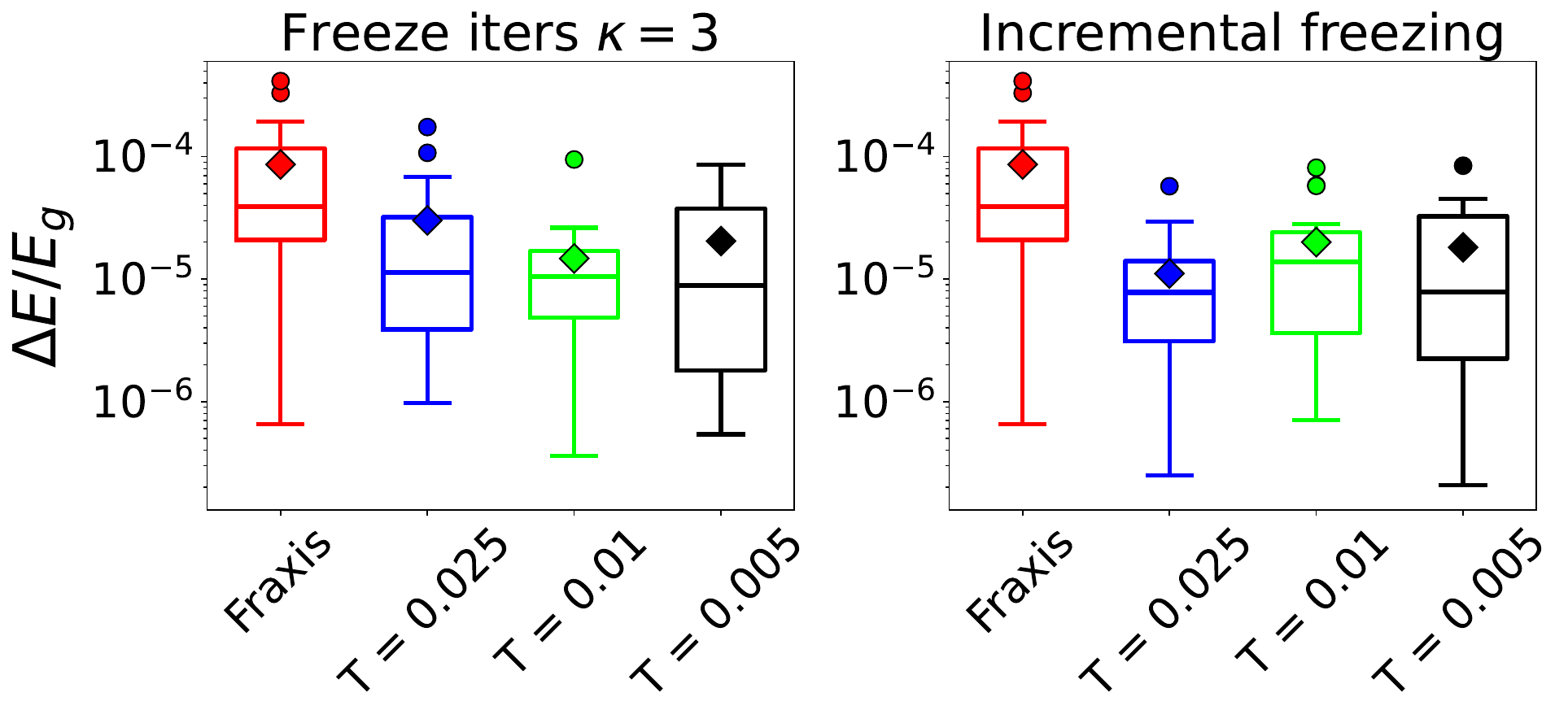}
    \cprotect\caption{Results for 5-qubit Heisenberg model with base \verb|Fraxis| (red) and gate freezing method with the matrix norm distance metric for $L=5n$ layers. Gate freezing threshold was set to $T = 0.025$ (blue), $0.01$ (green), $0.005$ (black). Freeze iterations $\kappa=3$ and incremental gate freezing were used.}
    \label{fraxis_matrix_heisenberg}
\end{figure}

\begin{figure}
    \includegraphics[width=0.99\linewidth]{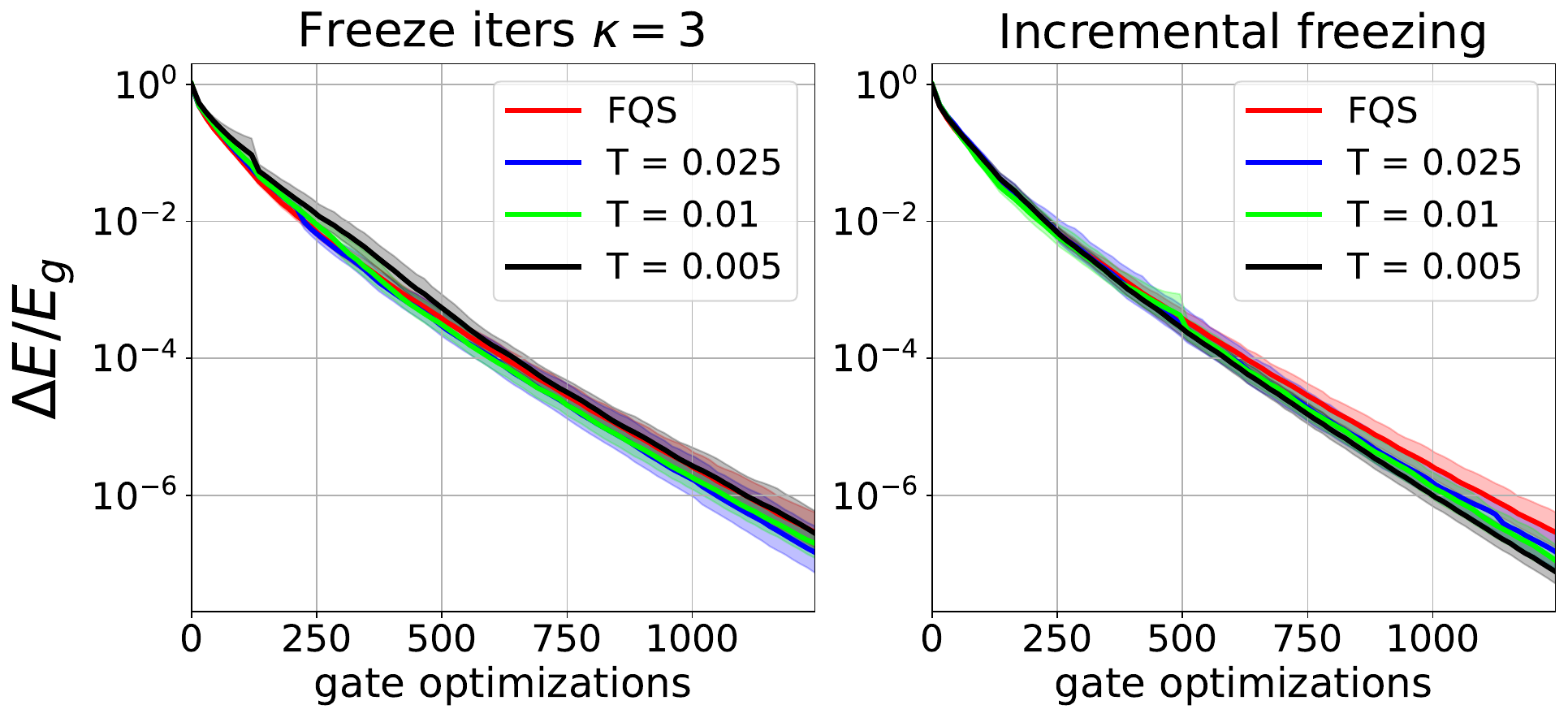}
    \includegraphics[width=0.99\linewidth]{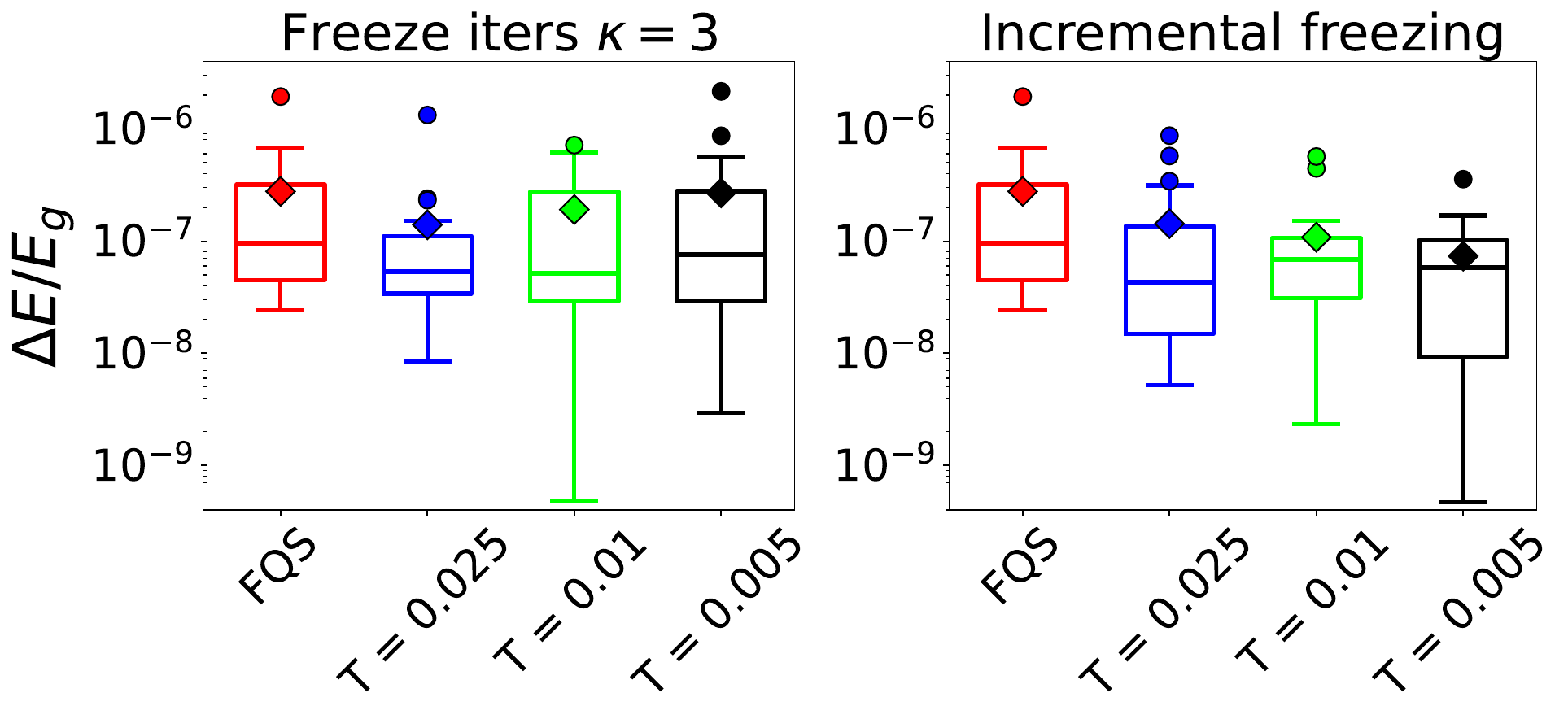}
    \cprotect\caption{Results for 5-qubit Heisenberg model with base \verb|FQS| (red) and gate freezing method with the matrix norm distance metric for $L=5n$ layers. Gate freezing threshold was set to $T = 0.025$ (blue), $0.01$ (green), $0.005$ (black). Freeze iterations $\kappa=3$ and incremental gate freezing were used.}
    \label{fqs_matrix_heisenberg}
\end{figure}

The results for \verb|Fraxis| are shown in Fig.~\ref{fraxis_matrix_heisenberg}. These results are nearly identical to the results obtained with the parameter-based metric distance between consecutive parameter values. The general trend towards the ground state in Fig.~\ref{fraxis_matrix_heisenberg} with the matrix norm metric is similar to that in Fig.~\ref{fraxis_heisenberg}, where a parameter-based distance metric is used. With fixed gate freezing iterations $\kappa=3$, all thresholds $T$ exhibit similar performance in both convergence and boxplot distribution of runs at the end of the optimization of 10 iterations. The same applies to incremental gate freezing; there is no noticeable difference between the fixed gate freezing results and the results in Fig.~\ref{fraxis_heisenberg}.

Finally, the results for \verb|FQS| using the matrix norm distance metric are shown in Fig.~\ref{fqs_matrix_heisenberg}. Here, the base \verb|FQS| has the same convergence of the mean with the freezing iterations $\kappa=3$. However, with thresholds $T=0.01$ and $T=0.005$, some runs can find a much better solution according to the boxplots in Fig.~\ref{fqs_matrix_heisenberg}. The incremental freezing method performs slightly better than the fixed gate freezing method. Particularly, for $T=0.005$, the convergence of the mean is somewhat tighter, and several runs achieve a solution closer to the ground state according to the boxplots.

\subsubsection{Parameter Threshold Crossings Without Gate Freezing} \label{gate_freeze_iters_section}

In this subsection, we analyze how frequently the parameters of individual gates change less than the threshold $T$ in consecutive iterations. In these experiments, we do not apply gate freezing. Instead, we run base versions of \verb|Rotosolve|, \verb|Fraxis|, and \verb|FQS|, and record how often the consecutive parameter updates fall below a given threshold $T$ during the optimization for each gate. For each optimizer, we performed 50 independent runs, and the number of iterations was set to 50 for \verb|Rotosolve|, 50 for \verb|Fraxis|, and 30 for \verb|FQS|, respectively. The $d$-th element of the vector $\bm{\kappa}_d$ is increased by one if the change in the $d$-th parameter between consecutive iterations is smaller than the threshold $T$. We then compute the mean for each $\bm{\kappa}_d$ and plot the results as a two-dimensional grid over 50 runs, where rows indicate the qubit index and columns indicate the layer index to which the gate parameter belongs. The grid can be interpreted as representing the positions of the gates in the circuit. That is, the leftmost column corresponds to the first gates optimized during each iteration, while the rightmost column represents the last gates optimized in the circuit. Red cells indicate higher averages, while blue cells correspond to lower averages for each $\bm{\kappa}_d$. For all optimizers, the parameter-based distance metric was used.

\begin{figure}
    \centering
    \includegraphics[width=0.99\linewidth]{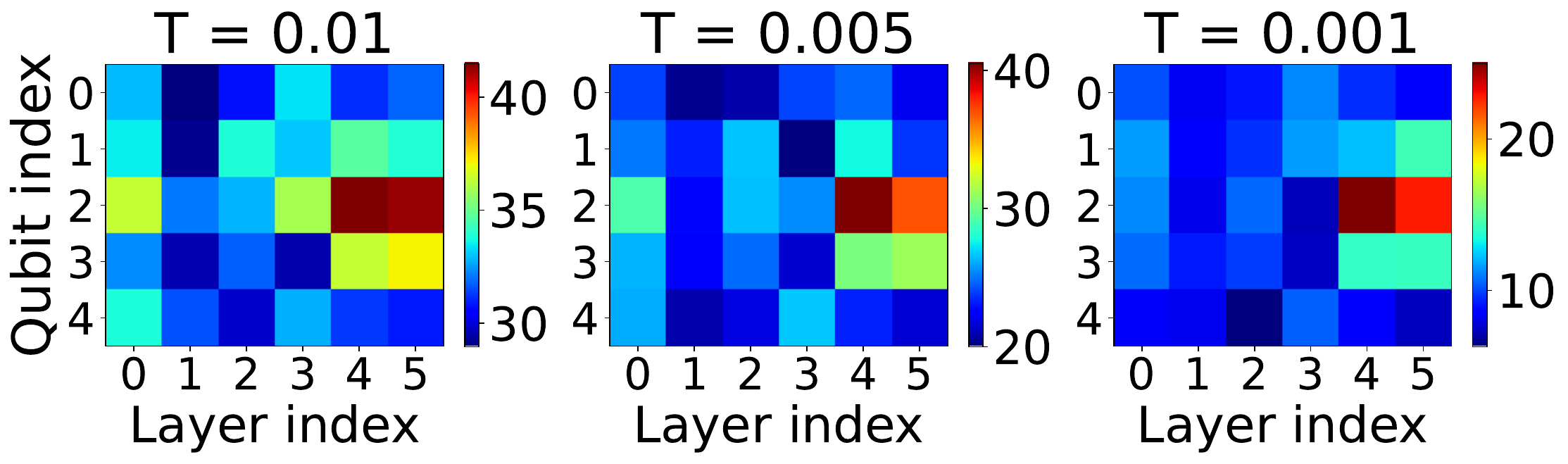}
    \cprotect\caption{Averages for  $\bm{\kappa}_d$ across 50 runs, where the change of $d$-th parameter between consecutive iterations falls below the threshold $T$. The base \verb|Rotosolve| optimizer was used with a parameter-based distance metric, and the threshold $T$ was set to have values $T = 0.01$ (left), $ 0.005$ (mid), and $0.001$ (right). Red cells of the grid indicate a higher average and blue cells a lower average value for $\bm{\kappa}_d$.}
    \label{roto_index_inc_freeze_results}
\end{figure}

\begin{figure}
    \centering
    \includegraphics[width=0.99\linewidth]{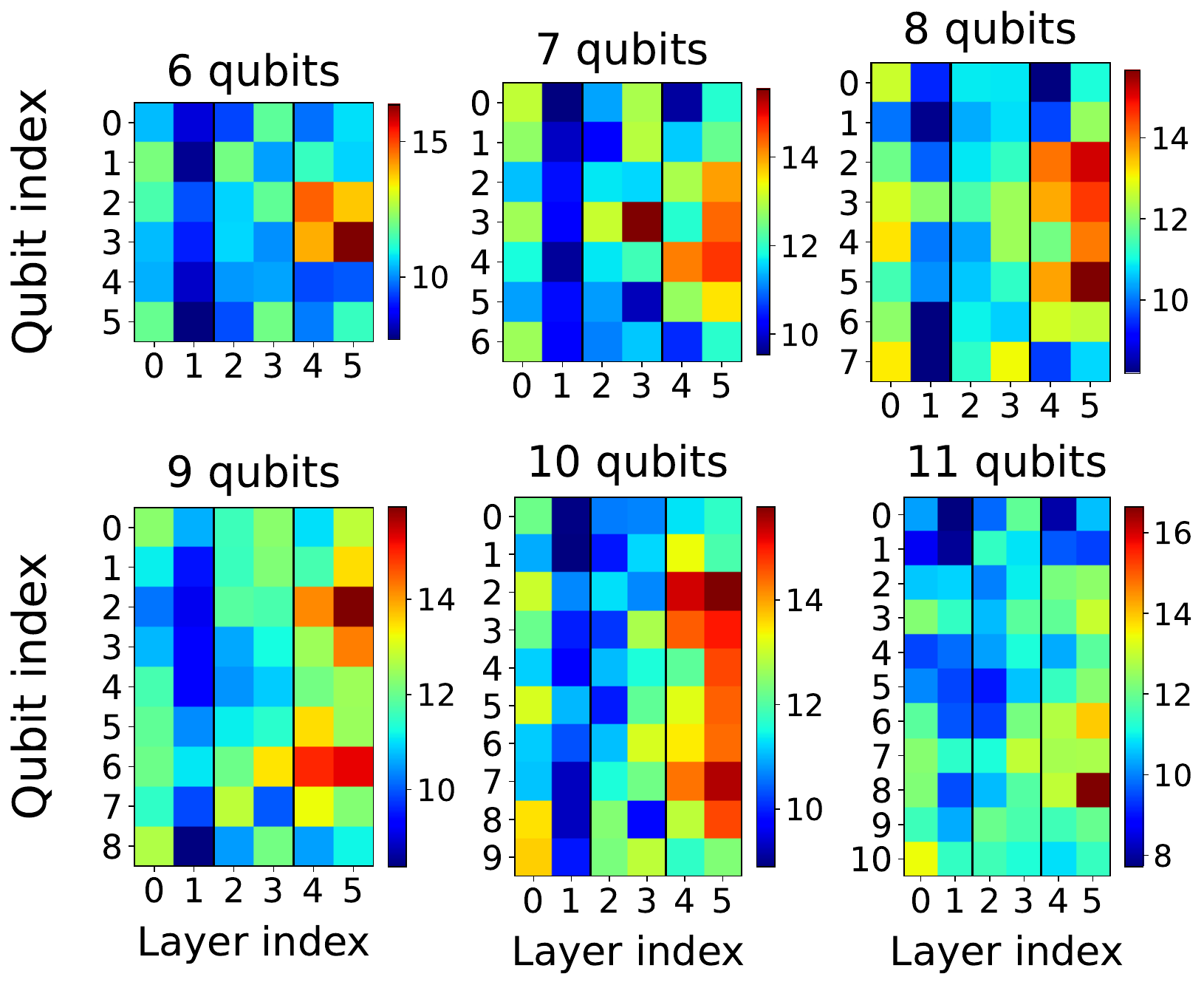}
    \cprotect\caption{Averages for  $\bm{\kappa}_d$ across 50 runs, where the change of $d$-th parameter between consecutive iterations falls below the threshold $T$. The base \verb|Rotosolve| optimizer was used with a parameter-based distance metric, and the threshold $T$ was set to $T = 0.01$. Red cells of the grid indicate a higher average and blue cells a lower average value for $\bm{\kappa}_d$. The number of qubits ranges from 6 to 11, incrementing by one, and the number of layers was set to 3.}
    \label{roto_index_inc_freeze_results_scale}
\end{figure}

\begin{figure}
    \centering
    \includegraphics[width=0.99\linewidth]{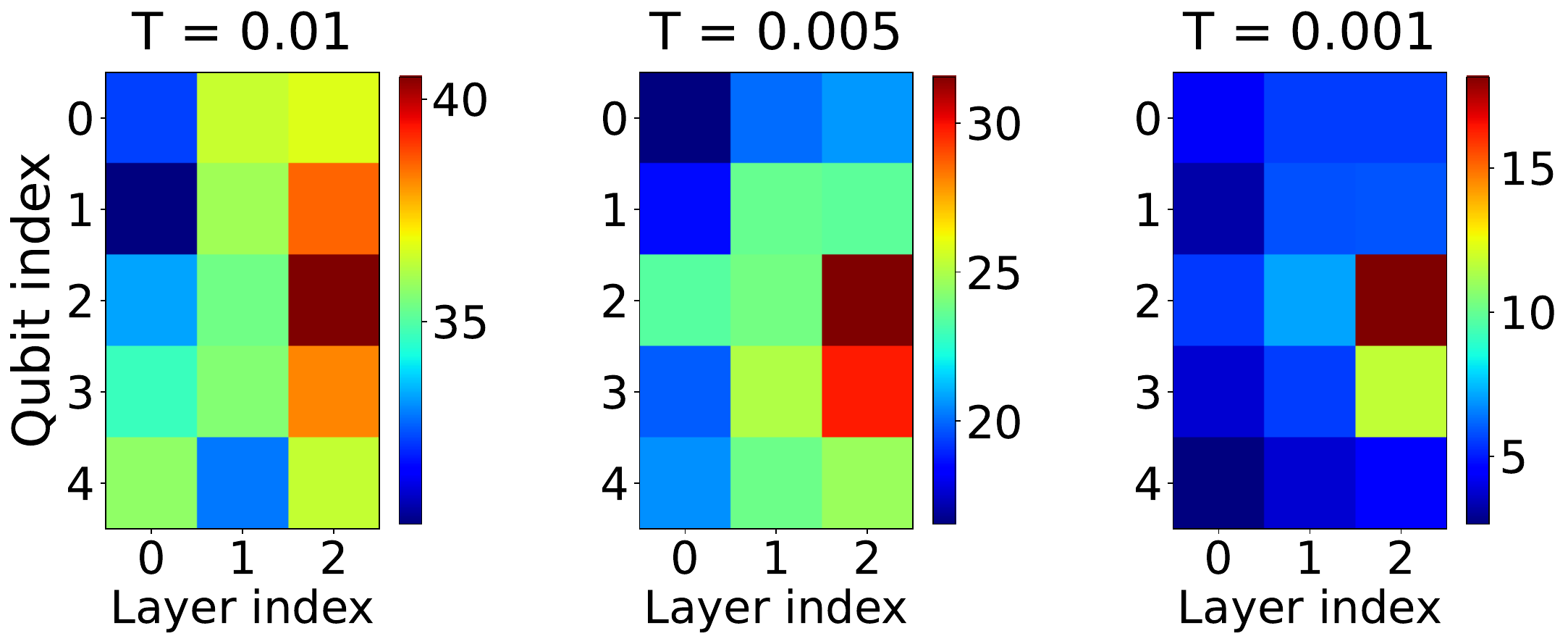}
    \cprotect\caption{Averages for  $\bm{\kappa}_d$ across 50 runs, where the change of $d$-th parameter between consecutive iterations falls below the threshold $T$. The base \verb|Fraxis| optimizer was used with a parameter-based distance metric, and the threshold $T$ was set to have values $T = 0.01$ (left), $ 0.005$ (mid), and $0.001$ (right). Red cells of the grid indicate a higher average and blue cells a lower average value for $\bm{\kappa}_d$.}
    \label{fraxis_index_inc_freeze_results}
\end{figure}

\begin{figure}
    \centering
    \includegraphics[width=0.99\linewidth]{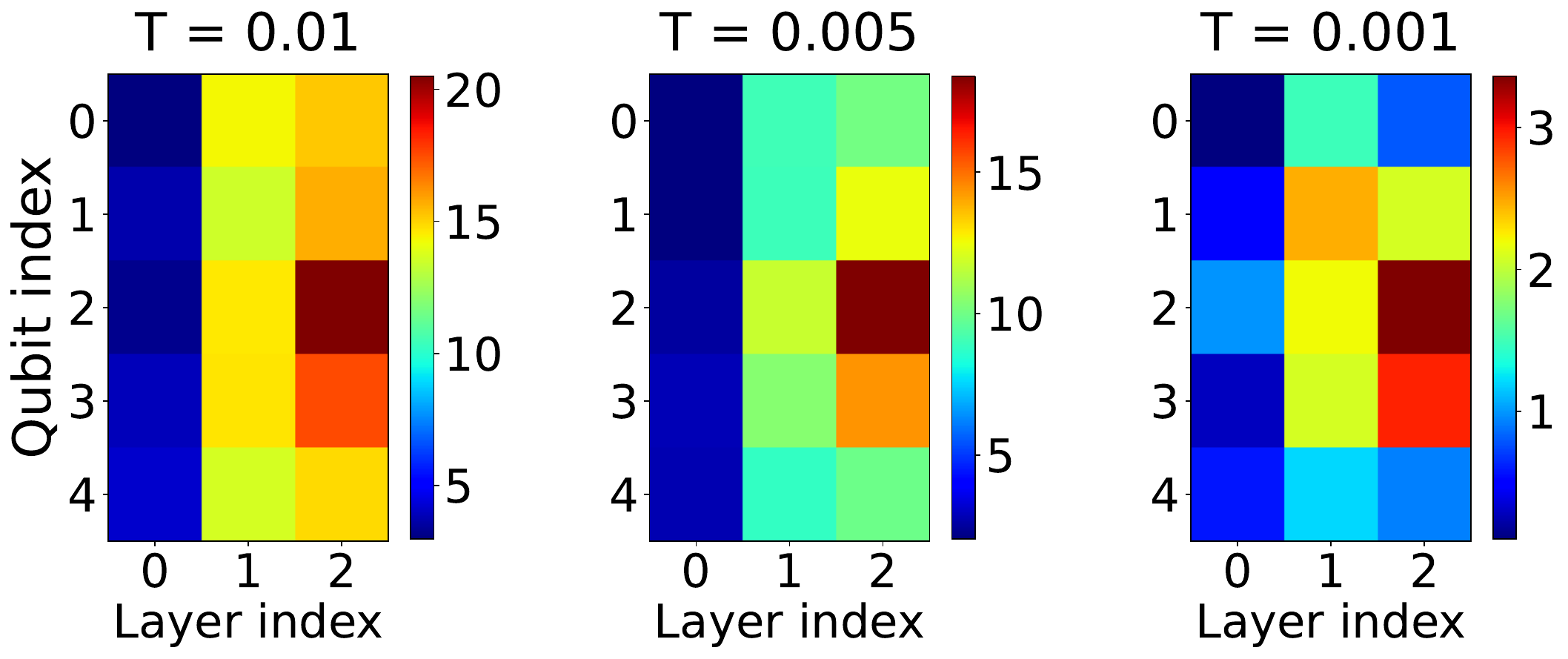}
   \cprotect\caption{Averages for  $\bm{\kappa}_d$ across 50 runs, where the change of $d$-th parameter between consecutive iterations falls below the threshold $T$. The base \verb|FQS| optimizer was used with a parameter-based distance metric, and the threshold $T$ was set to have values $T = 0.01$ (left), $ 0.005$ (mid), and $0.001$ (right). Red cells of the grid indicate a higher average and blue cells a lower average value for $\bm{\kappa}_d$.}
    \label{fqs_index_inc_freeze_results}
\end{figure}

\begin{figure*}
    \centering
    \includegraphics[width=0.85\linewidth]{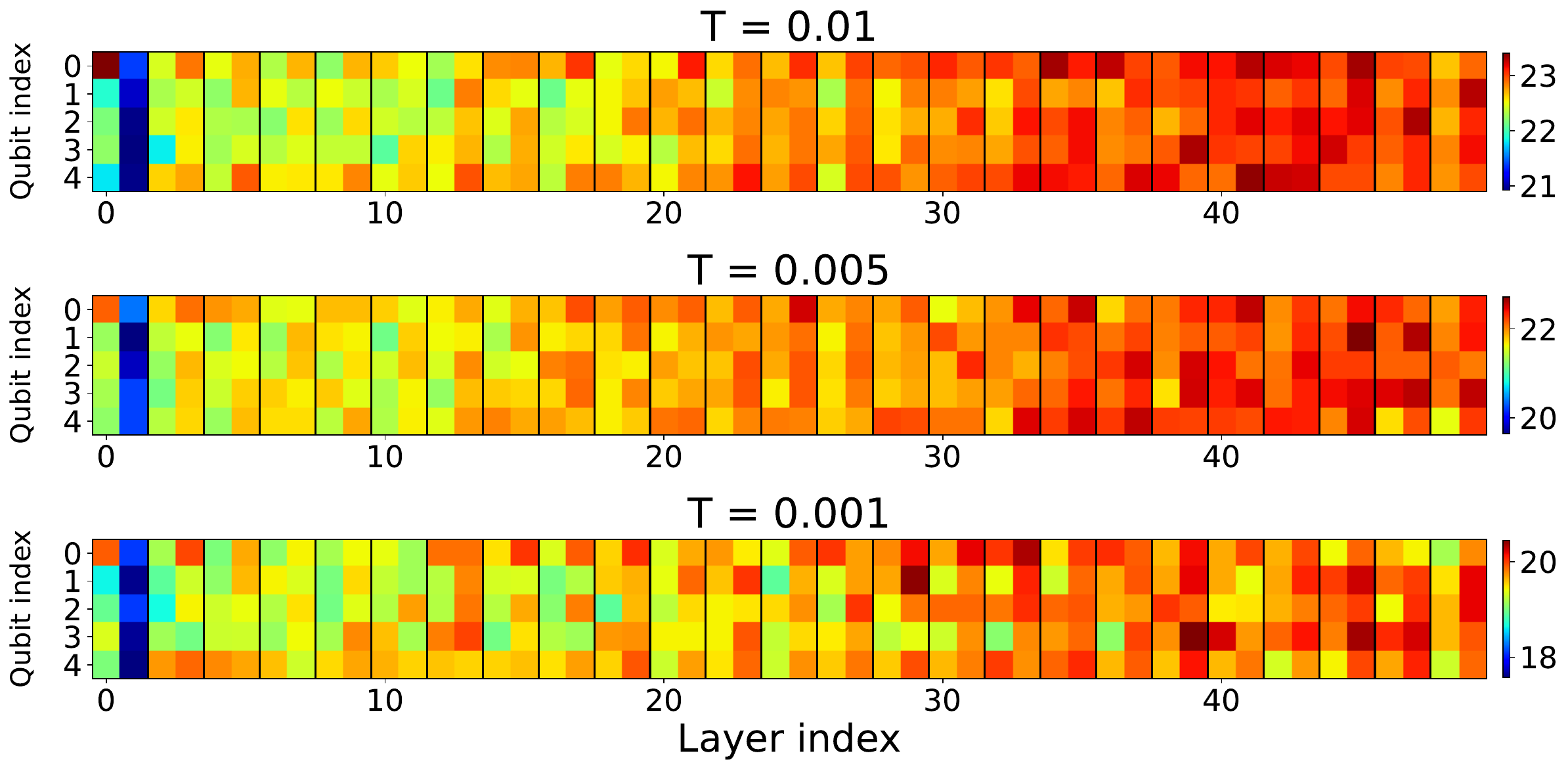}
    \cprotect\caption{Averages for  $\bm{\kappa}_d$ across 50 runs, where the change of $d$-th parameter between consecutive iterations falls below the threshold $T$. The base \verb|Rotosolve| optimizer was used with a parameter-based distance metric, and the threshold $T$ was set to have values $T = 0.01$ (top), $ 0.005$ (mid), and $0.001$ (bottom). Red cells of the grid indicate a higher average and blue cells a lower average value for $\bm{\kappa}_d$. The number of layers was set to $L=5n$.}
    \label{roto_index_inc_freeze_results_deep_circuit}
\end{figure*}

\begin{figure}
    \centering
    \includegraphics[width=0.95\linewidth]{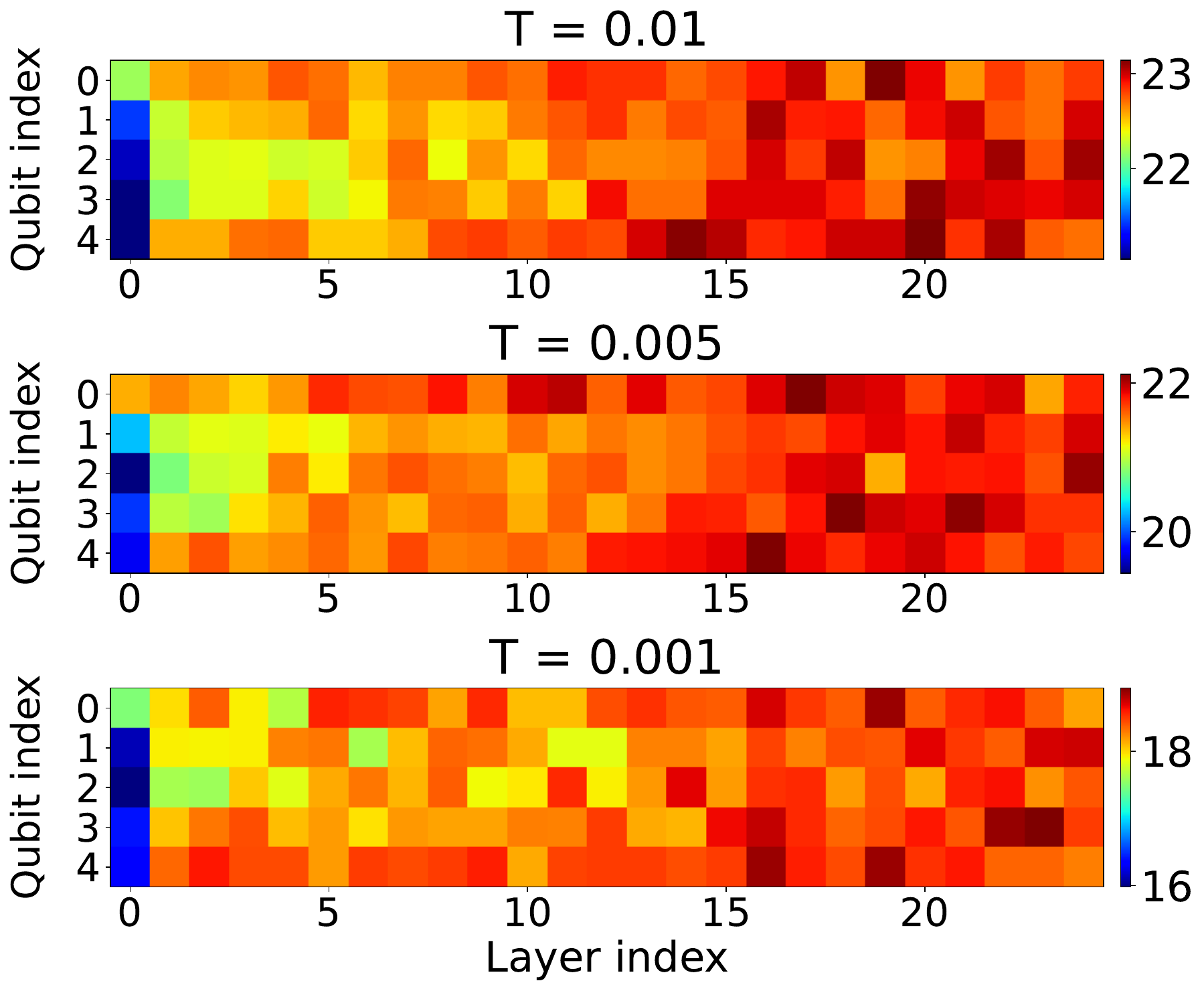}
    \cprotect\caption{Averages for  $\bm{\kappa}_d$ across 50 runs, where the change of $d$-th parameter between consecutive iterations falls below the threshold $T$. The base \verb|Fraxis| optimizer was used with a parameter-based distance metric, and the threshold $T$ was set to have values $T = 0.01$ (top), $ 0.005$ (mid), and $0.001$ (bottom). Red cells of the grid indicate a higher average and blue cells a lower average value for $\bm{\kappa}_d$. The number of layers was set to $L=5n$.}
    \label{fraxis_index_inc_freeze_results_deep_circuit}
\end{figure}

\begin{figure}
    \centering
    \includegraphics[width=0.95\linewidth]{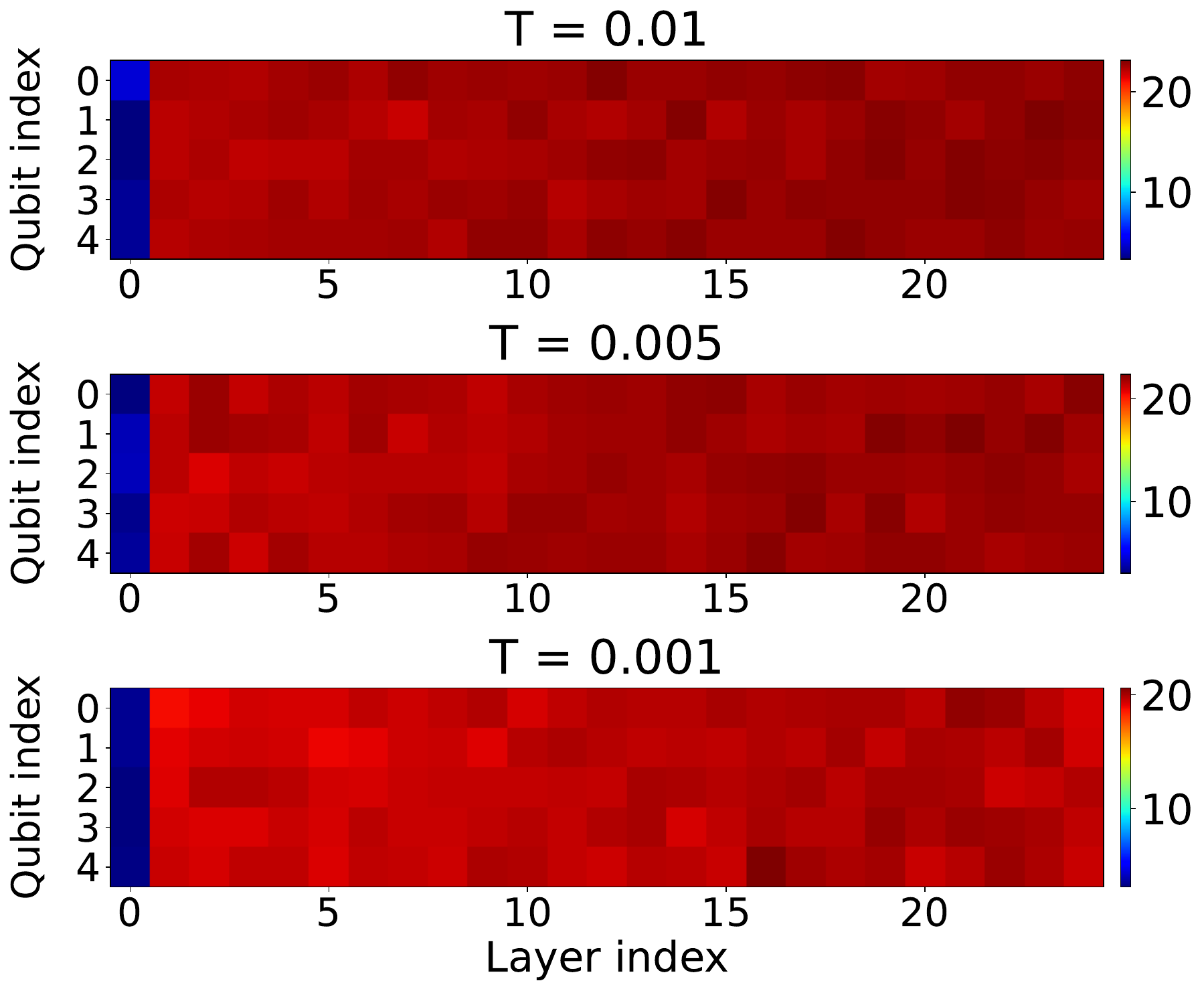}
   \cprotect\caption{Averages for  $\bm{\kappa}_d$ across 50 runs, where the change of $d$-th parameter between consecutive iterations falls below the threshold $T$. The base \verb|FQS| optimizer was used with a parameter-based distance metric, and the threshold $T$ was set to have values $T = 0.01$ (top), $ 0.005$ (mid), and $0.001$ (bottom). Red cells of the grid indicate a higher average and blue cells a lower average value for $\bm{\kappa}_d$. The number of layers was set to $L=5n$.}
    \label{fqs_index_inc_freeze_results_deep_circuit}
\end{figure}

\begin{figure}
    \centering
    \includegraphics[width=0.99\linewidth]{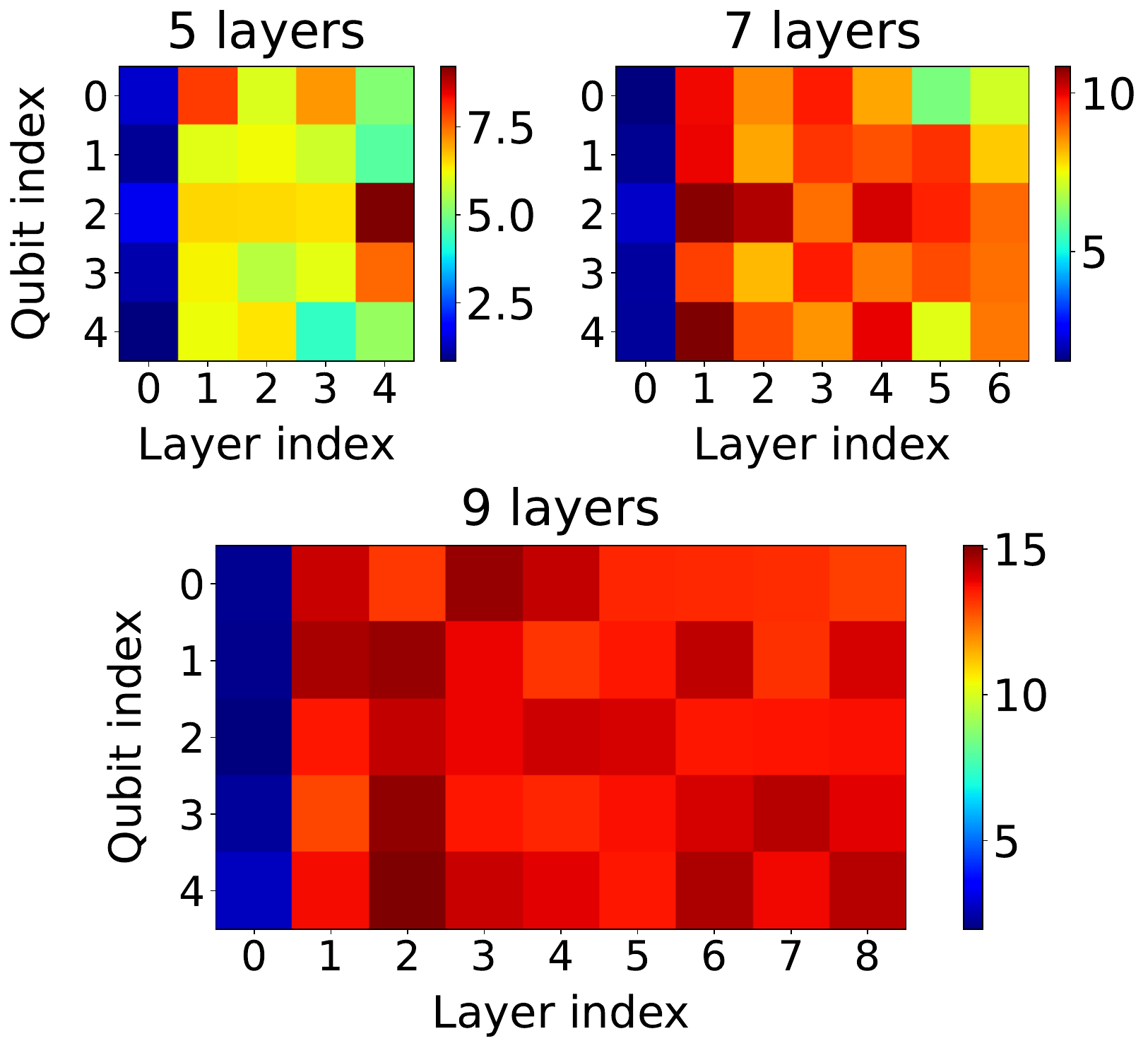}
   \cprotect\caption{Averages for  $\bm{\kappa}_d$ across 50 runs, where the change of $d$-th parameter between consecutive iterations falls below the threshold $T$. The base \verb|FQS| optimizer was used with a parameter-based distance metric, and the threshold $T$ was set to have values $T = 0.01$, $ 0.005$, and $0.001$. Red cells of the grid indicate a higher average and blue cells a lower average value for $\bm{\kappa}_d$. The number of layers was set to $L=5,7,9$.}
    \label{fqs_index_inc_freeze_results_circuit_depths}
\end{figure}

We show the results for \verb|Rotosolve| in Fig.~\ref{roto_index_inc_freeze_results}. Here, we have run a total of 50 runs with 3 layers. In the plots, each ansatz layer consists of $R_X$ and $R_Y$ gates that are applied to each qubit. This increases the number of parameters in the circuit by twice the number of actual layers. In Fig.~\ref{roto_index_inc_freeze_results}, we see that the lower the threshold value $T$, the parameterized gates in the last layer are more likely to change less than $T$ between consecutive iterations compared to earlier layers. This suggests that it may be advantageous to focus on freezing the gates in the last few layers. A notable observation is that the gates with the highest average are in the middle-qubit gates in the last layer. They have a significantly higher average of $\bm{\kappa}_d$ than any other gate in the circuit. We remark that in each subplot, the color scale values are relative to its own subplot. 

In addition, we repeated the experiment with qubits ranging from 6 to 11, incrementing by one, while keeping the number of layers and threshold fixed. The number of layers was set to 3, and the threshold $T$ to $T=0.01$. In Fig.~\ref{roto_index_inc_freeze_results_scale}, we used 25 iterations for each run, and a total of 50 runs were performed. As in the 5-qubit case, the gates that acquire the highest average for $\bm{\kappa}_d$ are focused on the last layer, and in most cases, around its middle qubits.

We repeated the same experiment on \verb|Fraxis| with 3 layers and 50 runs for a 5-qubit system. The results are shown in Fig.~\ref{fraxis_index_inc_freeze_results}. As in the case of \verb|Rotosolve|, we see similar results that with a lower threshold, the gates in the last layer have proportionally higher average values for $\bm{\kappa}_d$ compared to earlier layers. We also notice the same behavior for the last row in all subplots of Fig.~\ref{fraxis_index_inc_freeze_results} as in Fig.~\ref{roto_index_inc_freeze_results}. Again, the last layer obtains the highest averages for $\bm{\kappa}_d$, and the middle qubit gate has the highest average of all. For all tested values of $T$, the gates of the first layer of the circuit obtain lower average values for $\bm{\kappa}_d$ compared to the rest of the layers.
 
For \verb|FQS|, the results are shown in Fig.~\ref{fqs_index_inc_freeze_results}. In all subplots, the gates in the first layer have noticeably lower average values for $\bm{\kappa}_d$ than the rest of the gates. As in the case of \verb|Fraxis|, the highest average values of $\bm{\kappa}_d$ are concentrated in the last layer, especially near the middle qubit gate. Overall, the behavior of \verb|FQS| is similar to the \verb|Fraxis|.

\begin{figure*}
    \centering
    \includegraphics[width=0.85\linewidth]{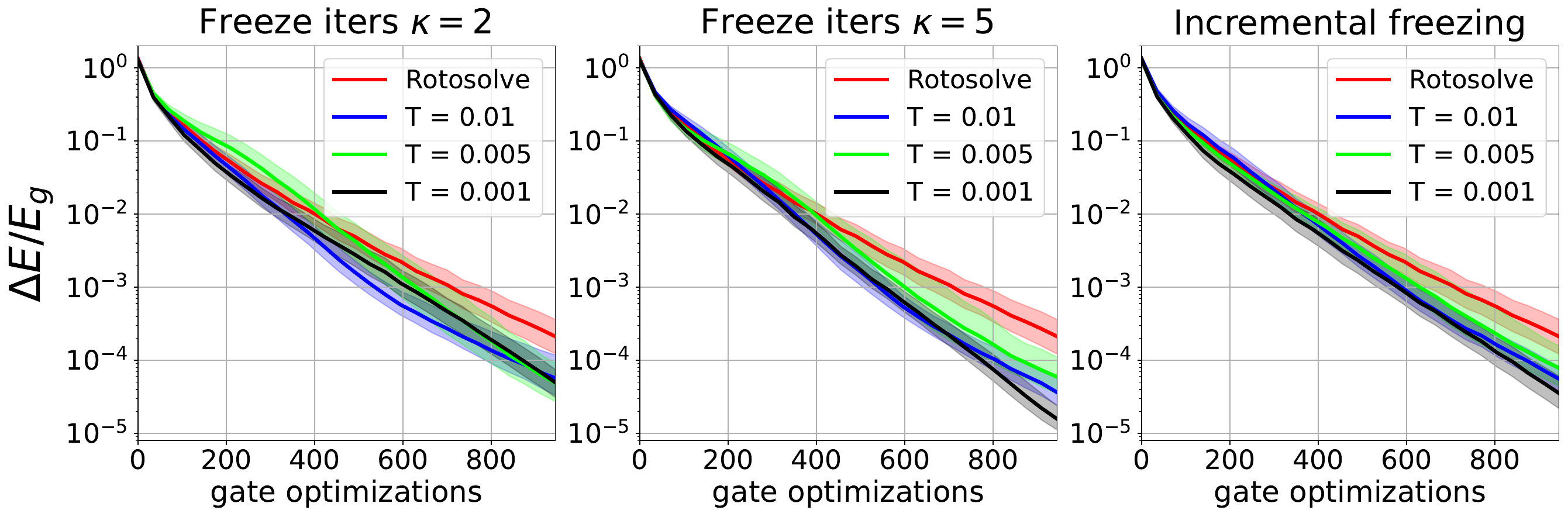}
    \includegraphics[width=0.85\linewidth]{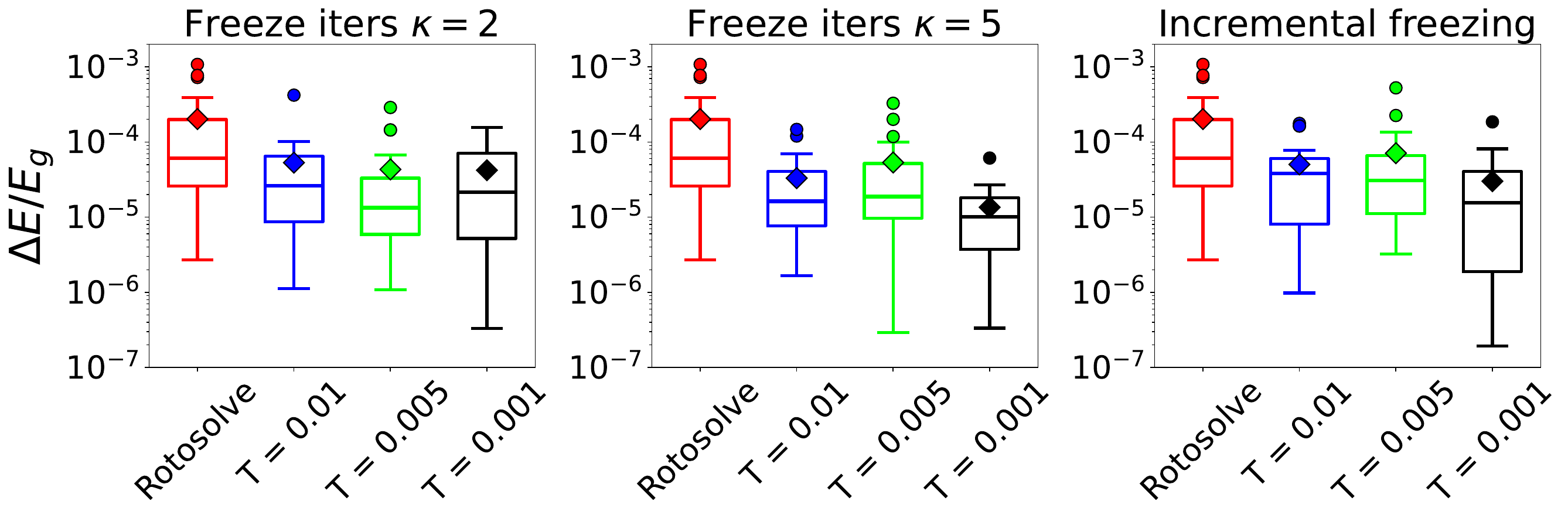}
   \cprotect\caption{Results for 4-qubit Fermi-Hubbard model with base \verb|Rotosolve| (red) and gate freezing method with the parameter-based distance metric for $L=3n$ layers. Gate freezing threshold was set to $T = 0.01$ (blue), $0.005$ (green), $0.001$ (black). Freeze iterations $\kappa=2, 5$ and incremental gate freezing were used.}
    \label{rotosolve_hubbard_results}
\end{figure*}
 
We present the results for \verb|Rotosolve|, \verb|Fraxis|, and \verb|FQS| with deep PQCs in Figs.~\ref{roto_index_inc_freeze_results_deep_circuit},~\ref{fraxis_index_inc_freeze_results_deep_circuit}, and~\ref{fqs_index_inc_freeze_results_deep_circuit}, respectively. Here, we set the number of layers to $L=5n$ and the number of qubits to 5. We used a total of 25 iterations and 50 runs for each optimizer. With deep circuits, we observe how the averages for $\bm{\kappa}_d$ grow steadily towards the end of the circuits. This is emphasized with the \verb|Fraxis| and \verb|Rotosolve| optimizers. The most distinctive behavior is observed for \verb|FQS|, where the gates in all layers except for the first layer exhibit similar average values for $\bm{\kappa}_d$. We note that this happens regardless of the gate freezing threshold, as seen in Fig.~\ref{fqs_index_inc_freeze_results_deep_circuit}. Another observation is that for \verb|Rotosolve| and \verb|Fraxis|, the first gate in optimization order, that is, the top-left gate, obtains a higher average for $\bm{\kappa}_d$ than the rest of the gates in the first layer, as seen in Figs.~\ref{roto_index_inc_freeze_results_deep_circuit} and~\ref{fraxis_index_inc_freeze_results_deep_circuit}, respectively.

Furthermore, in Fig.~\ref{fqs_index_inc_freeze_results_circuit_depths}, the average values for $\bm{\kappa}_d$ with \verb|FQS| are examined with a fixed threshold of $T=0.005$ and using 5, 7, and 9 layers for the 5-qubit system. For 5 layers, the results are similar to Fig.~\ref{fqs_index_inc_freeze_results}, where the number of layers is set to 3. At 7 layers, the contrast between the first layer and the remaining gates in the circuit starts to emerge. For 9 layers, the results are nearly identical to Fig.~\ref {fqs_index_inc_freeze_results_deep_circuit} with the deep circuits. This indicates a peculiar feature for \verb|FQS|: beyond a certain circuit depth and fixed threshold $T$, the average values for $\bm{\kappa}_d$ remain largely stable except for the first layer, even if more layers are added to the circuit. 

In addition, we provide more results to this subsection in the Appendices~\ref{appendix_threshold_counts} and~\ref{appendix_threshold_counts_ansatzes} with the circuit depth set to 3 layers. In Appendix~\ref{appendix_threshold_counts}, we examine the average values for $\bm{\kappa}_d$, using \verb|Fraxis| and \verb|FQS| with the matrix norm metric. In Appendix~\ref{appendix_threshold_counts_ansatzes}, we repeat the experiment in this subsection, using the parameter-based distance metric for all optimizers and different ansatz circuits provided in Appendix~\ref{appendix_circuits}. 

\begin{figure}
    \centering
    \includegraphics[width=0.99\linewidth]{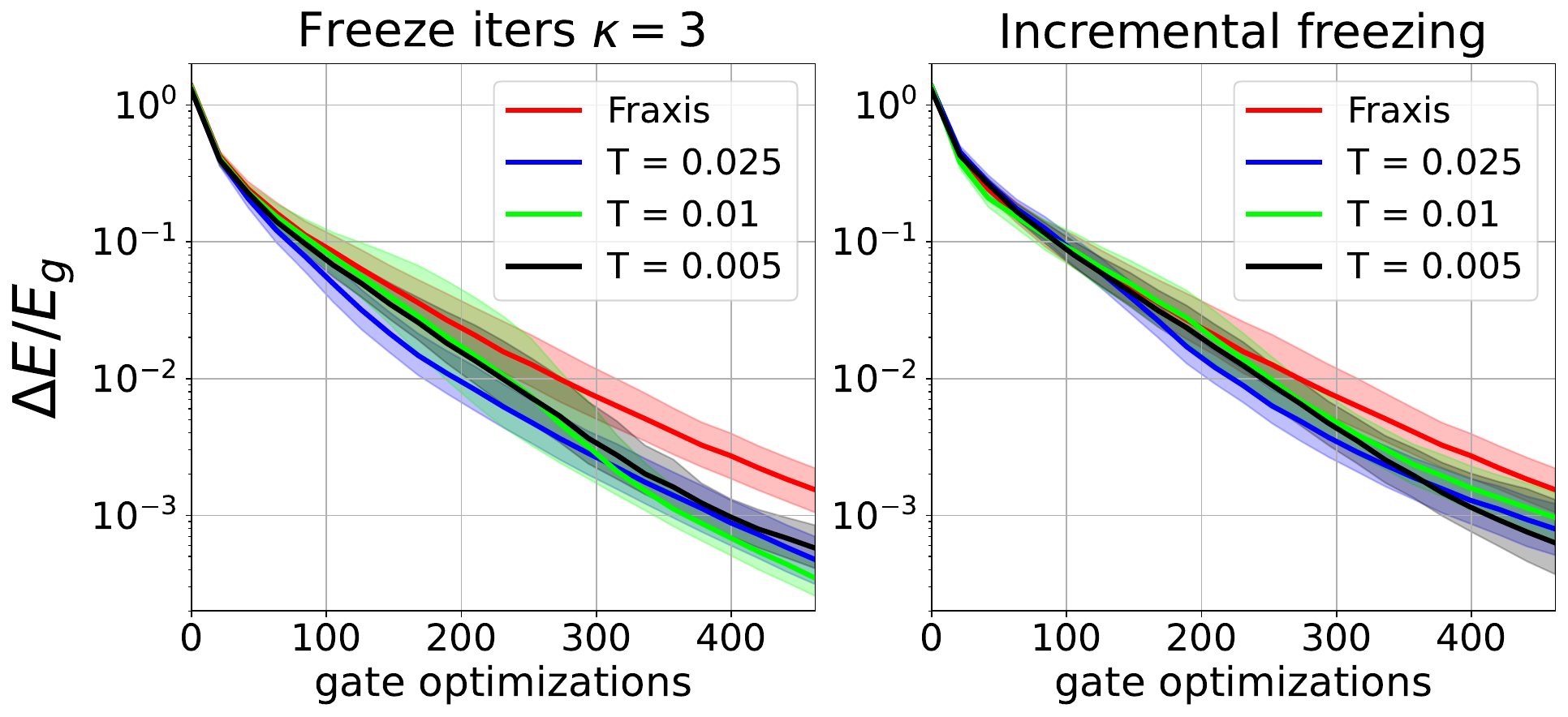}
    \includegraphics[width=0.99\linewidth]{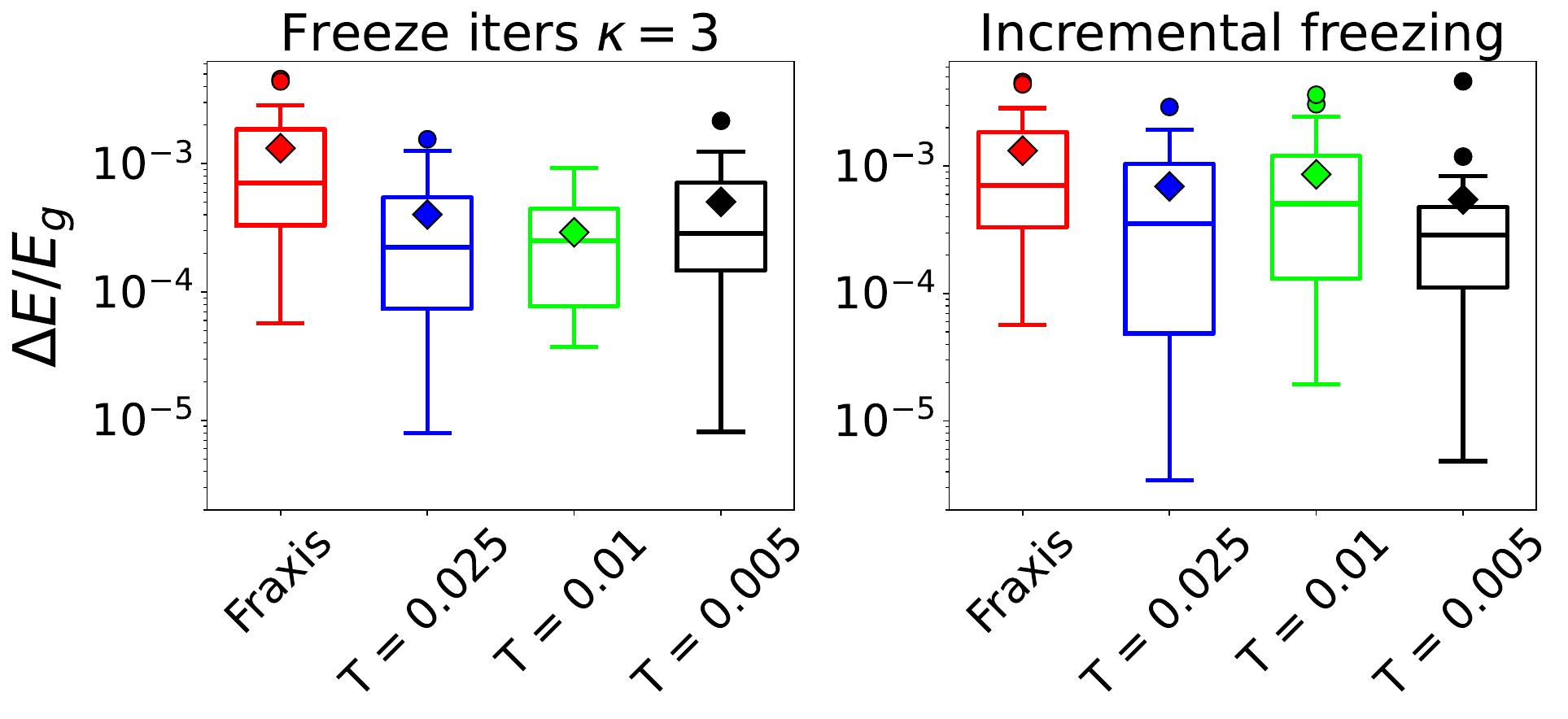}
   \cprotect\caption{Results for 4-qubit Fermi-Hubbard model with base \verb|Fraxis| (red) and gate freezing method with the parameter-based distance metric for $L=3n$ layers. Gate freezing threshold was set to $T = 0.025$ (blue), $0.01$ (green), $0.005$ (black). Freeze iterations $\kappa=3$ and incremental gate freezing were used.}
    \label{fraxis_hubbard_results}
\end{figure}

\begin{figure}
    \centering
    \includegraphics[width=0.99\linewidth]{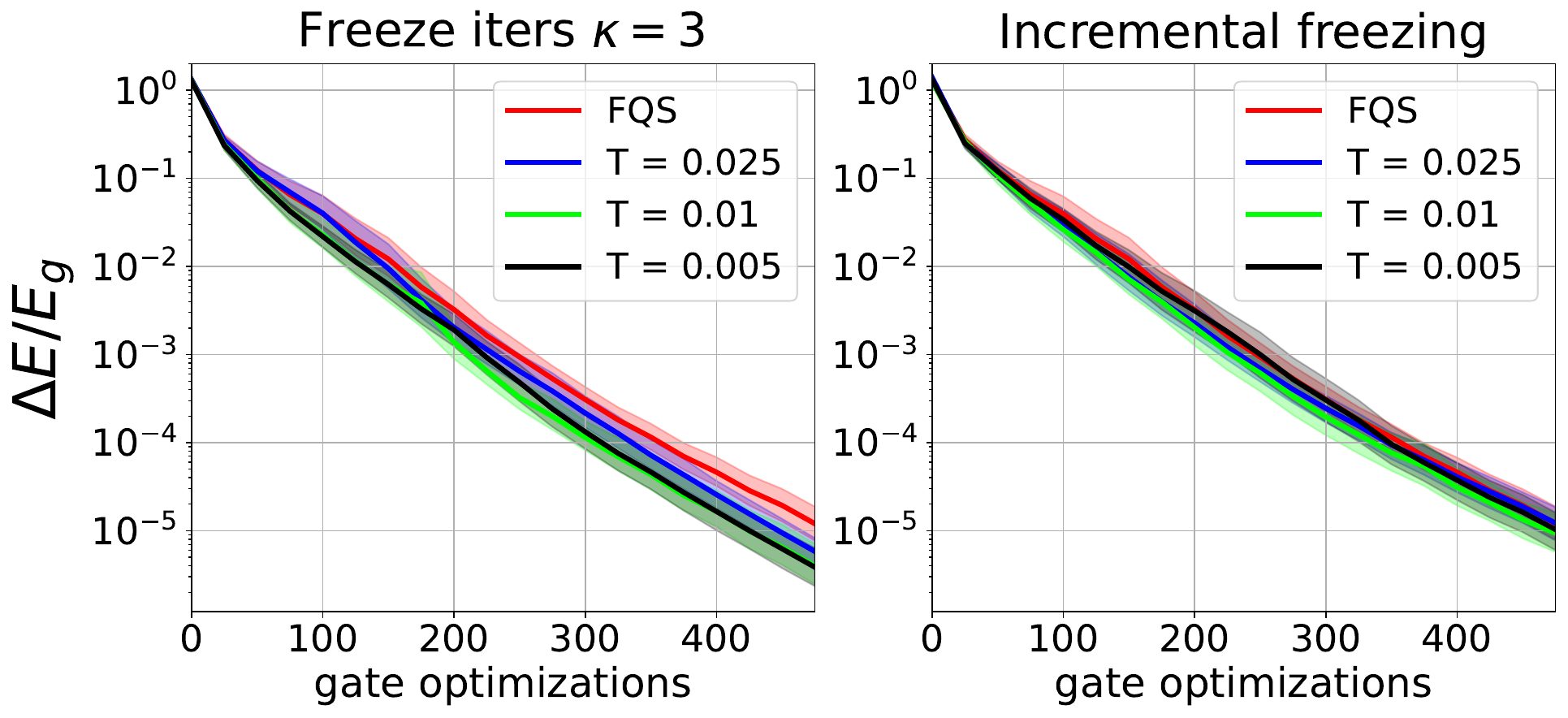}
    \includegraphics[width=0.99\linewidth]{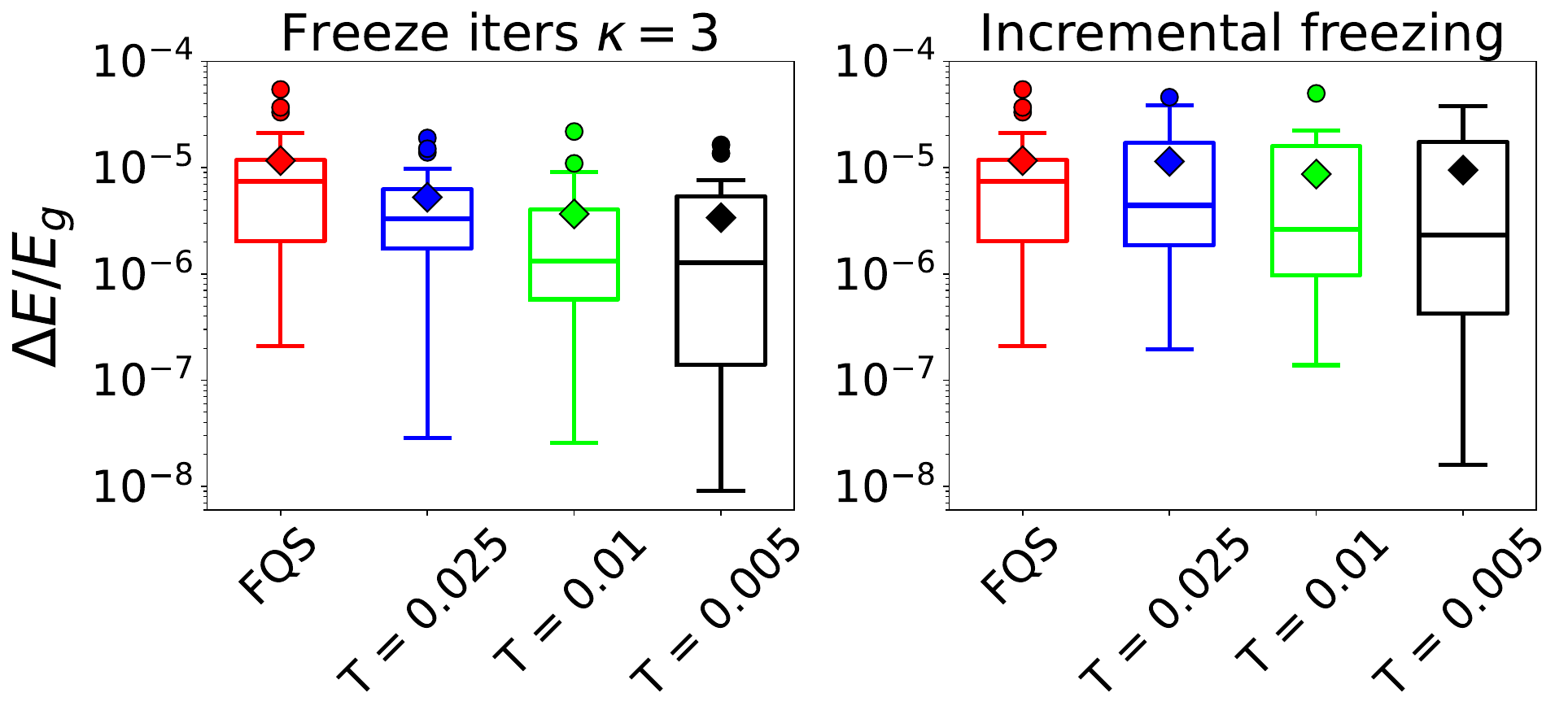}
   \cprotect\caption{Results for 4-qubit Fermi-Hubbard model with base \verb|FQS| (red) and gate freezing method with the parameter-based distance metric for $L=3n$ layers. Gate freezing threshold was set to $T = 0.025$ (blue), $0.01$ (green), $0.005$ (black). Freeze iterations $\kappa=3$ and incremental gate freezing were used.}
    \label{fqs_hubbard_results}
\end{figure}

\begin{figure}
    \centering
    \includegraphics[width=0.99\linewidth]{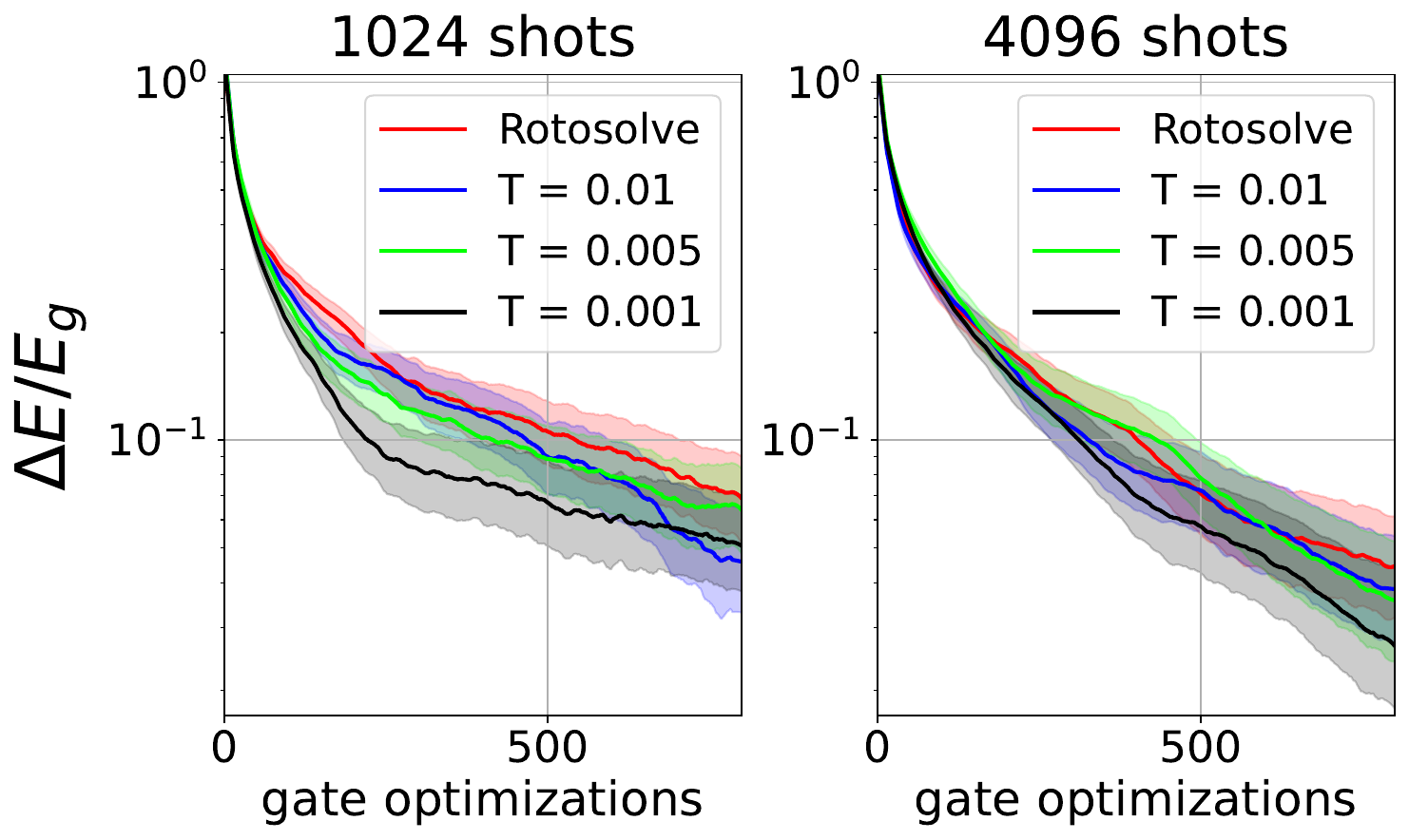}
    \includegraphics[width=0.99\linewidth]{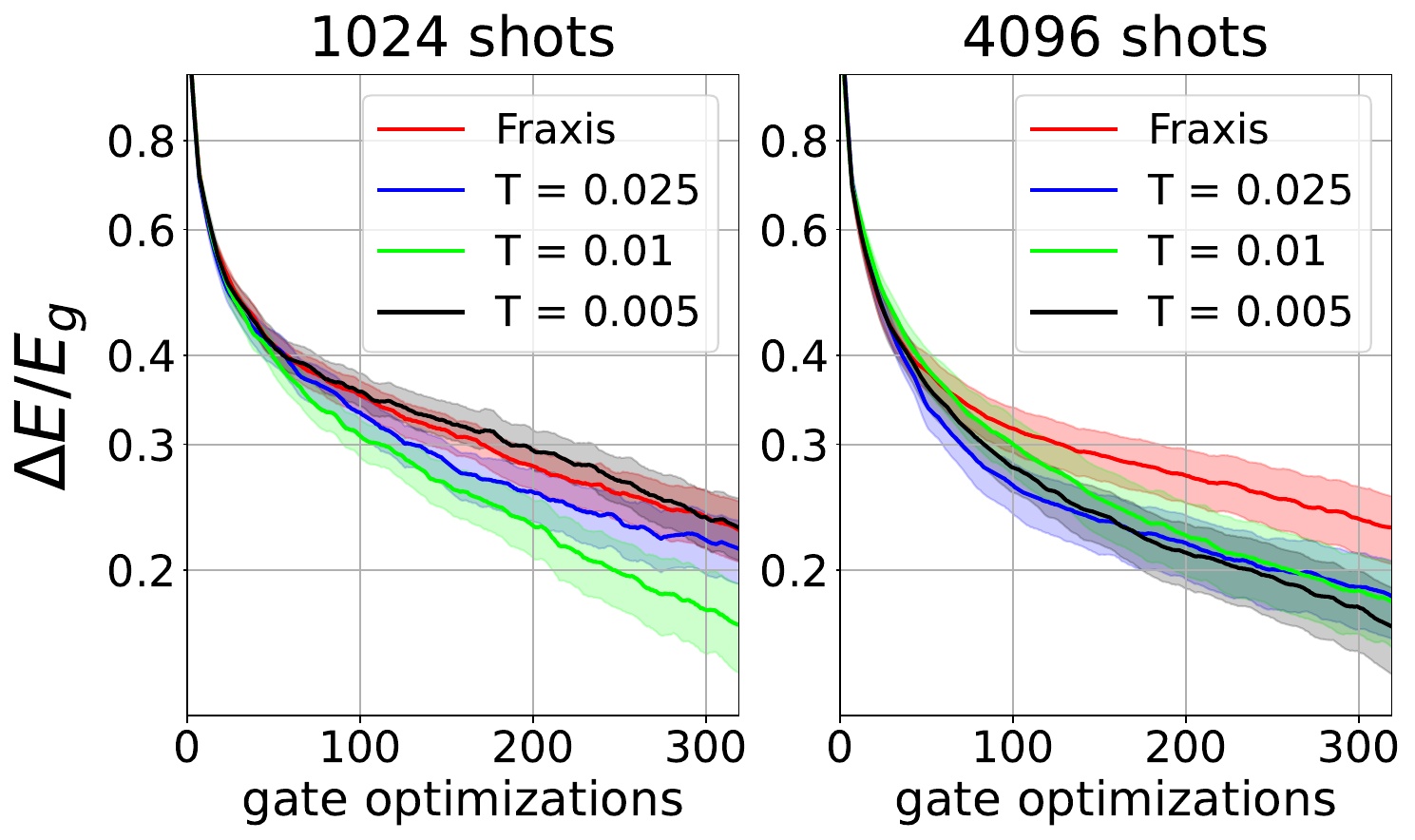}
    \includegraphics[width=0.99\linewidth]{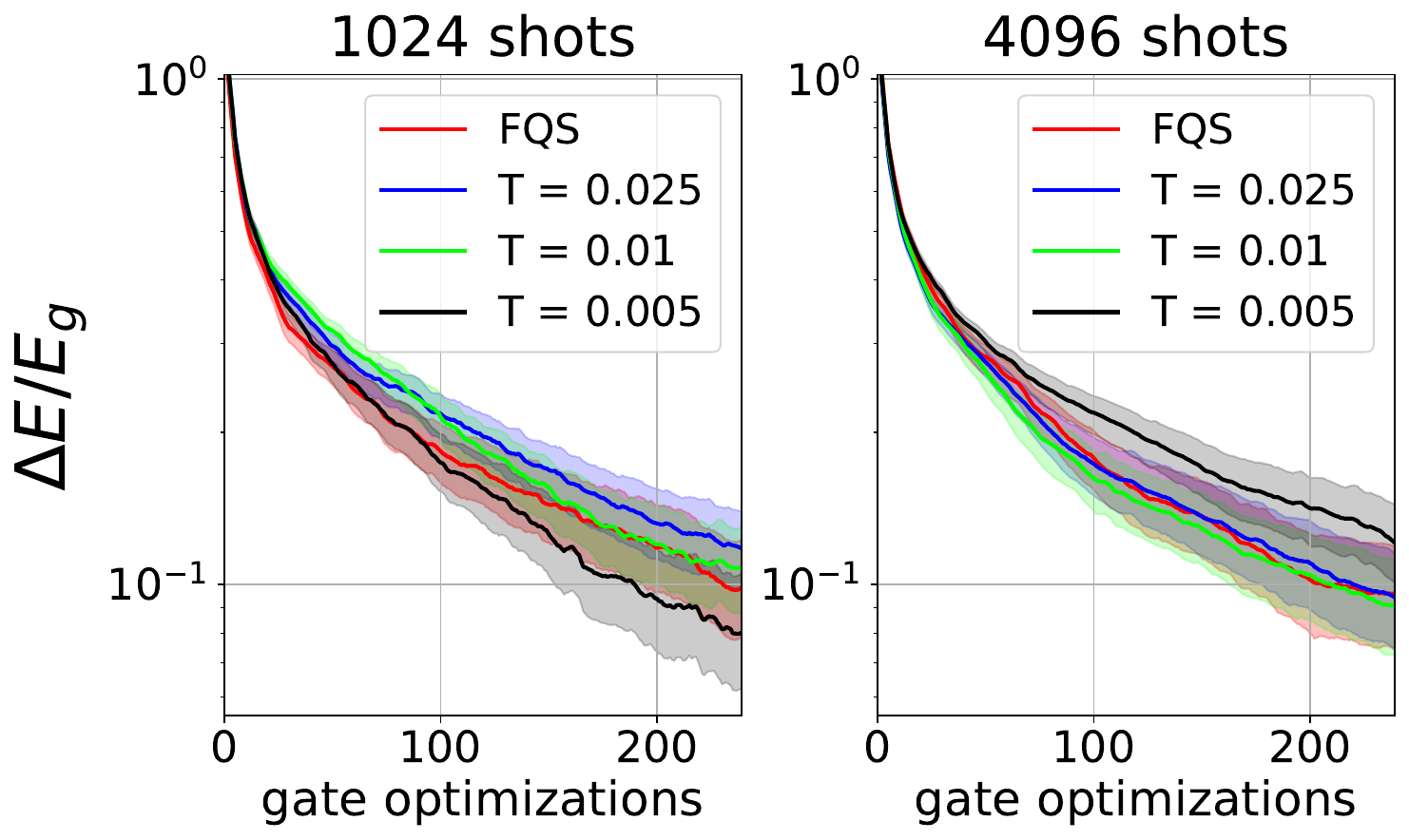}
    \cprotect\caption{Performance of the incremental gate freezing with 4-qubit Fermi-Hubbard model for \verb|Rotosolve|, \verb|Fraxis|, and \verb|FQS| optimizers. The number of layers was set to $L=4$ for all optimizers. 1024 and 4096 shots were used to approximate each Hamiltonian term to simulate the shot noise. Each line corresponds to a mean of 30 runs, and the shaded area 68\% confidence interval around the mean.}
    \label{2DFermiHubbard_noisy}
\end{figure}

\subsection{Fermi-Hubbard Model} \label{section_fermi_hubbard_results}
 
In this section, we show our results for the Fermi-Hubbard model \cite{fermi_hubbard_model}, which describes how fermions interact in a lattice. The model is used to study the electronic and magnetic properties of materials, as well as in experiments with a fermionic quantum gas in the optical lattice~\cite{fermi_hubbard_applications}. We intend to solve the ground state of this model for a $1\times2$ lattice with a 4-qubit system. The Fermi-Hubbard model Hamiltonian is defined as follows \cite{hubbard_hamiltonian}
\begin{equation}
    H = -t \sum_{<i,j>,\sigma} (\hat{c}_{i,\sigma}^\dagger \hat{c}_{j,\sigma} + \text{h.c.}) + U_C\sum_i \hat{n}_{i,\uparrow}\hat{n}_{i,\downarrow}.
\end{equation}
Here, the first term denotes the kinetic term, $t$ is the tunneling matrix, and $<i,j>$ are the neighboring sites. The operators $ \hat{c}_{i,\sigma}^\dagger$ and $\hat{c}_{j,\sigma}$ are the fermionic creation and annihilation operators, respectively, and h.c. stands for the Hermitian conjugate. The operators $\hat{c}_{i,\sigma}^\dagger$ and $\hat{c}_{i,\sigma}$ correspond to adding or removing the fermion from the site $i$ with the spin state $\sigma$. The final term is the potential term, where $U_C$ is the strength of the interaction on site and the occupation number $\hat{n}_{i}$ is the number of particles on site $i$ with up or down states. In this work, we set $t = U_C = 0.5$ in all experiments. The corresponding Hamiltonian is extracted using the PennyLane package \cite{pennylane} with the corresponding lattice size, coefficients $t$ and $U_C$, and applying the Jordan-Wigner mapping~\cite{jordan_wigner} to the creation and annihilation operators. The mapping requires 2 qubits for each lattice site, which means that the lattice of size $1\times2$ is a 4-qubit system.

Next, we show our results for the optimizers \verb|Rotosolve|, \verb|Fraxis|, and \verb|FQS|. In all the following figures in this subsection, unless otherwise stated, each line in the top row represents a mean of 20 runs, and the shaded area is a 90\% confidence interval around the mean. In the bottom row, the box plots summarize the results at the end of 20 runs. The boxes span from the first quartile (Q1) to the third quartile (Q3); the horizontal bar within each box indicates the median, the diamond marker denotes the mean, and the whiskers represent values within 1.5 times the interquartile range (IQR). The ground state of the Hamiltonian is approximately $E_g=-0.78$, and we examined the relative error $\Delta E/E_g$ to the ground state, where $\Delta E = \expval{M} - E_g$.

\begin{figure}
    \centering
    \includegraphics[width=0.99\linewidth]{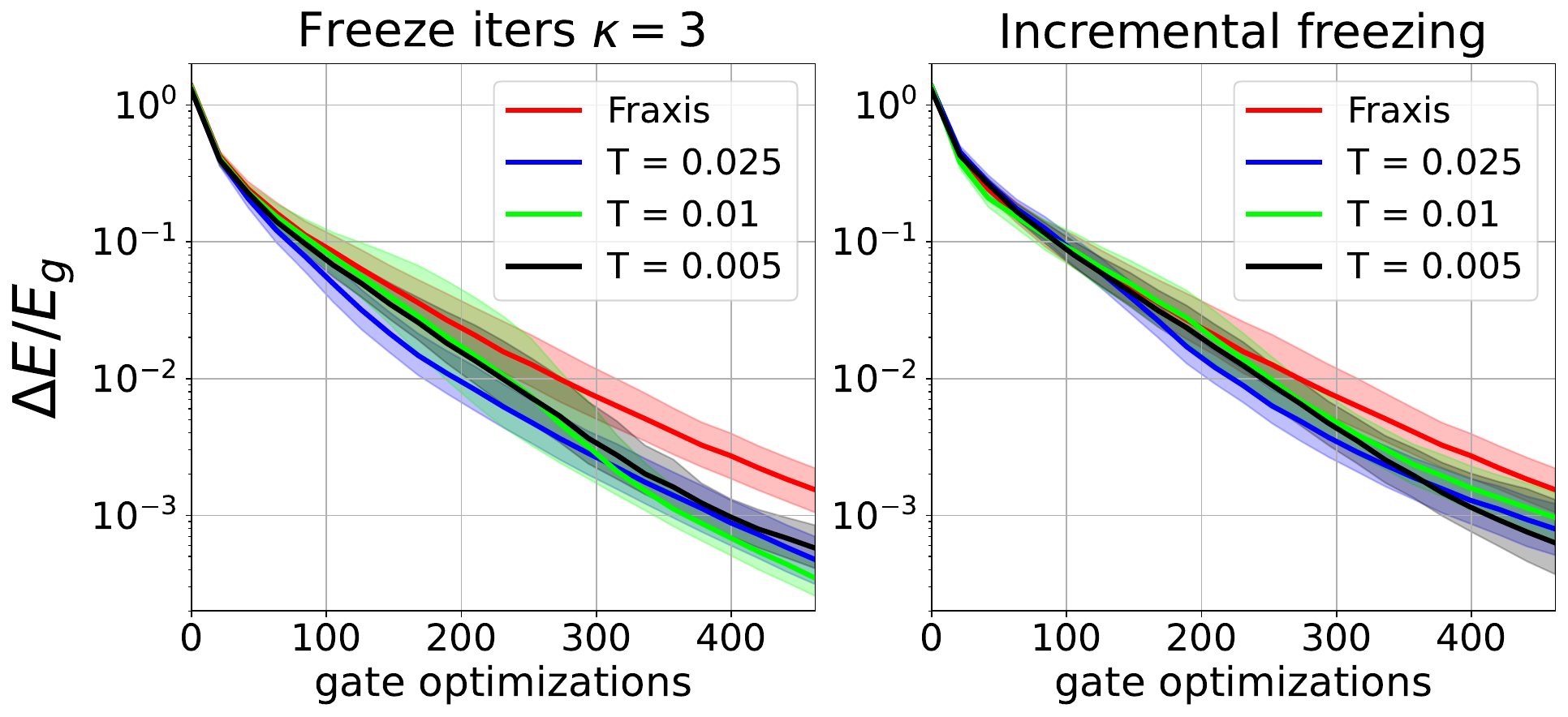}
    \includegraphics[width=0.99\linewidth]{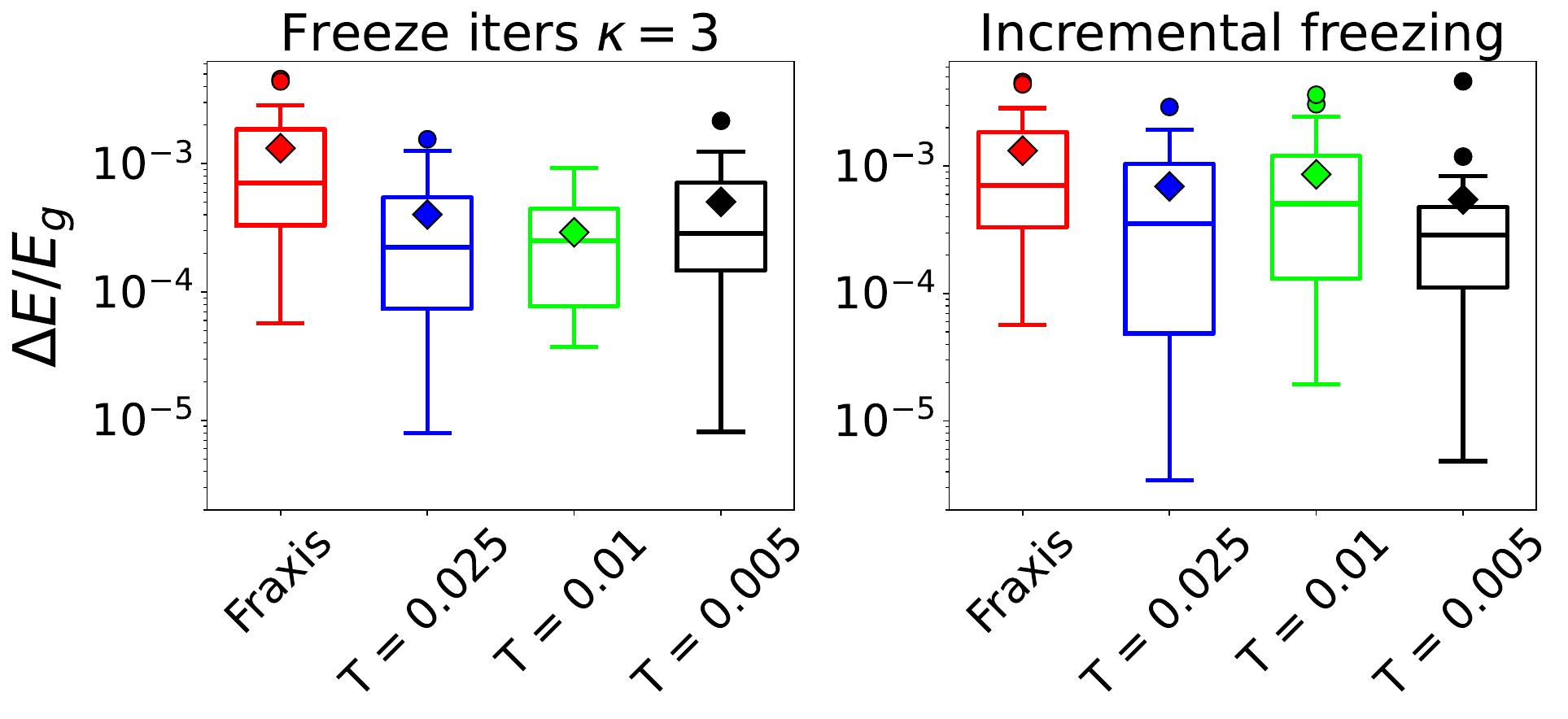}
   \cprotect\caption{Results for 4-qubit Fermi-Hubbard model with base \verb|Fraxis| (red) and gate freezing method with the matrix norm distance metric for $L=3n$ layers. Gate freezing threshold was set to $T = 0.025$ (blue), $0.01$ (green), $0.005$ (black). Freeze iterations $\kappa=3$ and incremental gate freezing were used.}
    \label{fraxis_hubbard_results_matrix_norm}
\end{figure}

\begin{figure}
    \centering
    \includegraphics[width=0.99\linewidth]{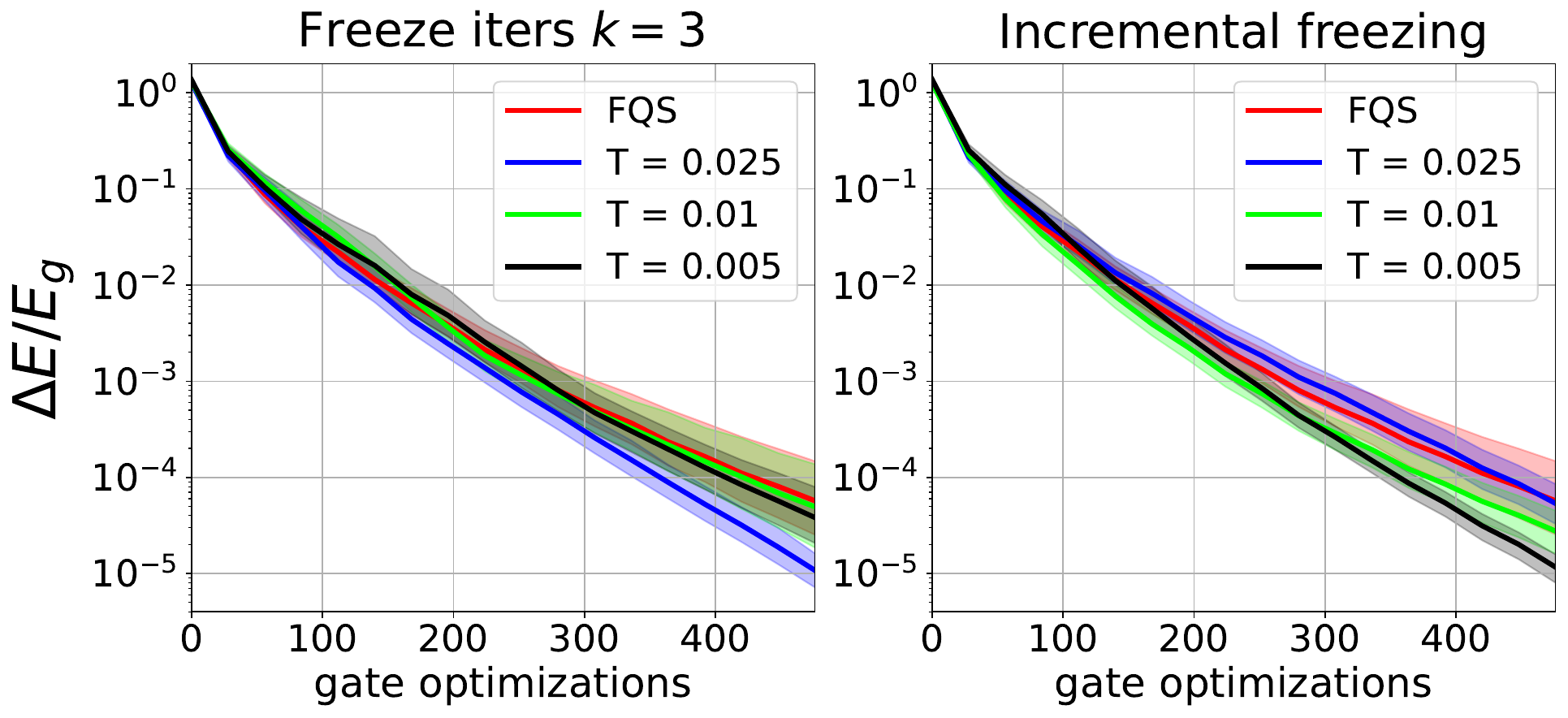}
    \includegraphics[width=0.99\linewidth]{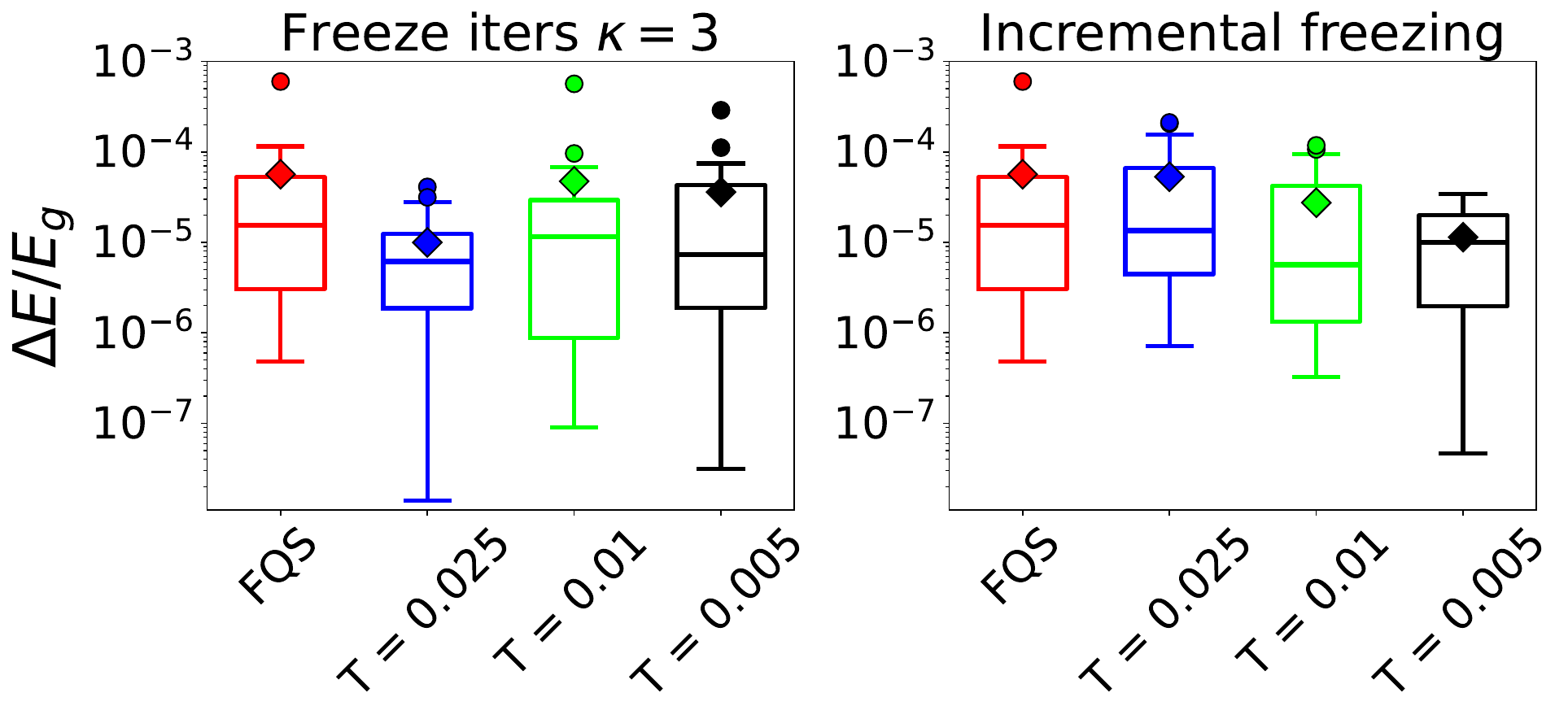}
   \cprotect\caption{Results for 4-qubit Fermi-Hubbard model with base \verb|FQS| (red) and gate freezing method with the matrix norm distance metric for $L=3n$ layers. Gate freezing threshold was set to $T = 0.025$ (blue), $0.01$ (green), $0.005$ (black). Freeze iterations $\kappa=3$ and incremental gate freezing were used.}
    \label{fqs_hubbard_results_matrix_norm}
\end{figure}

\begin{figure}
    \centering
    \includegraphics[width=0.99\linewidth]{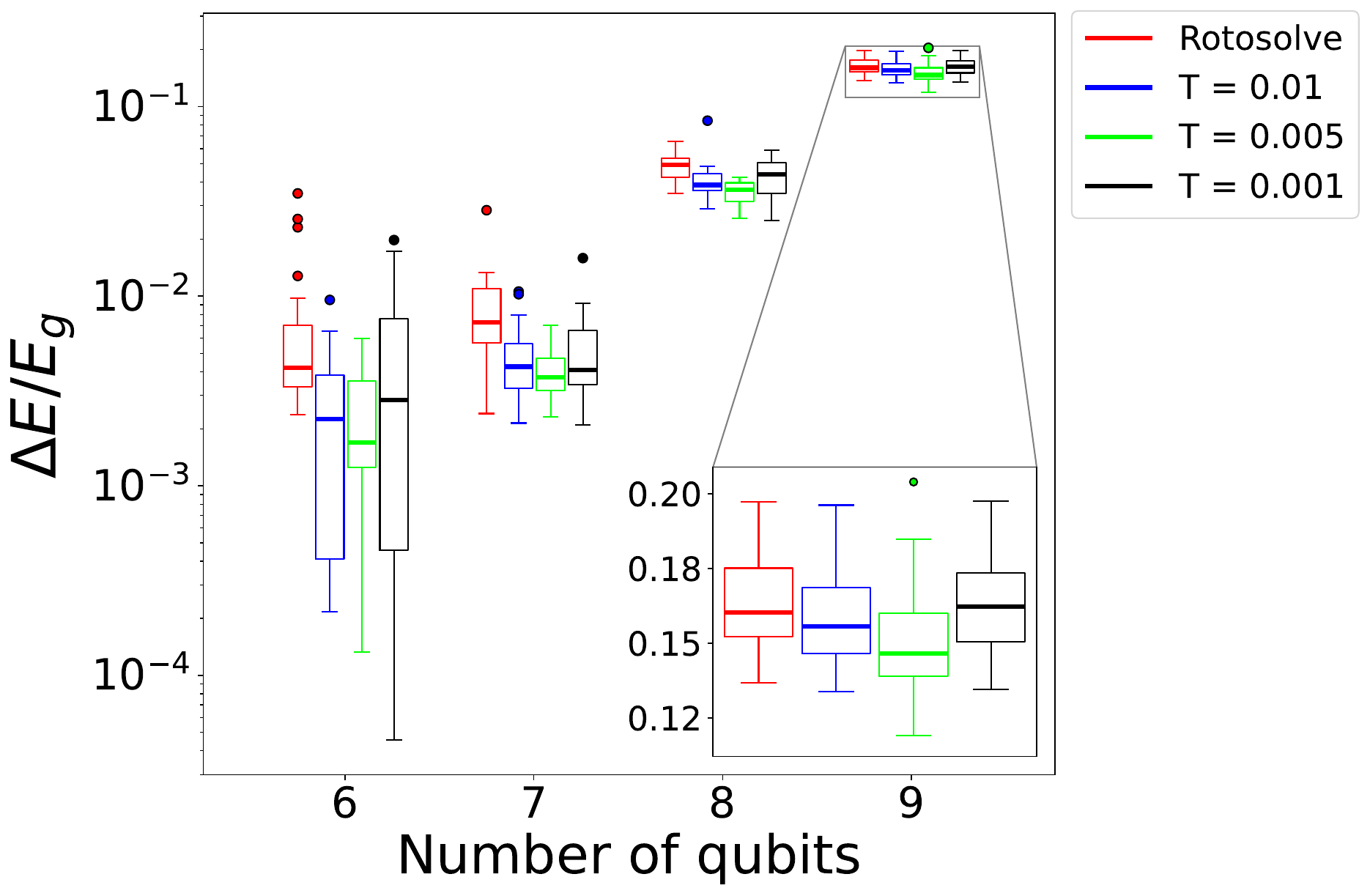}
    \cprotect\caption{Relative errors for the Heisenberg model for qubits ranging from 6 to 9, incrementing by 1. The box plots summarize the results for base \verb|Rotosolve| (red) and the incremental gate freezing method with gate freezing thresholds set to $T=0.01$ (blue), $0.005$ (green), and $0.001$ (black). The vertical axis denotes the relative error from the true ground state on a logarithmic scale, and the horizontal axis the number of qubits used in the circuit. The number of layers for each circuit with $n$ qubits was set to $L=5n$ according to Fig.~\ref{roto_gd_ansatz}. The boxes span from the first quartile (Q1) to the third quartile (Q3). The horizontal bar inside the box represents the median, and the whiskers represent values within 1.5 times the interquartile range (IQR). The points outside the range of whiskers are outliers.}
    \label{scalability_boxplot}
\end{figure}

\begin{figure*}
    \centering
    \includegraphics[width=0.65\linewidth]{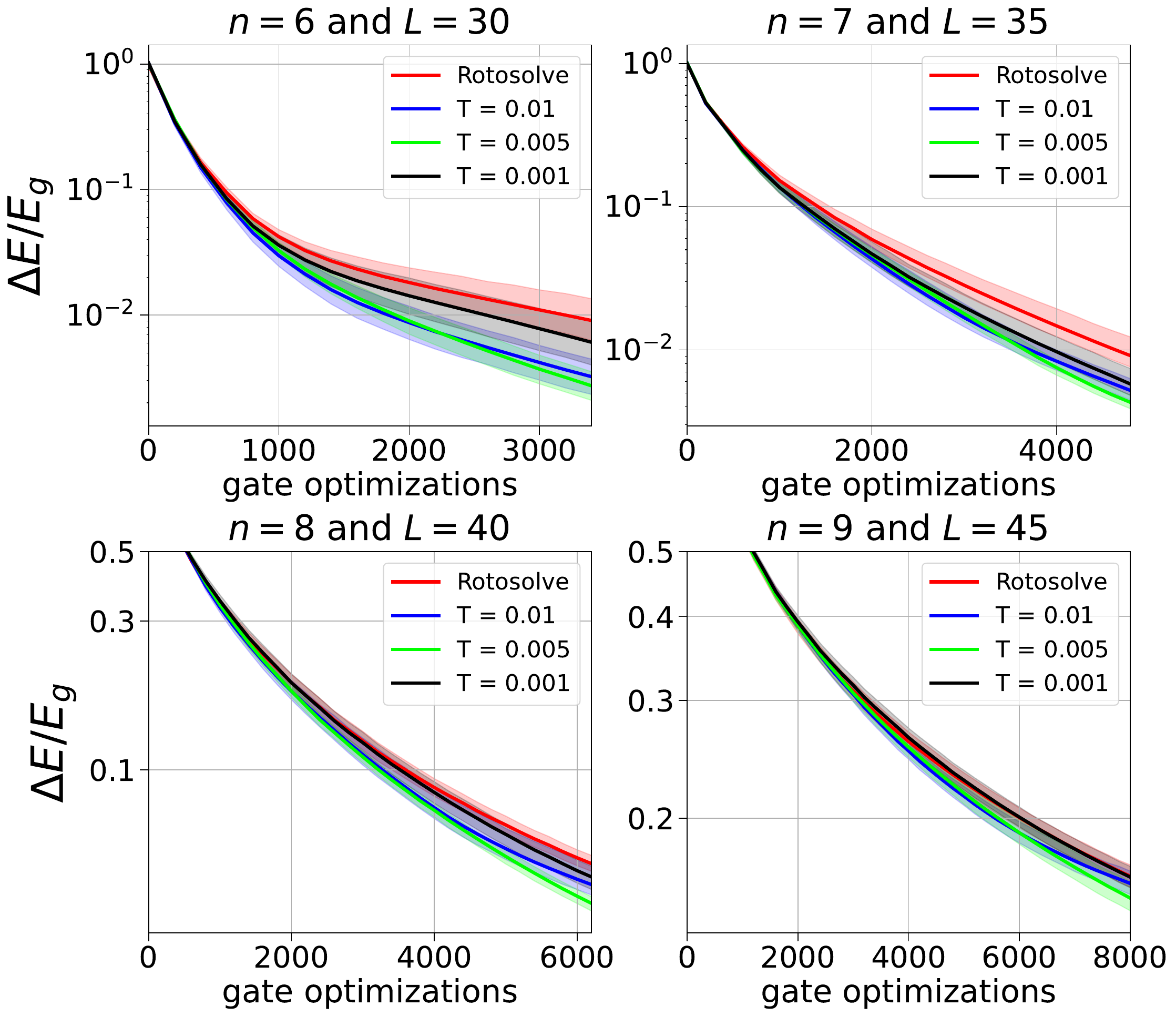}
   \cprotect\caption{Relative errors for the Heisenberg model for qubits ranging from 6 to 9, incrementing by 1. Base \verb|Rotosolve| (red) and gate freezing method with the parameter-based distance metric for $L=5n$ layers and ansatz from Fig.~\ref{roto_gd_ansatz} was used. Gate freezing threshold was set to $T = 0.01$ (blue), $0.005$ (green), $0.001$ (black), and incremental gate freezing was used. Each line represents the mean of 20 runs, and the shaded area is the 90\% confidence interval around the mean.}
    \label{scalability_RxRy_layer_means}
\end{figure*}

\begin{figure}
    \centering
    \includegraphics[width=0.99\linewidth]{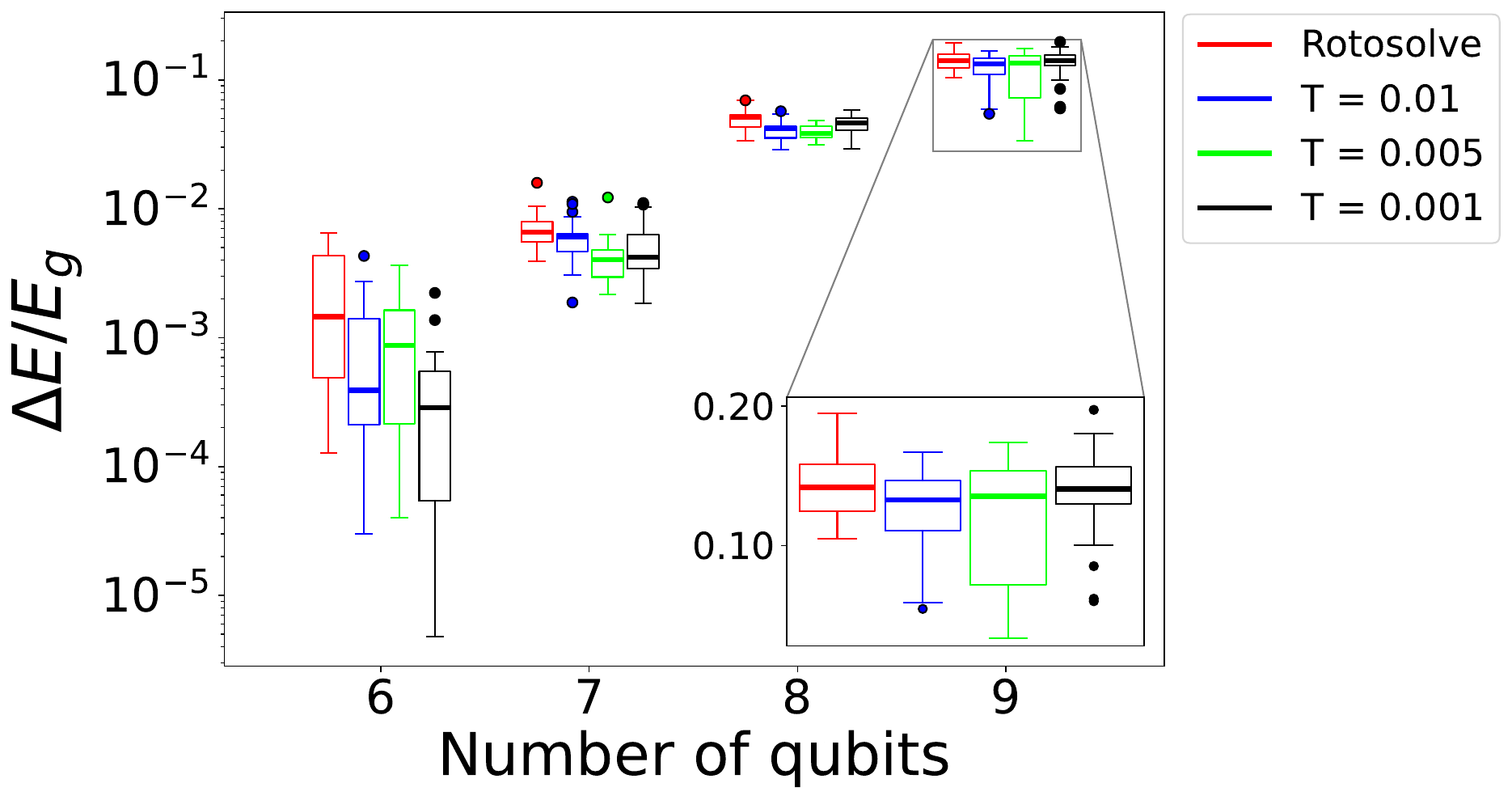}
    \cprotect\caption{Relative errors for the Heisenberg model for qubits ranging from 6 to 9, incrementing by 1. The box plots summarize the results for base \verb|Rotosolve| (red) and the incremental gate freezing method with gate freezing thresholds set to $T=0.01$ (blue), $0.005$ (green), and $0.001$ (black). The vertical axis denotes the relative error from the true ground state on a logarithmic scale, and the horizontal axis the number of qubits used in the circuit. The number of layers for each circuit with $n$ qubits was set to $L=10n$ according to Fig.~\ref{Ansatz_circuit_image}. The boxes span from the first quartile (Q1) to the third quartile (Q3). The horizontal bar inside the box represents the median, and the whiskers represent values within 1.5 times the interquartile range (IQR). The points outside the range of whiskers are outliers.}
    \label{scalability_boxplot_random_PQC}
\end{figure}

\begin{figure*}
    \centering
    \includegraphics[width=0.85\linewidth]{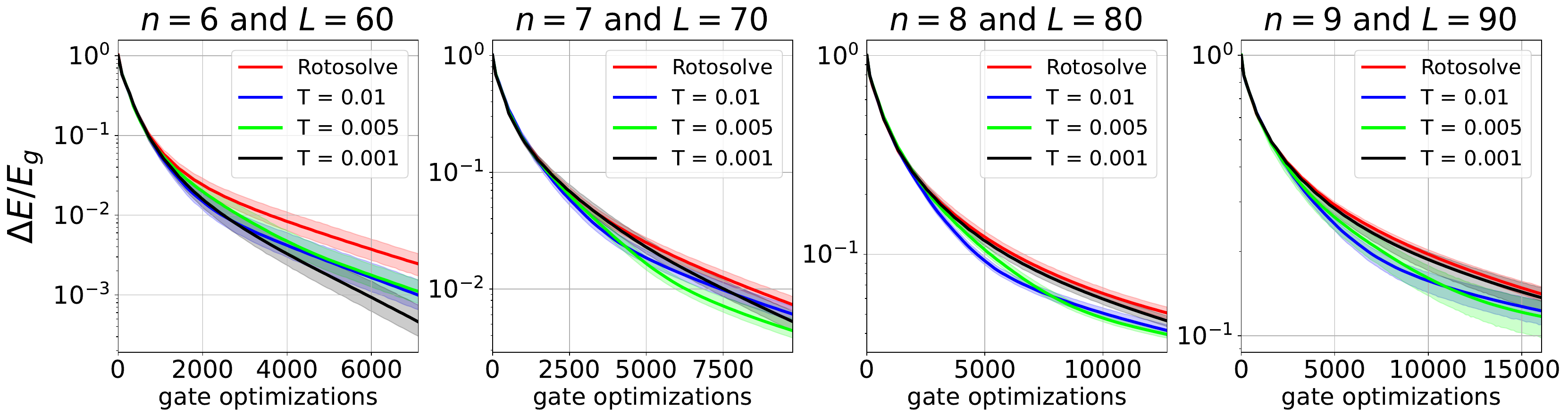}
   \cprotect\caption{Relative errors for the Heisenberg model for qubits ranging from 6 to 9, incrementing by 1. Base \verb|Rotosolve| (red) and gate freezing method with the parameter-based distance metric for $L=10n$ layers and ansatz from Fig.~\ref{Ansatz_circuit_image} was used. Gate freezing threshold was set to $T = 0.01$ (blue), $0.005$ (green), $0.001$ (black), and incremental gate freezing was used. Each line represents the mean of 20 runs, and the shaded area is the 90\% confidence interval around the mean.}
    \label{scalability_one_R_layer_means}
\end{figure*}

\subsubsection{Parameter Distances}

We tested fixed and incremental freezing for all optimizers with $L=3n$ layers and used the same ansatz circuit as for the Heisenberg model. The number of iterations in each run was set to 10. The gate freezing methods were executed until they reached a total number of gate iterations equal to that of one run of the base version of the corresponding optimizer.

The results for \verb|Rotosolve| are shown in Fig.~\ref{rotosolve_hubbard_results}. For fixed freeze iterations $\kappa=2$, all thresholds obtain similar convergences at the end of the optimization process. The boxplots for $\kappa=2$ in Fig.~\ref{rotosolve_hubbard_results} are much more weighted towards the ground state, and all thresholds have similar boxplots. When we set $\kappa=5$, the threshold value $T=0.001$ shows the fastest convergence speed and achieves one magnitude better mean value across 20 runs at the end of the optimization process. Thresholds $ T=0.005$ and $T=0.01$ have nearly identical results. As for the incremental gate freezing method, it performs as well as fixed gate freezing with $\kappa=5$ in all threshold values.

We present the results for \verb|Fraxis| optimizer in Fig.~\ref{fraxis_hubbard_results}. Again, we have similar results compared to the previous section, where the base \verb|Fraxis| has overall the worst performance compared to gate freezing methods, regardless of the threshold values $T$. Here, the overall trend for the convergence of the mean is noticeably better than base \verb|Fraxis|. Also, in some cases, the median value at the end of the optimization is nearly a magnitude better than the base \verb|Fraxis|. This is in the case for $T=0.025$ with fixed gate freezing.

The results for \verb|FQS| optimizer are shown in Fig.~\ref{fqs_hubbard_results}. All gate freezing methods with any threshold value and the base \verb|FQS| have similar convergences of the mean. However, the boxplots in the bottom row in Fig.~\ref{fqs_hubbard_results} exhibit a better range of results and median values for $T=0.005$ in both fixed and incremental gate freezing methods.

The results for shallow PQCs simulations with shot noise using 1024 and 4096 shots are shown in Fig.~\ref{2DFermiHubbard_noisy}. Here, we performed a total of 30 runs and used 4 layers for all optimizers. \verb|Fraxis| has the greatest improvement using incremental gate freezing, but \verb|Rotosolve| also benefits from gate freezing when the gate freezing threshold $T$ is well chosen. For \verb|FQS|, there is no noticeable improvement with gate freezing. We also observed that the improvement from gate freezing gradually decreases as the circuit depth increases. This indicates that the gate freezing methods are most useful for shallow circuits with only a few layers of optimizable gates when the shot noise is used.

\subsubsection{Matrix Norm Distances}

Now we present our results for \verb|Fraxis| and \verb|FQS| with matrix norm as distance metric. The experiments were performed in the same way as with the parameter-based distance metric. A total of 20 runs with 10 iterations for \verb|Fraxis| and  \verb|FQS| with freeze iterations set to $\kappa=3$. The incremental gate freezing algorithms were executed until they reached a total number of gate iterations equal to that of one run of the base version of the corresponding optimizer. In all gate freezing settings, the thresholds were set to $T=0.025, 0.01$, and $0.005$.

The results for \verb|Fraxis| with matrix norm are shown in Fig.~\ref{fraxis_hubbard_results_matrix_norm}. When the fixed gate freezing iterations are set to $\kappa=3$, the best mean convergence is obtained with $T=0.01$ or $T=0.025$. All thresholds have a faster convergence of the mean and a better confidence interval of 90\% around the mean. The performance of incremental freezing improves as the threshold value decreases, the mean and median values get better, and the boxplot gets narrower, indicating a more compact distribution of the runs at the end of the optimization.

The results in Fig.~\ref{fqs_hubbard_results_matrix_norm} indicate that the \verb|FQS| benefits with a higher threshold value $T=0.025$ with fixed gate freezing iterations and a lower threshold $T=0.005$ with incremental gate freezing. Here, in both cases, the mean converges faster towards the ground state, and by extrapolating the results, the base \verb|FQS| would need more gate optimizations than gate freezing methods with the threshold values mentioned above.

\subsection{Scalability}

In our final experiment, we investigated the performance of the proposed method by evaluating the impact of errors as a function of system size, by varying the number of qubits in circuits implementing the Heisenberg model Hamiltonian from Eq.~(\ref{Heisenberg_hamiltonian}). We applied the \verb|Rotosolve| optimizer to deep quantum circuits characterized with qubit counts ranging from 6 to 9, increments of one. The number of layers for each circuit was set to $L = 5n$, where $n$ denotes the number of qubits. The ansatz circuit used in this experiment is the same as the one depicted in Fig.~\ref{roto_gd_ansatz}. The incremental gate freezing method was employed using gate freezing thresholds of $T=0.01, 0.005, 0.001$. Each scenario involved a total of 20 runs, with optimization limited to 10 iterations per run. We examined the relative error compared to the ground state. The exact ground state of the system was computed using the QuSpin Python package \cite{Quspin_package}.

The results are presented in Fig.~\ref{scalability_boxplot} as box plots. The base version of \verb|Rotosolve| exhibits the highest relative error compared to the incremental gate freezing method by varying the qubit count. In comparison to the results in Sec.~\ref{Heisenberg_section}, we have similar results for 5 qubits. As we increased the number of qubits and layers in the quantum circuits, we observed a growing relative error relative to the reference ground state. It is important to note that the computational complexity of the optimization process intensifies significantly, given that the size of the system expands exponentially in the sense of Hilbert space dimensionality. Although the relative error increases with the size of the circuit, the incremental gate freezing method, in most cases, outperforms the base version of \verb|Rotosolve|. Additionally, we present the convergence of the mean for each system size in Fig.~\ref{scalability_RxRy_layer_means}. For all system sizes, we recognize a trend for the incremental freezing, where they approach the true ground state faster than the base \verb|Rotosolve|. As the number of qubits increases, the trend is harder to distinguish with a limited number of iterations. We also note that the relative error is significantly increased with 9 qubits compared to 6 qubits.

We then examined the performance of the incremental gate freezing method on PQCs with $L=10n$ layers, where $n$ is the number of qubits in the PQC. A similar approach was used in Ref.~\cite{mcclean2018barren} with random quantum circuits. Instead of using a fixed ansatz of $R_X$ and $R_Y$ gates depicted in Fig.~\ref{roto_gd_ansatz}, we used the ansatz circuit from Fig.~\ref{Ansatz_circuit_image}. At the beginning of each run, we randomly choose each unitary $R_d$ to be one of the rotation gates $R_X, R_Y, R_Z$, that is, $R_d \in\{R_X, R_Y, R_Z\}$. This allows us to create random deep PQCs and examine whether the choice of rotation gates has an effect on the results. A total of 20 runs were executed, with 20 iterations of optimization, and for each run, a random PQC was created for the optimization.

We present the results for random deep PQCs in Fig.~\ref{scalability_boxplot_random_PQC} as boxplots. The results are similar to Fig.~\ref{scalability_boxplot}, where the interquartile range of incremental gate freezing with some threshold $T$ in every system size is better than the base \verb|Rotosolve|. Also, with $n=6$ and $n=7$, all median values, regardless of the threshold of incremental freezing, exceed base \verb|Rotosolve|. Furthermore, we depict the convergence of the means with a 90\% confidence interval around the mean in Fig.~\ref{scalability_one_R_layer_means}. Here, the convergence is more clearly visualized, which highlights the practical application of the gate freezing method in deep random PQCs. The threshold value $T=0.005$ performs best with various numbers of qubits, followed by $T=0.01$. When the threshold is small, it does not necessarily scale the best, as in the smaller system sizes. That is, with 5 and 6 qubits, $T=0.001$ performs better than for 7, 8, or 9 qubits.

We emphasize that the observed improvements of the gate freezing with an increasing system size in both width and depth should not be interpreted as mitigation of barren plateaus. The gate freezing focuses on resource allocation and does not affect the global cost landscape in PQC optimization. Additionally, as the system size grows, more iterations are required to obtain a better convergence compared to the base version of \verb|Rotosolve|.

\section{Conclusions} \label{conclusions_section}

Our experiments show that gate freezing methods can improve the convergence speed and overall performance of existing optimizers \verb|Rotosolve|, \verb|Fraxis|, and \verb|FQS|. Our methods are based on measuring the distance between the previous and new parameter value of each gate after gate optimization, which provides valuable information that has not been extensively used before in the VQAs. Our methods cover both the matrix norm approach and the parameter-based distance between previous and current gate updates.

\begin{table}[h] 
\renewcommand{\arraystretch}{1.3}
\centering
\scalebox{1.05}{
 \begin{tabular}{|c|c | c | c |} 
 \hline
 Hamiltonian & Base Rotosolve & Relative Impr. & Best $T$  \\ \hline
 Heisenberg &  $1.73 \cdot 10^{-6}$  & $0.941$ & 0.01 \\ \hline
 Fermi-Hubbard & $6.11 \cdot 10^{-5}$ & $0.746$ & 0.001 \\ \hline
 \end{tabular}}
 \cprotect\caption{Collection of results for \verb|Rotosolve| and best incremental freezing method with threshold $T$ for 5-qubit Heisenberg and 4-qubit Fermi-Hubbard Hamiltonians. The median is used as a metric for both baseline \verb|Rotosolve| and relative improvement compared to the best performing $T$ of incremental freezing.}
 \label{table_results_rotosolve_overview}
\end{table}

\begin{table*} 
\renewcommand{\arraystretch}{1.3}
\centering
\scalebox{1.2}{
 \begin{tabular}{|c|c | c | c |c |} 
\hline
 Hamiltonian & Base Fraxis & Relative Improvement & Best $T$ & metric $\D$  \\ \hline
 Heisenberg &  $4.22 \cdot 10^{-5}$  & $0.815$ & 0.025 & Matrix norm \\ \hline
 Fermi-Hubbard & $7.05 \cdot 10^{-4}$ & $0.859$ & 0.01 & Parameter distance\\ \hline \addlinespace[0.3cm] \hline
  Hamiltonian & Base FQS & Relative Improvement & Best $T$ & metric $\D$  \\ \hline
 Heisenberg &  $9.62 \cdot 10^{-8}$  & $0.766$ & 0.005 & Parameter distance \\ \hline
 Fermi-Hubbard & $1.54 \cdot 10^{-5}$ & $0.849$ & 0.005 & Parameter distance\\ \hline
 \end{tabular}}
 \cprotect\caption{Collection of results for \verb|Fraxis| and \verb|FQS| optimizers with the best threshold $T$ for 5-qubit Heisenberg and 4-qubit Fermi-Hubbard Hamiltonians using the incremental freezing method. The median is used as the metric for the baseline optimizers and relative improvement compared to the best performing $T$ of incremental freezing.}
 \label{table_results_fraxis_fqs_overview}
\end{table*}

The proposed freezing algorithms with fixed threshold $T$ and freeze iterations $\kappa$ often outperformed their base versions when optimizing the 5-qubit Heisenberg Hamiltonian for \verb|Rotosolve| and \verb|Fraxis|, whereas the improvements for \verb|FQS| were smaller. By incorporating gate freezing into existing VQAs, one can improve the performance by reallocating the resources to more poorly optimized gates. We observed this phenomenon with both statevector simulations in deep circuits and simulations with shot noise for shallow PQCs in some cases.

In addition, we tested a dynamic version of gate freezing, where the number of gate freezes for the specific gate increases by one whenever the gate is frozen. This also yielded favorable results for the freezing algorithms. With \verb|Rotosolve| and \verb|Fraxis|, this was the case, but for the high expressivity \verb|FQS| it only gave a minor advantage when optimizing the Fermi-Hubbard Hamiltonian. The results for the incremental gate freezing method with the Heisenberg and Fermi-Hubbard Hamiltonians are summarized in Table~\ref{table_results_rotosolve_overview} for \verb|Rotosolve| and in Table~\ref{table_results_fraxis_fqs_overview} for \verb|Fraxis| and \verb|FQS|, respectively. In both tables, we report the median values at the end of the optimization (see boxplots in Sec.~\ref{Heisenberg_section} and~\ref{section_fermi_hubbard_results}), and compare the best incremental gate freezing configuration defined by the threshold $T$ and metric $\D$ with the corresponding baseline optimizer. The relative improvement in the median value with respect to the baseline optimizer is also reported. 

We also examined how often the parameter update of each gate falls below a given threshold $T$ between consecutive iterations using a parameter-based metric. We computed the average value of the gate freeze iteration vector $\bm{\kappa}_d$ for each gate, where $\bm{\kappa}_d$ is incremented by one if the change of the $d$-th parameter between consecutive iterations is smaller than $T$. We observed several notable trends; the average value for $\bm{\kappa}_d$ is higher in the later layers compared to the earlier layers. The center qubit in the last layer has the highest average for all optimizers, regardless of the freezing threshold $T$ for the 5-qubit system with 3 layers. This observation is related to the construction of the ansatz circuit; additional results are provided in the Appendix~\ref{appendix_threshold_counts_ansatzes}. With different ansätze, we observe different gates that obtain higher average values for $\bm{\kappa}_d$ in the circuit. We obtained similar results with various qubit counts while keeping the number of layers fixed when using \verb|Rotosolve|. Additionally, we investigated the averages for each $\bm{\kappa}_d$ on deep circuits with $L=5n$ layers. Another notable observation concerns the \verb|FQS| optimizer, once a certain circuit depth is exceeded while keeping the threshold $T$ fixed, nearly all gates except those in the first layer obtain high average values for $\bm{\kappa}_d$. The average values of $\bm{\kappa}_d$ for \verb|Rotosolve| and \verb|Fraxis| increase progressively with circuit depth in deep PQCs.

Then, we experimented with our methods on the 4-qubit Fermi-Hubbard model on a $1\times2$ lattice with a statevector simulator and simulations with 1024 and 4096 shots. Again, gate freezing improved the results relative to the base versions of the optimizers, although the improvement was not consistent for the parameter-based distance metric. The best improvement was observed for \verb|Rotosolve| and \verb|Fraxis|, while \verb|FQS| showed smaller improvements with gate freezing. In the presence of shot noise, the clearest improvements were observed for \verb|Fraxis|, while \verb|Rotosolve| showed smaller improvements depending on the threshold of the gate freezing. For \verb|FQS|, there was no noticeable improvement in the presence of shot noise.

Finally, we examined how the incremental gate freezing method behaves with an increasing system size using the \verb|Rotosolve| optimizer. The number of qubits was varied from 6 to 9 qubits, incrementing by one, and the ansatz circuit in Fig.~\ref{roto_gd_ansatz} with $L=5n$ layers was used. In addition, we evaluated the performance of the incremental gate freezing method while optimizing random deep PQCs with $L=10n$ layers using the ansatz circuit depicted in Fig.~\ref{Ansatz_circuit_image}. The results of the relative error to the ground state indicate that the gate freezing method is beneficial when the complexity of the circuit increases linearly with both depth and width. However, this comes at the cost of a larger number of iterations being required to observe the improvements in performance. We note that the gate freezing method does not resolve or mitigate barren plateaus that emerge in deep and random PQCs, as the gate freezing methods presented in this work focus on resource allocation to improve existing optimizers \verb|Rotosolve|, \verb|Fraxis|, and \verb|FQS|.

The results for the number of gate freezing iterations in Sec.~\ref{gate_freeze_iters_section} provide information on how the parameters evolve on average during the optimization process. This suggests that there may be more effective ways to create the gate optimization sequence. We intend to explore this behavior further and how to utilize it in the PQC optimization. In future work, one could also explore how the gate freezing methods could be implemented in other VQAs, such as variational trace distance estimation~\cite{variational_trace_distance_estimation}, variational imaginary-time evolution~\cite{variational_imaginary_time_evolution}, and variational quantum singular value decomposition~\cite{variational_quantum_singular_value_decomposition}. Furthermore, gate freezing methods on the gradient-free optimizers could be studied further in the context of exponential hardness of optimization from the locality~\cite{exponential_hardness_locality_QNN}, where the entire variation range of the cost function vanishes exponentially when optimizing parameters of any local quantum gates. This becomes evident as the number of qubits increases. 

To conclude this work, we emphasize that our method could be extended to any $n$-qubit gates using the matrix norm as a distance metric for gate freezing. Our work can also be extended to other types of VQAs by following the idea in Sec.~\ref{param_sec} if parameter-based gate freezing is desired.

\section{Data availability}
The source code and data to generate the figures in the paper are available at Ref.~\cite{code_repository}. 

\section*{Author Contributions}

J.V.P. developed the code for the gate freezing algorithms, conducted the numerical experiments, and prepared the figures. J.V.P., M.R., and L.Y. contributed to the theoretical analysis and interpretation of results.  J.V.P. wrote the original manuscript. L.Y., A.M., M.R., and I.T. reviewed the manuscript and provided critical feedback that helped shape the research. I.T. supervised the project and was responsible for funding acquisition and project management. 

\begin{acknowledgments}
We acknowledge funding by Business Finland for the project 8726/31/2022 CICAQU. J.V.P. received funding from InstituteQ's doctoral school. 
\end{acknowledgments}

\appendix

\section{Ansatz Circuits} \label{appendix_circuits}

Here, we provide ansatz circuits used in additional experiments in the Appendix~\ref{appendix_threshold_counts_ansatzes}. Ansatz type B is illustrated in Fig.~\ref{ansatz_circuit_B}, type C in Fig.~\ref{ansatz_circuit_C}, and type D in Fig.~\ref{ansatz_circuit_D}.

\begin{figure}
    \[
    \Qcircuit @C=0.95em @R=.9em {
    & & \mbox{$L$ layers} & & & & &  \\
    & & & & & & & & & \\
     \lstick{\ket{0}_1} & \gate{R_X\Bigl(\theta_{ln + 1}\Bigr)} & \gate{R_Y\Bigl(\theta_{ln + 6}\Bigr)} &  \ctrl{0} & \qw & \qw & \qw & \qw & \meter\\
     \lstick{\ket{0}_2} &  \gate{R_X\Bigl(\theta_{ln + 2}\Bigr)} & \gate{R_Y\Bigl(\theta_{ln + 7}\Bigr)} & \ctrl{-1} & \ctrl{1} & \qw  &  \qw & \qw & \meter\\
     \lstick{\ket{0}_3} &  \gate{R_X\Bigl(\theta_{ln + 3}\Bigr)}  & \gate{R_Y\Bigl(\theta_{ln + 8}\Bigr)} & \qw & \ctrl{0} & \ctrl{0}  & \qw & \qw & \meter\\
     \lstick{\ket{0}_4} &  \gate{R_X\Bigl(\theta_{ln + 4}\Bigr)}  & \gate{R_Y\Bigl(\theta_{ln + 9}\Bigr)} & \qw & \qw & \ctrl{-1}  & \ctrl{0} & \qw & \meter\\
     \lstick{\ket{0}_5} &  \gate{R_X\Bigl(\theta_{ln + 5}\Bigr)} & \gate{R_Y\Bigl(\theta_{ln + 10}\Bigr)} & \qw & \qw & \qw & \ctrl{-1} & \qw & \meter  \gategroup{3}{2}{7}{7}{1.2em}{--} 
    }
    \]
    \\
    \[
    \Qcircuit @C=1.5em @R=.9em {
    & & \mbox{$L$ layers} & & & & &  \\
    & & & & & & & & & \\
     \lstick{\ket{0}_1} & \gate{R_{ln+1}\Bigl(\theta_{ln + 1}\Bigr)} &  \ctrl{0} & \qw & \qw & \qw & \qw & \meter\\
     \lstick{\ket{0}_2} & \gate{R_{ln+2}\Bigl(\theta_{ln + 2}\Bigr)}  & \ctrl{-1} & \ctrl{1} & \qw  &  \qw & \qw & \meter\\
     \lstick{\ket{0}_3} & \gate{R_{ln+3}\Bigl(\theta_{ln + 3}\Bigr)}   & \qw & \ctrl{0} & \ctrl{0}  & \qw & \qw & \meter\\
     \lstick{\ket{0}_4} & \gate{R_{ln+4}\Bigl(\theta_{ln + 4}\Bigr)}   & \qw & \qw & \ctrl{-1}  & \ctrl{0} &\qw & \meter\\
     \lstick{\ket{0}_5} & \gate{R_{ln+5}\Bigl(\theta_{ln + 5}\Bigr)}  & \qw & \qw & \qw & \ctrl{-1} & \qw & \meter  \gategroup{3}{2}{7}{6}{1.2em}{--} 
    }
    \]
    \cprotect\caption{Ansatz circuits B1 (top) and B2 (bottom) with a cascade entangling layer.}
    \label{ansatz_circuit_B}
\end{figure}

\begin{figure}
    \centering
    \[
    \Qcircuit @C=0.95em @R=.9em {
    & & \mbox{$L$ layers} & & & & & &  \\
    & & & & & & & & & & \\
     \lstick{\ket{0}_1} & \gate{R_X\Bigl(\theta_{ln + 1}\Bigr)} & \gate{R_Y\Bigl(\theta_{ln + 6}\Bigr)} &  \ctrl{0} & \qw & \qw & \qw & \ctrl{4} & \qw & \meter\\
     \lstick{\ket{0}_2} &  \gate{R_X\Bigl(\theta_{ln + 2}\Bigr)} & \gate{R_Y\Bigl(\theta_{ln + 7}\Bigr)} & \ctrl{-1} & \ctrl{1} & \qw  & \qw & \qw & \qw & \meter\\
     \lstick{\ket{0}_3} &  \gate{R_X\Bigl(\theta_{ln + 3}\Bigr)}  & \gate{R_Y\Bigl(\theta_{ln + 8}\Bigr)} & \qw & \ctrl{0} & \ctrl{0}  & \qw & \qw & \qw & \meter\\
     \lstick{\ket{0}_4} &  \gate{R_X\Bigl(\theta_{ln + 4}\Bigr)}  & \gate{R_Y\Bigl(\theta_{ln + 9}\Bigr)} & \qw & \qw & \ctrl{-1}  & \ctrl{0} & \qw & \qw & \meter\\
     \lstick{\ket{0}_5} &  \gate{R_X\Bigl(\theta_{ln + 5}\Bigr)} & \gate{R_Y\Bigl(\theta_{ln + 10}\Bigr)} & \qw & \qw & \qw & \ctrl{-1} & \ctrl{-1} & \qw & \meter  \gategroup{3}{2}{7}{8}{1.2em}{--} 
    }
    \]
    \\[0.05cm]
    \[
    \Qcircuit @C=1.4em @R=.9em {
    & & \mbox{$L$ layers} & & & & & & \\
    & & & & & & & & & & \\
     \lstick{\ket{0}_1} & \gate{R_{ln+1}\Bigl(\theta_{ln + 1}\Bigr)} &  \ctrl{0} & \qw & \qw & \qw & \ctrl{4} & \qw & \meter\\
     \lstick{\ket{0}_2} & \gate{R_{ln+2}\Bigl(\theta_{ln + 2}\Bigr)}  & \ctrl{-1} & \ctrl{1} & \qw  &  \qw & \qw & \qw & \meter\\
     \lstick{\ket{0}_3} & \gate{R_{ln+3}\Bigl(\theta_{ln + 3}\Bigr)}   & \qw & \ctrl{0} & \ctrl{0}  & \qw & \qw & \qw & \meter\\
     \lstick{\ket{0}_4} & \gate{R_{ln+4}\Bigl(\theta_{ln + 4}\Bigr)}   & \qw & \qw & \ctrl{-1}  & \ctrl{0} & \qw & \qw & \meter\\
     \lstick{\ket{0}_5} & \gate{R_{ln+5}\Bigl(\theta_{ln + 5}\Bigr)}  & \qw & \qw & \qw & \ctrl{-1} & \ctrl{0} & \qw & \meter  \gategroup{3}{2}{7}{7}{1.2em}{--} 
    }
    \]
    \cprotect\caption{Ansatz circuits C1 (top) and C2 (bottom) with a cyclic entangling layer.}
    \label{ansatz_circuit_C}
\end{figure}

\begin{figure}
    \centering
    \[
    \Qcircuit @C=0.8em @R=.9em {
    & & \mbox{$L$ layers} & & & & & & \\
    & & & & & & & & & & \\
     \lstick{\ket{0}_1} & \gate{R_X\Bigl(\theta_{ln + 1}\Bigr)} & \gate{R_Y\Bigl(\theta_{ln + 6}\Bigr)} &  \ctrl{0} & \ctrl{2} & \ctrl{3} & \ctrl{4} & \qw & \meter\\
     \lstick{\ket{0}_2} &  \gate{R_X\Bigl(\theta_{ln + 2}\Bigr)} & \gate{R_Y\Bigl(\theta_{ln + 7}\Bigr)} & \ctrl{-1} & \qw & \qw  &  \qw & \qw & \meter\\
     \lstick{\ket{0}_3} &  \gate{R_X\Bigl(\theta_{ln + 3}\Bigr)}  & \gate{R_Y\Bigl(\theta_{ln + 8}\Bigr)} & \qw & \ctrl{0} & \qw  & \qw & \qw & \meter\\
     \lstick{\ket{0}_4} &  \gate{R_X\Bigl(\theta_{ln + 4}\Bigr)}  & \gate{R_Y\Bigl(\theta_{ln + 9}\Bigr)} & \qw & \qw & \ctrl{-1}  & \qw & \qw & \meter\\
     \lstick{\ket{0}_5} &  \gate{R_X\Bigl(\theta_{ln + 5}\Bigr)} & \gate{R_Y\Bigl(\theta_{ln + 10}\Bigr)} & \qw & \qw & \qw & \ctrl{-1} & \qw & \meter  \gategroup{3}{2}{7}{7}{1.2em}{--} 
    }
    \]
    \\
    \[
    \Qcircuit @C=1.4em @R=.9em {
    & & \mbox{$L$ layers} & & & & &  \\
    & & & & & & & & & \\
     \lstick{\ket{0}_1} & \gate{R_{ln+1}\Bigl(\theta_{ln + 1}\Bigr)} &  \ctrl{0} & \ctrl{2} & \ctrl{3} & \ctrl{4} & \qw & \meter\\
     \lstick{\ket{0}_2} & \gate{R_{ln+2}\Bigl(\theta_{ln + 2}\Bigr)}  & \ctrl{-1} & \qw & \qw  &  \qw & \qw & \meter\\
     \lstick{\ket{0}_3} & \gate{R_{ln+3}\Bigl(\theta_{ln + 3}\Bigr)}   & \qw & \ctrl{-1} & \qw  & \qw & \qw & \meter\\
     \lstick{\ket{0}_4} & \gate{R_{ln+4}\Bigl(\theta_{ln + 4}\Bigr)}   & \qw & \qw & \ctrl{-1}  & \qw &\qw & \meter\\
     \lstick{\ket{0}_5} & \gate{R_{ln+5}\Bigl(\theta_{ln + 5}\Bigr)}  & \qw & \qw & \qw & \ctrl{-1} & \qw & \meter  \gategroup{3}{2}{7}{6}{1.2em}{--} 
    }
    \]
    \cprotect\caption{Ansatz circuits D1 (top) and D2 (bottom) with a one-qubit connector entangling layer.}
    \label{ansatz_circuit_D}
\end{figure}

\section{Gradient-Free Optimizers} \label{appendix_optimizers}

\subsection{Rotosolve Optimizer}

In this section, we generally describe how the gradient-free optimizer, \verb|Rotosolve|~\cite{Ostaszewski_2021} works. First, we consider that one ansatz circuit layer $U_l(\bm{\theta}_l)$ consists of $2n$ different parameterized single-qubit gates. In this setting, two parameterized single-qubit gates act on each qubit in sequence, $R_X$ and $R_Y$ gates, respectively, followed by the entanglement layer $W_l$. Then the unitary $U_l(\bm{\theta}_l)$ becomes 
\begin{widetext}
\begin{equation} \label{layer_unitary}
    U_l(\bm{\theta}_l) = W_l \left(\bigotimes_{k=1}^{n} e^{-i\theta_{n(2l+1)+k} H_{n(2l+1)+k} / 2} \right)\left(\bigotimes_{k'=1}^{n} e^{-i\theta_{2nl+k'} H_{2nl+k'} / 2} \right).
\end{equation}
\end{widetext}
Here, the Hermitian operators $H_{2nl+k'}$ and $H_{n(2l+1)+k}$ denote the Pauli $X$ and $Y$ operators according to Fig.~\ref{roto_gd_ansatz}, respectively.

The ansatz circuit has a total of $L$ layers. Then, the entire vector $\bm{\theta}$ consists of the $2Ln$ parameters $\newline \bm{\theta} = \left(\theta_1, \dots ,\theta_{2Ln} \right)$. An illustration of this circuit is shown in Fig.~\ref{roto_gd_ansatz}.

The $d$-th single qubit rotation gate $R_d ~ (d= 1,\dots,2Ln)$ with the parameter $\theta_d \in (-\pi, \pi ]$ is defined as \cite{nielsen2010quantum}
\begin{equation} \label{single_qubit_gate}
    R_d(\theta_d) = \cos\left(\frac{\theta_d}{2} \right) I - i \sin\left(\frac{\theta_d}{2} \right)H_d.
\end{equation}
Here, $H_d$ is the Hermitian unitary generator that satisfies the condition $H_d^2=I$ and $I$ is the $2\times2$ identity matrix. To operate the gate $R_d$ in the circuit on the $k$-th qubit, it needs to be represented as a $2^n\times2^n$ unitary matrix. We achieve this by applying a tensor product with $2\times 2$ identity matrices as follows:
\begin{eqnarray} \label{roto_gate}
    R_d '(\theta_d) =  && I^{\otimes(k-1)}  \otimes \left[ \cos\left(\frac{\theta_d}{2} \right)I - i\sin\left(\frac{\theta_d}{2} \right)H_d \right] \nonumber \\
    &&\otimes I^{\otimes (n-k)}.
\end{eqnarray}\\

By applying Eq.~(\ref{roto_gate}) to Eq.~(\ref{layer_unitary}), we obtain the following form for the cost function \cite{pankkonen2025improvingvariationalquantumcircuit}

\begin{widetext}
\begin{equation} \label{hat_M}
\begin{split}
        \langle \hat{M} \rangle = \text{Tr} & \Bigl( \hat{M} W_{L-1}  R_{2Ln}'(\theta_{2Ln}) \cdots R_{d+1}'(\theta_{d+1})R_{d}'(\theta_{d}) R_{d-1}'(\theta_{d-1}) \cdots R_{2}'(\theta_{2}) R_{1}'(\theta_{1}) \\
        & \times \rho_0 R_1'(\theta_1)^\dagger R_{2}'(\theta_{2})^\dagger \cdots R_{d-1}'(\theta_{d-1})^\dagger R_{d}'(\theta_{d})^\dagger R_{d+1}'(\theta_{d+1})^\dagger \cdots R_{2Ln}'(\theta_{2Ln})^\dagger W_{L-1}^\dagger\Bigr).
\end{split}
\end{equation}
\end{widetext}
Now, we define the quantum circuits that come before and after the $d$-th parameterized gate $R_d'$ as $V_1$ and $V_2$, respectively. This allows us to write Eq.~(\ref{hat_M}) in more compact form
\begin{equation}
    \langle \hat{M} \rangle = \text{Tr} \Bigl( \hat{M} V_1 R_d'(\theta_d)V_2 \rho_0 V_2 ^\dagger R_d'(\theta_d)^\dagger V_1^\dagger \Bigr).
\end{equation}
By further applying the cyclic property of the trace operation, we can absorb the quantum circuits $V_1$ and $V_2$ into the new variables $M$ and $\rho$ as follows
\begin{eqnarray}
     \rho & \equiv V_2 \rho_0 V_2 ^\dagger,\label{new_rho} \\
     M  &\equiv V_1^\dagger \hat{M} V_1. \label{new_M}
\end{eqnarray}
Finally, we obtain the relevant form of the cost function similar to Refs.~\cite{Ostaszewski_2021, fraxis, pankkonen2025improvingvariationalquantumcircuit}
\begin{equation} \label{new_cost_function}
    \langle M \rangle = \text{Tr} \Bigl( M  R_d'(\theta_d) \rho R_d'(\theta_d)^\dagger  \Bigr).
\end{equation}\\
By using Eq.~(\ref{single_qubit_gate}), the cost function becomes \cite{Ostaszewski_2021}
\begin{eqnarray} \label{roto_traces}
    \langle M \rangle_{\theta_d} \ &&= \cos^2\left(\frac{\theta_d}{2} \right) \text{Tr}(M \rho ) \nonumber \\
    &&+ \ i  \cos\left(\frac{\theta_d}{2} \right)  \sin\left(\frac{\theta_d}{2} \right) \text{Tr}(M [\rho, H_d'] )  \nonumber \\
    &&+\ \sin^2\left(\frac{\theta_d}{2} \right) \text{Tr}(M H_d'\rho H_d').
\end{eqnarray}
Here, $H_d'$ is a $2^n \times 2^n$ matrix that consists of a tensor product of the Hermitian generator $H_d$ and $2\times 2$ identity matrices. This follows straight from Eq.~(\ref{roto_gate}). By evaluating the cost function at angles $\theta_d =-\pi/2, 0, \pi/2$ and $\pi$, one can solve the values for the traces in Eq.~(\ref{roto_traces}). After that, the cost function $\langle M \rangle_{\theta_d}$ can be written as $\langle M \rangle_{\theta_d} = A \sin(\theta_d + B) + C$, where the coefficients $A, B$ and $C$ are combinations of measurements at different angles $\theta_d$ \cite{Ostaszewski_2021}. The minimum of the cost function is located at $\theta_d^* = -\frac{\pi}{2} - B + 2\pi m$, where $m\in \mathbb{Z}$ and the coefficient $B$ defined as
\begin{eqnarray}
    B &= \arctan2 \bigl( 2 \langle M \rangle_{\phi} - \langle M \rangle_{\phi+\frac{\pi}{2}} - \langle M \rangle_{\phi-\frac{\pi}{2}}, \nonumber \\
    & \langle M \rangle_{\phi+\frac{\pi}{2}} - \langle M \rangle_{\phi-\frac{\pi}{2}}\bigr) - \phi.
\end{eqnarray}
Here, $\phi \in \mathbb{R}$ and it can be chosen arbitrarily. We chose to set it to zero in this work. The minimum value for the angle $\theta_d$ can finally be written as follows (see Ref.~\cite{Ostaszewski_2021})
\begin{eqnarray}
    \theta_d ^* &&= \underset{\theta_d}{\arg \min} \langle M \rangle_{\theta_d} \nonumber \\
    &&=  - \frac{\pi}{2} - \arctan2 (2  \langle M \rangle_{\phi} -  \langle M \rangle_{\phi+ \frac{\pi}{2}} - \langle M \rangle_{\phi - \frac{\pi}{2}}, \nonumber \\ && \quad \quad \quad \quad \quad \quad \quad \quad  \langle M \rangle_{\phi+ \frac{\pi}{2}} - \langle M \rangle_{\phi - \frac{\pi}{2}} ) +2\pi m. \qquad \quad
\end{eqnarray}

\subsection{Free-Axis Selection Optimizer}

In this and the following sections, we will go through two matrix diagonalization optimizers for the single-qubit gate optimization, \verb|Fraxis|~\cite{fraxis} and \verb|FQS|~\cite{fqs}. The \verb|Fraxis| optimizer works like a \verb|Rotosolve| optimizer: it is a sequential single-qubit gate optimizer, which optimizes the PQC one gate at a time and uses circuit evaluations to determine the optimal rotation axis of the gate with a fixed rotation angle $\theta_d=\pi$.

Instead of a fixed axis of rotation (i.e. $R_X, R_Y$ or $R_Z$), we represent the $d$-th rotation gate $R_d \ (d=1,\ldots ,Ln)$ with a rotation angle $\theta_d$ and a unit vector $\hat{\bm{n}}_d$ as follows \cite{fraxis}
\begin{eqnarray}
    R_d(\theta_d) \ &&= e^{-i \frac{\theta_d}{2} \hat{\bm{n}}_d \cdot \bm{\sigma}} = \cos\left(\frac{\theta_d}{2} \right)I - i \sin\left(\frac{\theta_d}{2} \right) \hat{\bm{n}} \cdot \bm{\sigma} \nonumber \\ && \equiv R_{\hat{\bm{n}}}(\theta_d),
\end{eqnarray}
where $\bm{\sigma} = (X,Y,Z)$ are the Pauli matrices and ${\hat{\bm{n}}_d} = (n_{x,d}, n_{y,d}, n_{z,d})$ are the components of the unit vector.

Now, like the \verb|Rotosolve| gates, we express the $d$-th gate as a $2^n \times 2^n$ matrix. We apply the tensor product of the unit $2\times2$ matrices to the $d$-th gate that acts on a particular qubit, hence
\begin{eqnarray} \label{fraxis_gate}
    R_{\hat{\bm{n}}_d}'(\theta_d) = \   &&I^{\otimes(k-1)}  \otimes \left[ \cos\left(\frac{\theta_d}{2} \right)I - i\sin\left(\frac{\theta_d}{2} \right)(\hat{\bm{n}} \cdot \bm{\sigma}) \right] \nonumber \\ && \otimes I^{\otimes(n-k)} .
\end{eqnarray}

Applying the same procedure as in the previous section for \verb|Rotosolve|, that is, we first apply Eq.~(\ref{fraxis_gate}) to Eq.~(\ref{U_one_layer}). Here, we remark that we use the ansatz circuit from Fig.~\ref{Ansatz_circuit_image}. After that, a similar form of the cost function is obtained as in Eq.~(\ref{hat_M}). Then, we define the quantum circuits that enter before and after the $d$-th gate as $V_1$ and $V_2$, respectively. Here, the values of $V_1$ and $V_2$ depend on the values of $\hat{\bm{n}}_k$ and $\theta_k$ where $k\neq d$. Using the same definitions as in Eqs.~(\ref{new_rho}) and (\ref{new_M}), we arrive at the same form for the cost function as in Eq.~(\ref{new_cost_function})
\begin{eqnarray}
    \langle M \rangle_{\hat{\bm{n}}_d, \theta_d} =  \text{Tr} \Bigl( M R_{\hat{\bm{n}}_d}'(\theta_d) \rho  R_{\hat{\bm{n}}_d}'(\theta_d)^\dagger \Bigr).
\end{eqnarray}
The \verb|Fraxis| is optimized using Lagrange multipliers to minimize the cost function $ \langle M \rangle_{\hat{\bm{n}_d}, \theta_d}$ w.r.t. the axis unit vector $\hat{\bm{n}}$~\cite{fraxis}. We follow the notation from Ref.~\cite{fraxis}, where the subscript $d$ is omitted to improve readability. We also set $\theta_d = \pi$ for the rest of the section. 

First, we want to find the optimal parameters $\hat{\bm{n}}^*$ and Lagrange multiplier $\lambda^*$, which minimize the objective function
\begin{equation}\label{lagrange_multipliers}
    (\hat{\bm{n}}^*, \lambda^*) = \argmin_{\hat{\bm{n}}, \lambda} f(\hat{\bm{n}}, \lambda), 
\end{equation}
where 
\begin{equation}
     f(\hat{\bm{n}}, \lambda) \equiv \langle M \rangle_{\hat{\bm{n}}} - \lambda(n_x^2 + n_y^2 + n_z^2 - 1).
\end{equation}
Differentiating the objective function in Eq.~(\ref{lagrange_multipliers}) w.r.t. $n_x, n_y$ and $n_z$ yields the following set of equations~\cite{fraxis}
\begin{eqnarray}
    \frac{\partial f}{\partial n_x} \ &&= 2n_x \text{Tr}(MX\rho X) + n_y\text{Tr}(MX\rho Y + MY\rho X) \nonumber \\&& + \  n_z \text{Tr}(MX\rho Z + MZ\rho X) -2\lambda n_x, \label{lagrange_nx} \\[0.3cm]
    \frac{\partial f}{\partial n_y} \ &&= 2n_y \text{Tr}(MY\rho Y) + n_x\text{Tr}(MX\rho Y + MY\rho X) \nonumber \\&& + \  n_z \text{Tr}(MY\rho Z + MZ\rho Y) -2\lambda n_y , \\[0.3cm]
    \frac{\partial f}{\partial n_z} \ &&= 2n_z \text{Tr}(MZ\rho Z) + n_x\text{Tr}(MX\rho Z + MZ\rho X) \nonumber \\&& + \  n_y \text{Tr}(MY\rho Z + MZ\rho Y) -2\lambda n_z. \label{lagrange_nz}
\end{eqnarray}\\[0.1cm]
The optimal $\hat{\bm{n}}^* = (n_x^*, n_y^*, n_z^*)^T$ satisfies the following condition 
\begin{equation}\label{lagrange_solution}
    \frac{\partial f}{\partial n_x} = \frac{\partial f}{\partial n_y} = \frac{\partial f}{\partial n_z} = 0.
\end{equation}
Furthermore, using the definitions from Ref.~\cite{fraxis}, we write the cross-term traces (i.e. $\text{Tr}(MX\rho Y)$) as follows
\begin{eqnarray}
    \text{Tr}(MX\rho Y + MY\rho X) \ &&= 2r_{(x+y)} - r_x - r_y, \\
    \text{Tr}(MX\rho Z + MZ\rho X) \ &&= 2r_{(x+z)} - r_x - r_z, \\
    \text{Tr}(MY\rho Z + MZ\rho Y) \ &&= 2r_{(y+z)} - r_y - r_z,
\end{eqnarray}
where $r_x, r_y, r_z, r_{(x+y)}, r_{(x+y)}$ and $r_{(y+z)}$ are defined as follows
\begin{eqnarray}
    r_x \ &&= \text{ Tr}(MX\rho X), \\
    r_y \ &&= \text{ Tr}(MY\rho Y), \\
    r_z \ &&= \text{ Tr}(MZ\rho Z),\\
    r_{(x+y)} \  &&= \text{ Tr}\left(M \left(\frac{X+Y}{\sqrt{2}} \right)\rho \left(\frac{X+Y}{\sqrt{2}} \right) \right), \\
    r_{(x+z)} \ &&= \text{ Tr}\left(M \left(\frac{X+Z}{\sqrt{2}} \right)\rho \left(\frac{X+Z}{\sqrt{2}} \right) \right), \\
    r_{(y+z)} \  &&= \text{ Tr}\left(M \left(\frac{Y+Z}{\sqrt{2}} \right)\rho \left(\frac{Y+Z}{\sqrt{2}} \right) \right). 
\end{eqnarray}\\

Finally, with these definitions, we can form the linear system of equations from Eqs.~(\ref{lagrange_nx})--(\ref{lagrange_nz}) and Eq.~(\ref{lagrange_solution})

\begin{widetext} 
\begin{equation}\label{fraxis_matrix_eq}
\begin{pmatrix}
2r_x & \qquad  2r_{(x+y)} - r_x - r_y & \qquad 2r_{(x+z)} - r_x - r_z \\
2r_{(x+y)} - r_x - r_y &\qquad  2r_y &\qquad   2r_{(y+z)} - r_y - r_z \\
2r_{(x+z)} - r_x - r_z &\qquad  2r_{(y+z)} - r_y - r_z &\qquad  2r_z
\end{pmatrix}
\begin{pmatrix}
n_x^*   \\
n_y^*  \\
n_z^*
\end{pmatrix}
= 2\lambda^* \textbf{I}
\begin{pmatrix}
n_x^*   \\
n_y^*  \\
n_z^*
\end{pmatrix}.
\end{equation}
\end{widetext}
Here, we remark that this represents the simplified Free-Axis Selection from Ref.~\cite{fraxis}, where the rotation angle $\theta_d$ is fixed to $\pi$.

After forming the matrix from the measurements in Eq.~(\ref{fraxis_matrix_eq}), the optimal axis is calculated by solving the three eigenvalues and eigenvectors in Eq.~(\ref{fraxis_matrix_eq}). The optimal axis $\n^*$ is the eigenvector that corresponds to the lowest eigenvalue. For a comprehensive description, see Ref.~\cite{fraxis}.

\subsection{Free Quaternion Selection Optimizer}

Free Quaternion Selection (\verb|FQS|) optimizer \cite{fqs} is a matrix-diagonalization-based optimizer that optimizes a unit quaternion $\q_d = (q_{0}, q_1, q_2, q_3)$ rather than the unit axis $\n$ as in the case of \verb|Fraxis|. The \verb|FQS| expands the idea of \verb|Fraxis| where we represent the single-qubit gate entirely in parameterized quaternion form. That is, instead of using the axis $\n$, the identity element in Eq.~(\ref{fraxis_gate}) is also included. We start by expressing the unit axis $\n$ is spherical coordinates as follows~\cite{fqs}
\begin{equation}
    \n = \n(\psi, \varphi) = (\cos \psi, \sin\psi\cos\varphi, \sin\psi \sin\varphi),
\end{equation}
where $\psi$ is the polar angle and $\varphi$ is the azimuth angle. Then, we define the extended Pauli matrix $\z \equiv (\varsigma_I, \varsigma_x, \varsigma_y, \varsigma_z) = (I, -iX, -iY, -iZ)$. Now we can express the single-qubit gate as \cite{pankkonen2025improvingvariationalquantumcircuit, fqs}
\begin{eqnarray} \label{fqs_gate}
    R_{\hat{\bm{n}}_d(\psi, \varphi)}'(\theta) \  && = I^{\otimes(k-1)}  \otimes \left[ \q_d(\theta, \psi, \varphi) \cdot \z \right] \otimes I^{\otimes(n-k)}  \nonumber \\ && \equiv R'(\q_d).
\end{eqnarray}
We remark that the subscript $d$ has been dropped from the angular variables $\theta, \psi$, and $\varphi$ to clarify the notation. 

We can now focus on the optimization of the single-qubit gate. The same procedure that we applied to \verb|Rotosolve| and \verb|Fraxis| also works on the \verb|FQS| optimizer. Again, applying the Eq.~(\ref{fqs_gate}) to Eq.~(\ref{U_one_layer}) leads to Eq.~(\ref{hat_M}). Then, define the quantum circuits before and after the $d$-th parameterized gate $R'(\q_d)$ as $V_1$ and $V_2$, respectively. After that, using the cyclic property of the trace operation and definitions from Eqs.~(\ref{new_rho}) and~(\ref{new_M}), we arrive at the quadratic form for the cost function $\langle M \rangle_{\q_d}$ as
\begin{equation}
    \langle M \rangle_{\q_d} = \text{Tr}\left(M R'(\q_d) \rho R'(\q_d)^\dagger \right).
\end{equation}

Since the components of a unit quaternion $\q$ do not affect the matrix products, we focus on the terms $\text{Tr}(M\z_d '\rho \z_d'^\dagger)$. Here, we have used $\z' = I^{\otimes(k-1)} \otimes \z \otimes I^{\otimes(n-k)} $. We now construct the matrix $S = (S_{\mu\nu})$ which can be expressed as follows~\cite{fqs}
\begin{equation}
    S_{\mu\nu} = \frac{1}{2} \text{Tr}\left[M(\varsigma_\mu' \rho \varsigma_\nu'^{\dagger} + \varsigma_\nu' \rho \varsigma_\mu'^{\dagger}) \right].
\end{equation}
The matrix $S$ is a real $4\times4$ symmetric matrix with all real entries.

Thus, we obtain the quadratic form for the cost function in terms of $\q_d$
\begin{equation}
    \langle M \rangle_{\q_d} = \q_d^T S \q_d.
\end{equation}
Similarly to \verb|Fraxis|, the unit quaternion $\q_d$ is optimized by solving the eigenvectors and eigenvalues of the matrix $S$ formed from the measured expectation values $\langle M \rangle$. The optimal $\q_d^*$ is the eigenvector of $S$ that corresponds to the lowest eigenvalue.

\section{Additional Experiments -- Parameter Threshold Crossings Without Gate Freezing} \label{appendix_threshold_counts}

In this section, we provide additional experiments relating to Sec. \ref{gate_freeze_iters_section}. We use the base versions of \verb|Fraxis| and \verb|FQS| with matrix norm-based gate freezing to compute the distance between the gate parameters in consecutive iterations.

\begin{figure}
    \centering
    \includegraphics[width=0.99\linewidth]{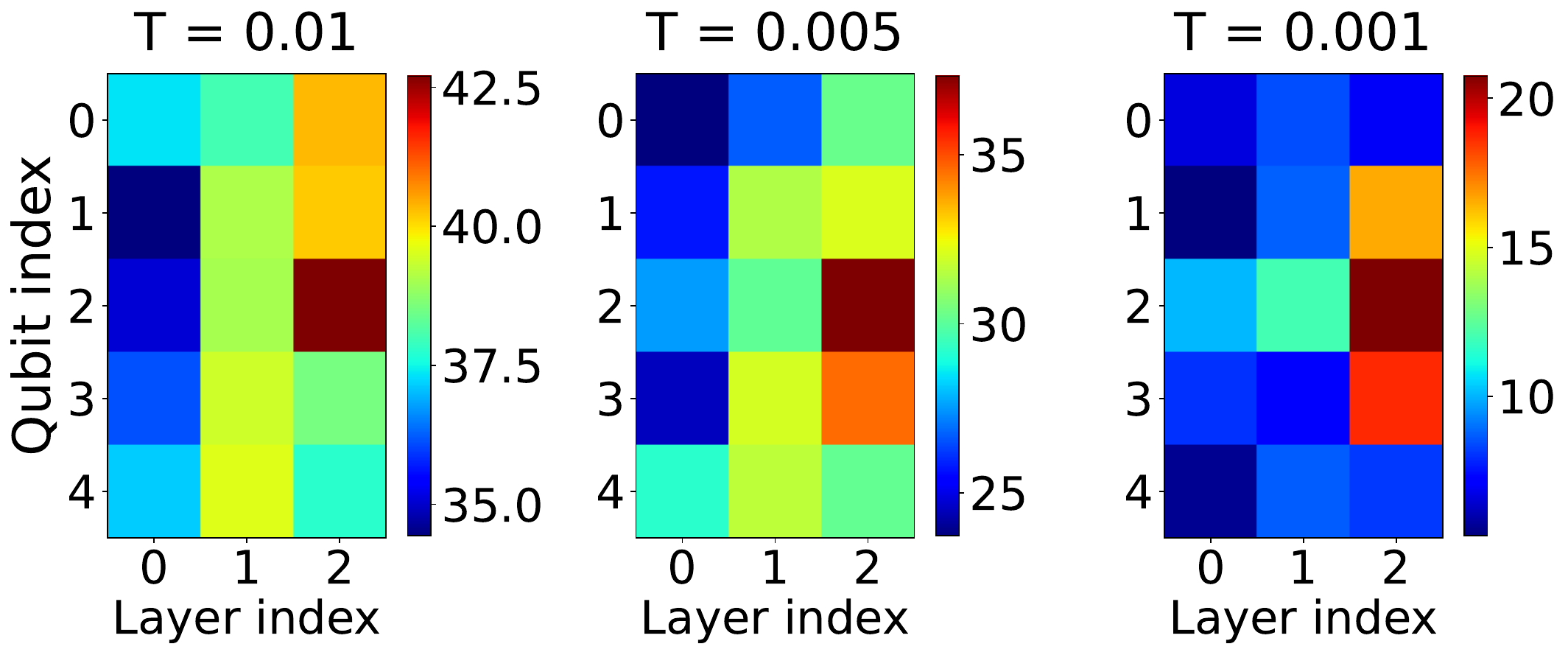}
    \cprotect\caption{Averages for  $\bm{\kappa}_d$ across 50 runs, where the change of $d$-th parameter between consecutive iterations falls below the threshold $T$. The base \verb|Fraxis| optimizer was used with a matrix norm distance metric, and the threshold $T$ was set to have values $T = 0.01$ (left), $ 0.005$ (mid), and $0.001$ (right). Red cells of the grid indicate a higher average and blue cells a lower average value for $\bm{\kappa}_d$.}
    \label{fraxis_index_no_freeze_results}
\end{figure}

\begin{figure}
    \centering
    \includegraphics[width=0.99\linewidth]{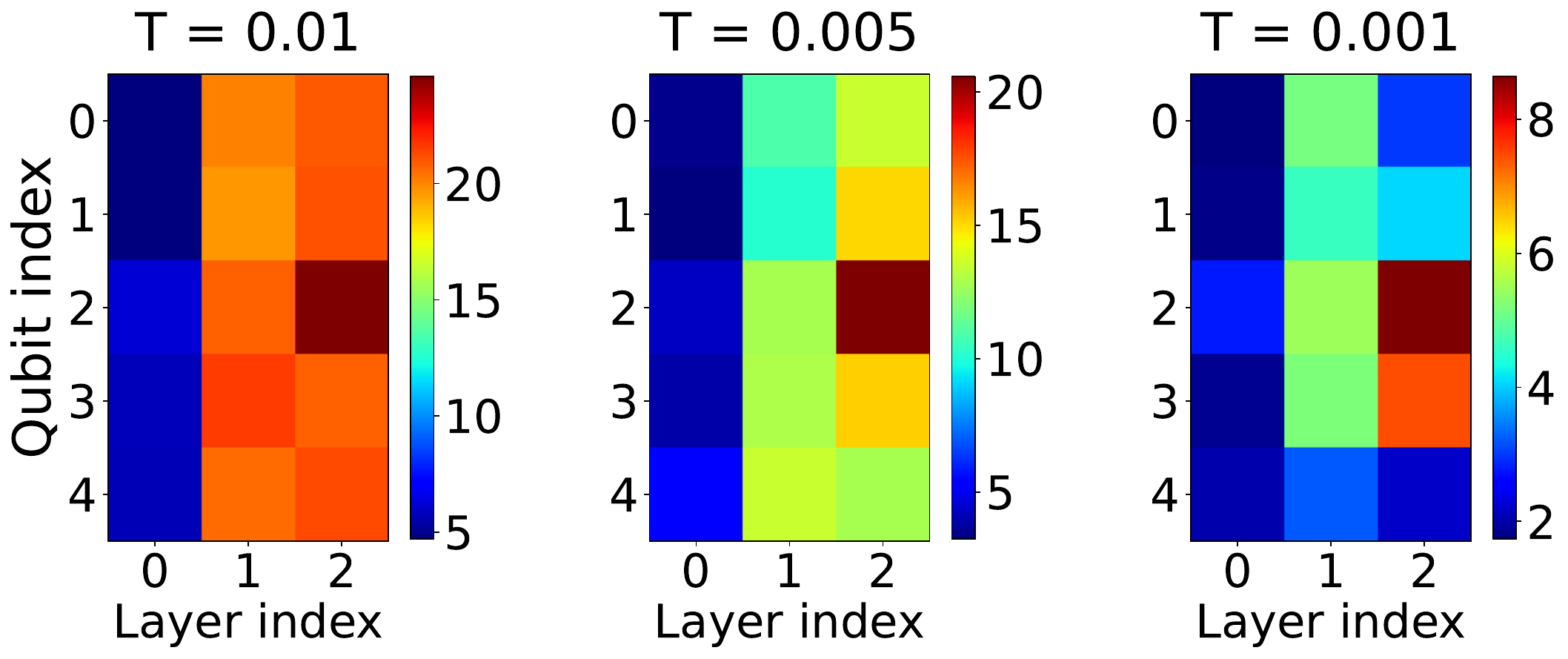}
    \cprotect\caption{Averages for  $\bm{\kappa}_d$ across 50 runs, where the change of $d$-th parameter between consecutive iterations falls below the threshold $T$. The base \verb|FQS| optimizer was used with a matrix norm distance metric, and the threshold $T$ was set to have values $T = 0.01$ (left), $ 0.005$ (mid), and $0.001$ (right). Red cells of the grid indicate a higher average and blue cells a lower average value for $\bm{\kappa}_d$.}
    \label{fqs_index_no_freeze_results}
\end{figure}

We have similar results to those for the gate freezing case with parameter-based distance metric in Sec.~\ref{gate_freeze_iters_section}. For \verb|Fraxis|, in Fig.~\ref{fraxis_index_no_freeze_results} we see that when we use the matrix norms to compute the distances between gate parameters, the middle qubits gate in the last layer obtains the highest average value for $\bm{\kappa}_d$ with all thresholds $T$. If we compare $T=0.001$ with the parameter-based distance metric from Sec.~\ref{gate_freeze_iters_section}, the matrix norm exhibits a more symmetrical pattern. Otherwise, it does not seem to be affected by whether we choose the matrix norm or parameter-based distance metric. For the \verb|FQS| optimizer in Fig.~\ref{fqs_index_no_freeze_results} and Sec.~\ref{gate_freeze_iters_section}, there is no difference for the matrix norm versus the parameter-based distance metric used. 

\section{Additional experiments -- Parameter Threshold Crossings Without Gate Freezing with Other Circuit Ansätze} \label{appendix_threshold_counts_ansatzes}

In this section, we provide additional results for the gate freezing iterations, similar to Sec.~\ref{gate_freeze_iters_section} but with different ansatz circuits (types B, C, and D). The circuits are provided in the Appendix~\ref{appendix_circuits}. For all optimizers, we use the parameter-based distance metric.

\begin{figure}
    \centering
    \includegraphics[width=0.99\linewidth]{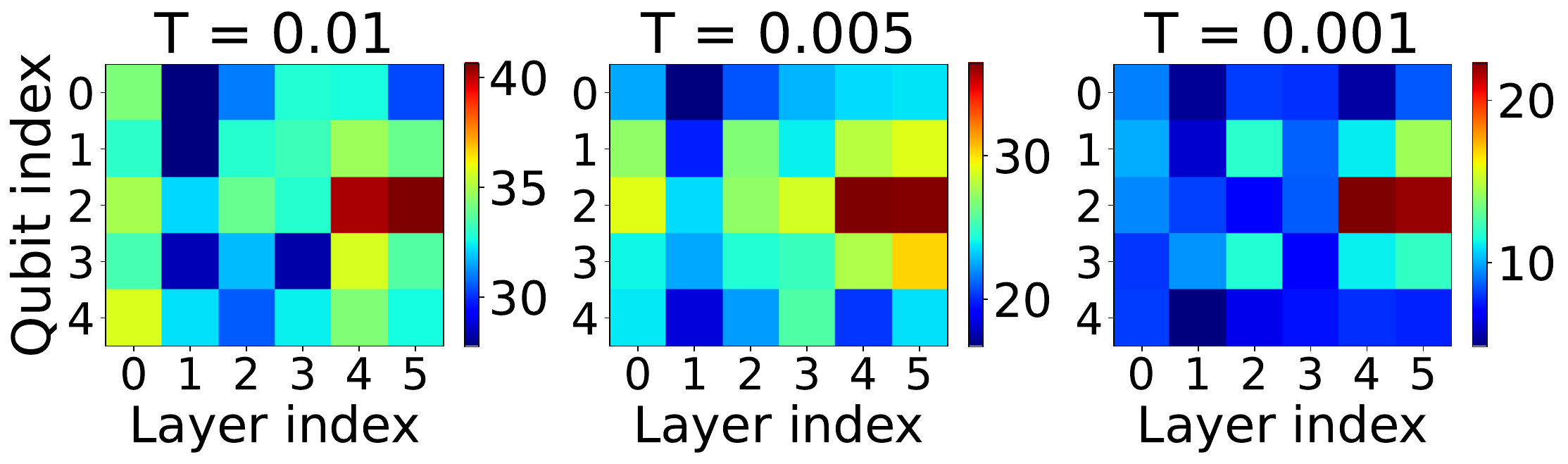}
    \includegraphics[width=0.99\linewidth]{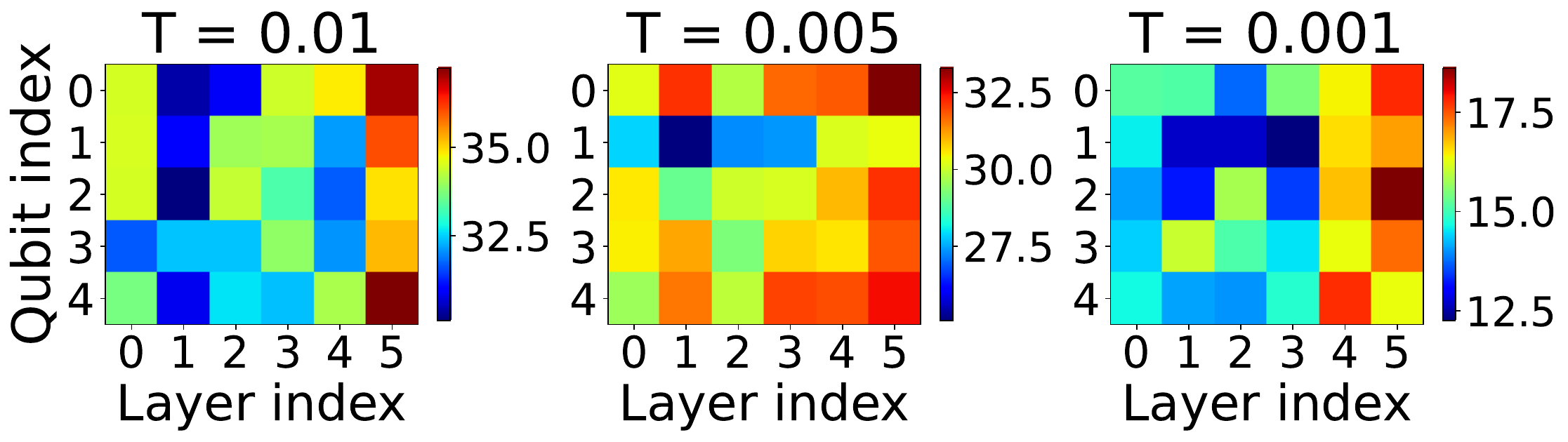}
    \includegraphics[width=0.99\linewidth]{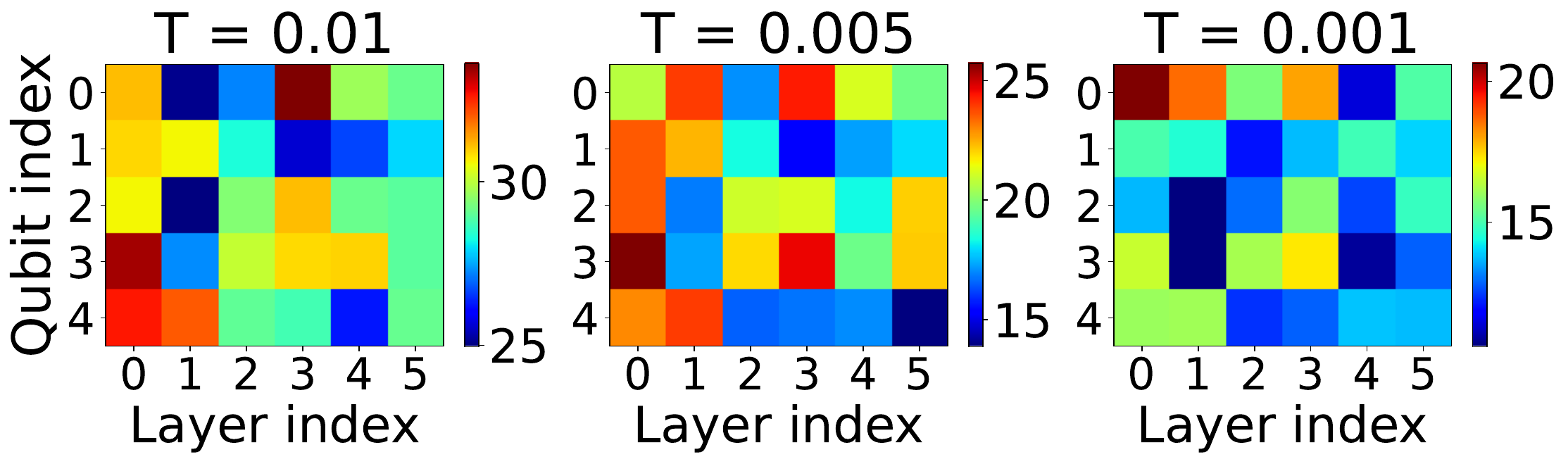}
    \cprotect\caption{Averages for  $\bm{\kappa}_d$ across 50 runs, where the change of $d$-th parameter between consecutive iterations falls below the threshold $T$. The base \verb|Rotosolve| optimizer and parameter-based distance metric were used. The gate freezing threshold $T$ was set to have values $T = 0.01$, $ 0.005$, and $0.001$, which are shown in the left, middle, and right columns, respectively. Each row corresponds to ansatz circuits B1, C1, and D1 from Appendix~\ref{appendix_circuits}, respectively, and each block in the grid represents the mean of the gate freeze iterations of 50 runs. Red cells of the grid indicate a higher average and blue cells a lower average value for $\bm{\kappa}_d$.}
    \label{rotosolve_additional_ansatz}
\end{figure}

The additional results for \verb|Rotosolve| are shown in Fig.~\ref{rotosolve_additional_ansatz}. Each row corresponds to the results for different ansatz circuits B1 (cascade), C1 (cyclic entangling), and D1 (one-qubit connector) from Appendix~\ref{appendix_circuits}, respectively. For ansatz circuit B1, we have similar results as in Sec.~\ref{gate_freeze_iters_section}. If the threshold is set to $T=0.001$, the B1 and C1 share a similar structure that both have the highest average for $\bm{\kappa}_d$ in the last layer, but for the C1 ansatz, they are spread out and not concentrated for the single qubit as in B1. Ansatz D1 seems to have no clear structure when compared to B1 and C1. Furthermore, no clear pattern emerges, except that some gates on the first qubit have higher averages for $\bm{\kappa}_d$. Also, with $T=0.01$ and $T=0.005$ for the D1 Ansatz, the gates with higher averages for $\bm{\kappa}_d$ are more focused on the first layer of the circuit, respectively.

\begin{figure}
    \centering
    \includegraphics[width=0.99\linewidth]{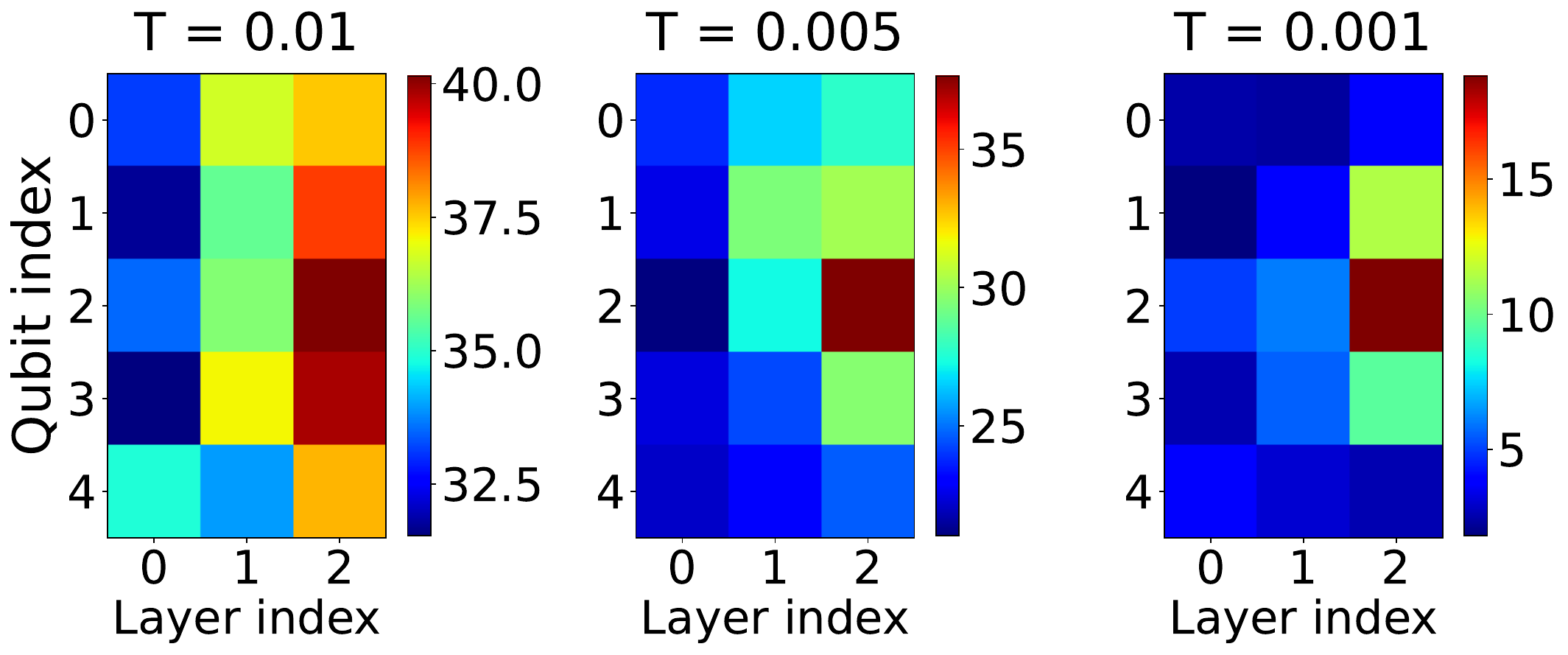}
    \includegraphics[width=0.99\linewidth]{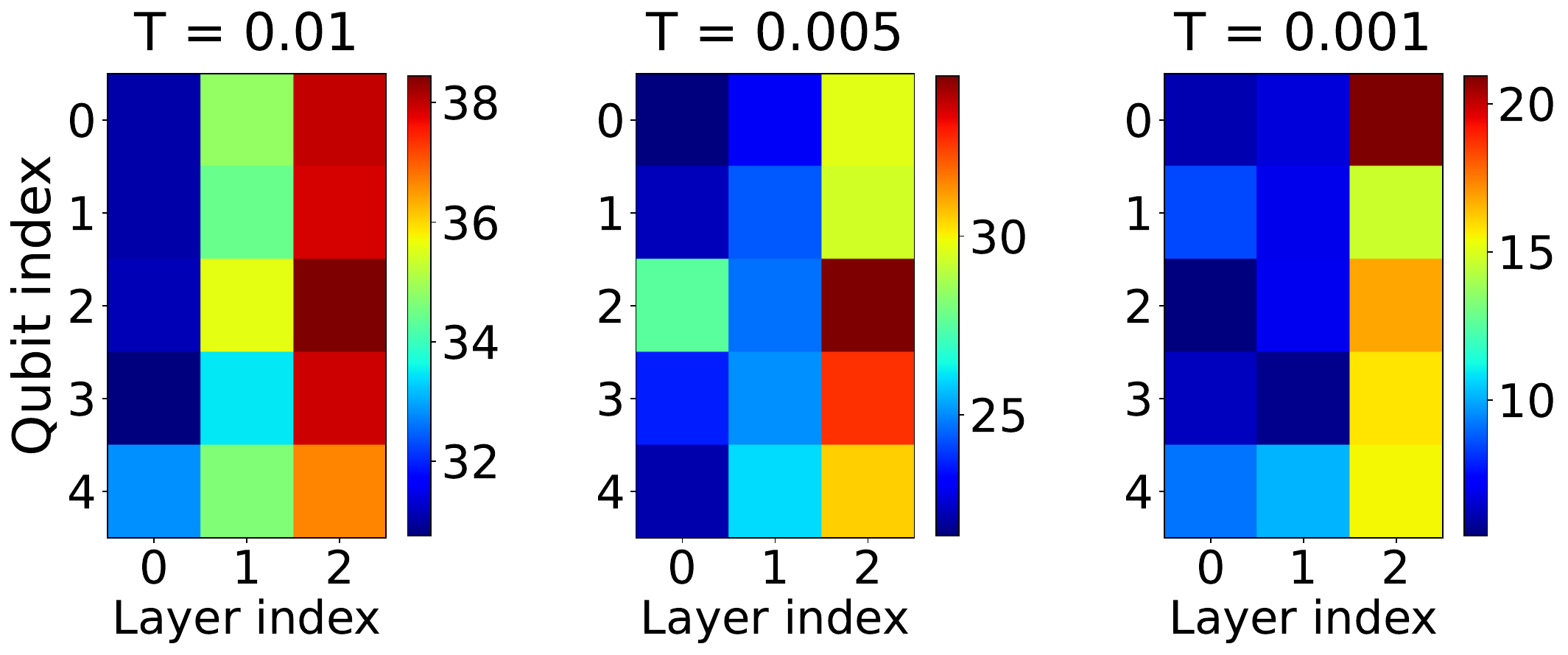}
    \includegraphics[width=0.99\linewidth]{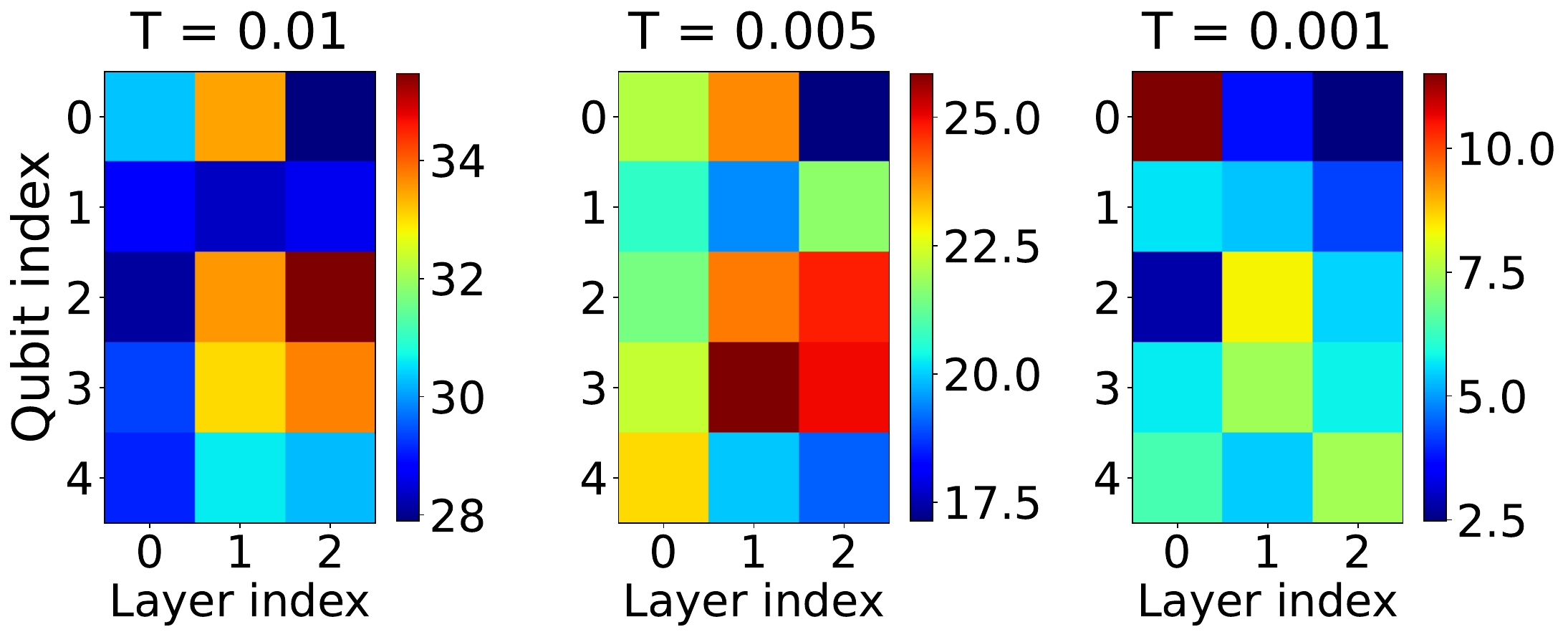}
    \cprotect\caption{Averages for  $\bm{\kappa}_d$ across 50 runs, where the change of $d$-th parameter between consecutive iterations falls below the threshold $T$. The base \verb|Fraxis| optimizer and parameter-based distance metric were used. The gate freezing threshold $T$ was set to have values $T = 0.01$, $ 0.005$, and $0.001$. They are shown in the left, middle, and right columns, respectively. Each row corresponds to ansatz circuits B2, C2, and D2 from Appendix~\ref{appendix_circuits}, respectively, and each block in the grid represents the mean of the gate freeze iterations of 50 runs. Red cells of the grid indicate a higher average and blue cells a lower average value for $\bm{\kappa}_d$.}
    \label{fraxis_additional_ansatz}
\end{figure}

The results for \verb|Fraxis| are shown in Fig.~\ref{fraxis_additional_ansatz}. For $T=0.01$, all ansätze have about the same behavior, where the gates in the first layer have lower averages $\bm{\kappa}_d$ compared to the rest. When we set $T=0.005$, the results are similar, except for the ansatz D2. The gates in the second layer obtain a higher average $\bm{\kappa}_d$ than those in the first and last layers. They are more spread out compared to other ansatz circuits, as in \verb|Rotosolve|'s case. Finally, when we set $T=0.001$, the gates in the first and second layers have similar low average values for ansätze B2 and C2. The gate with the highest average for D2 is the first gate in the optimization sequence, the top left gate. The gates with the lowest averages are located in the last layer, primarily affecting qubits with lower indices.

\begin{figure}
    \centering
    \includegraphics[width=0.99\linewidth]{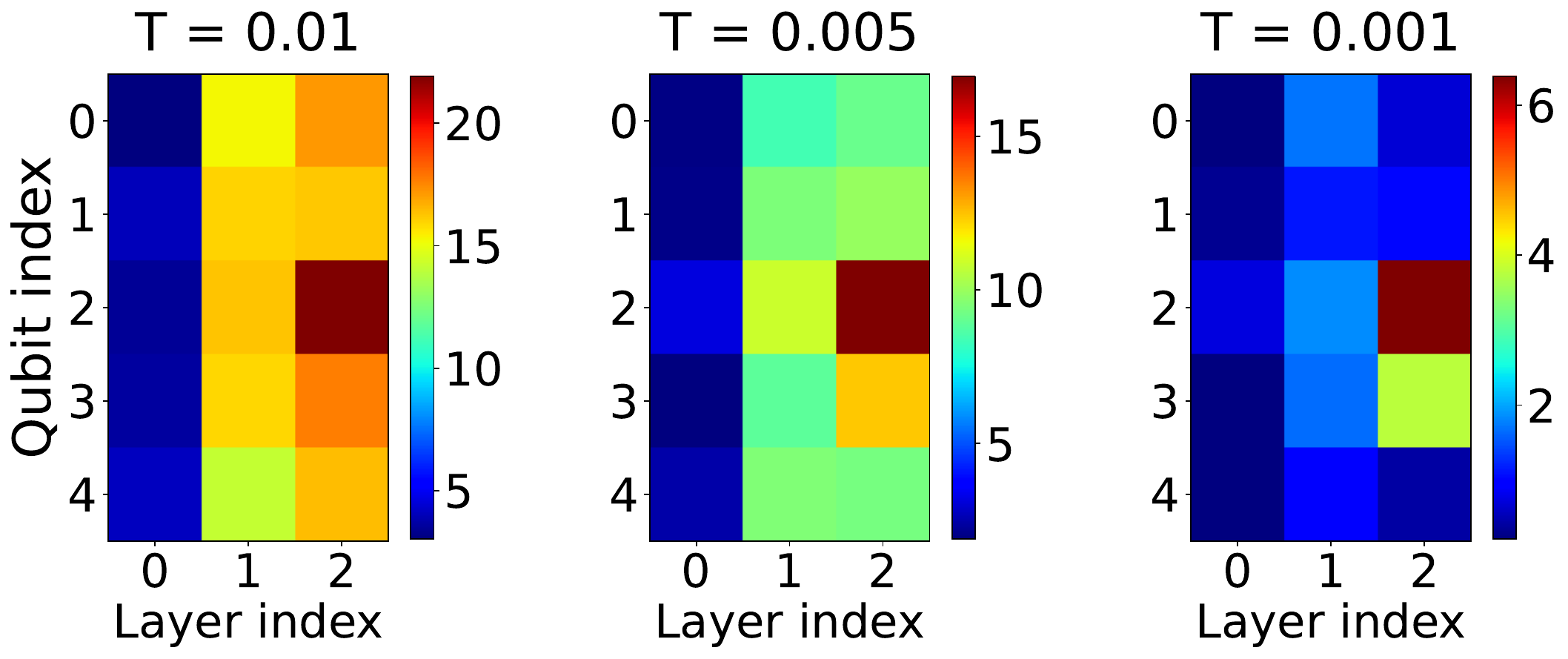}
    \includegraphics[width=0.99\linewidth]{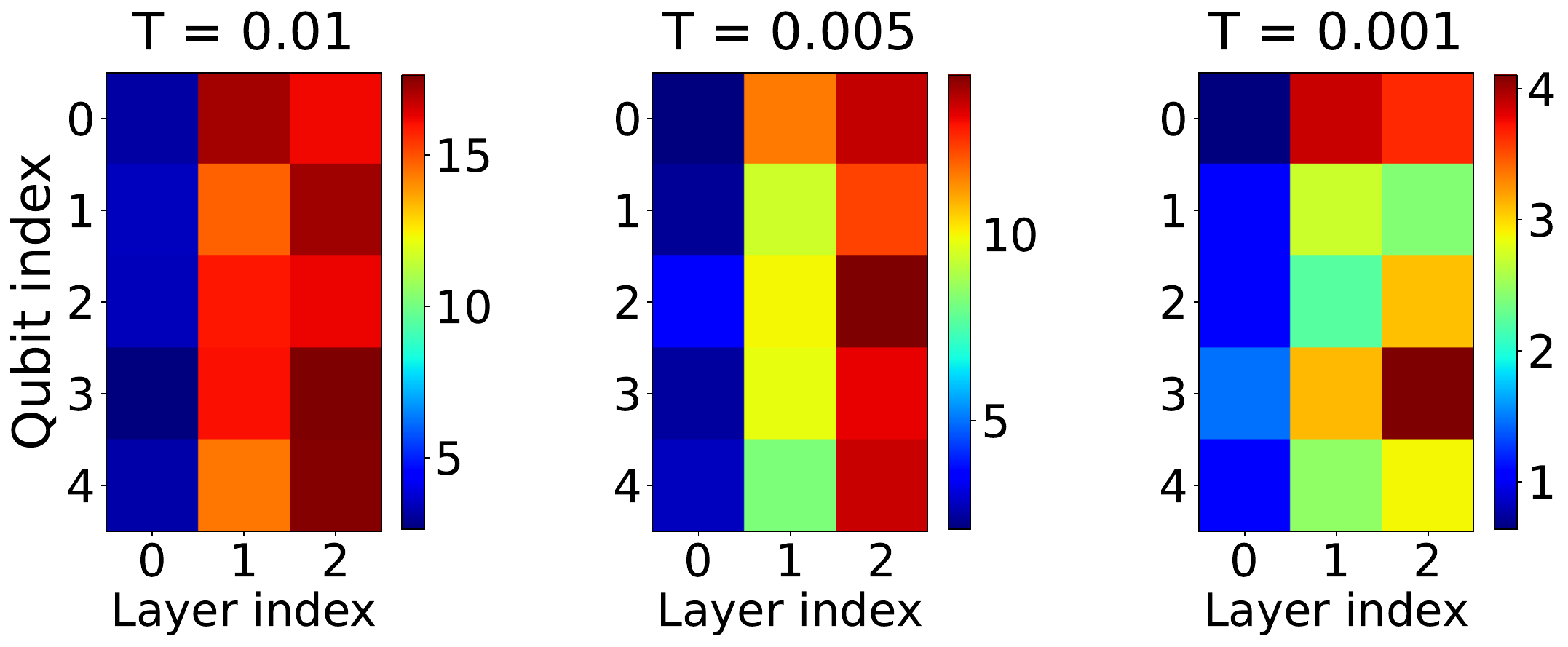}
    \includegraphics[width=0.99\linewidth]{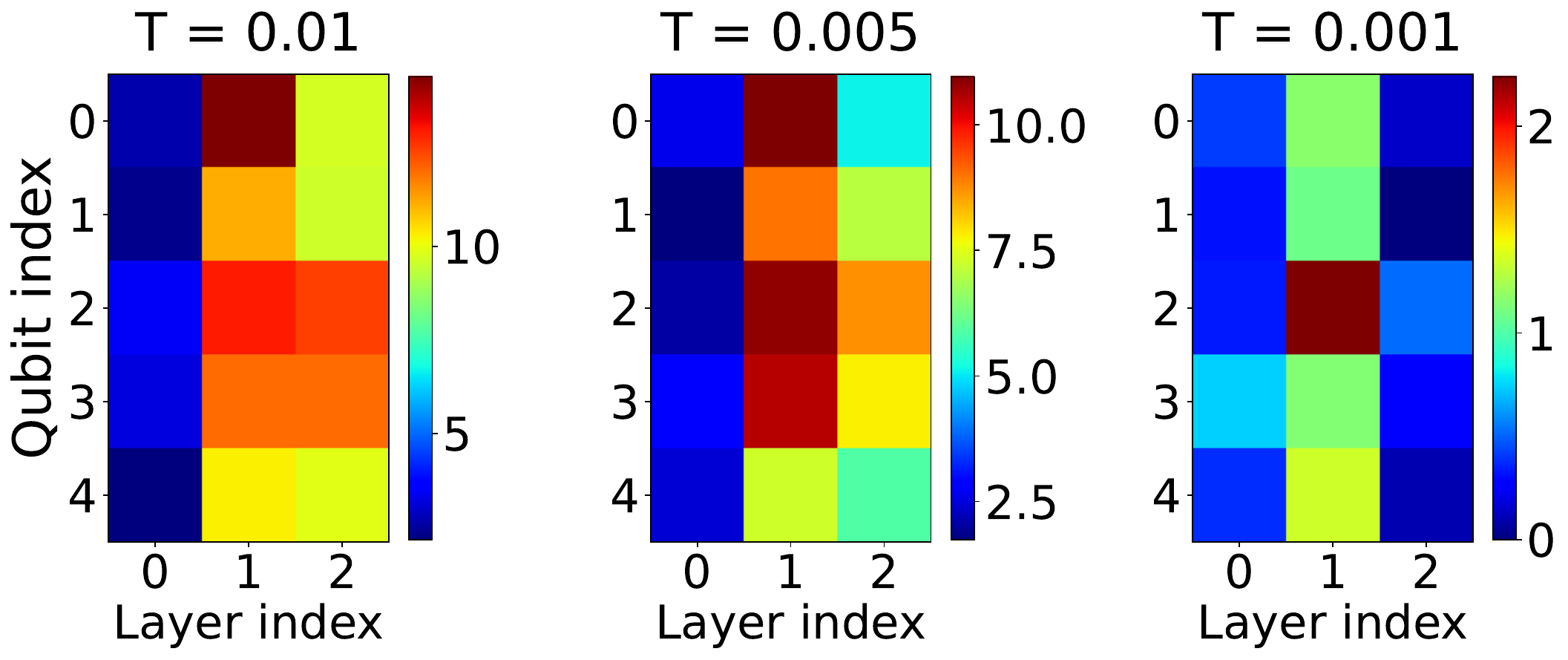}
    \cprotect\caption{Averages for  $\bm{\kappa}_d$ across 50 runs, where the change of $d$-th parameter between consecutive iterations falls below the threshold $T$. The base \verb|FQS| optimizer and parameter-based distance metric were used. The gate freezing threshold $T$ was set to have values $T = 0.01$, $ 0.005$, and $0.001$. They are shown in the left, middle, and right columns, respectively. Each row corresponds to ansatz circuits B2, C2, and D2 from Appendix~\ref{appendix_circuits}, respectively, and each block in the grid represents the mean of the gate freeze iterations of 50 runs. Red cells of the grid indicate a higher average and blue cells a lower average value for $\bm{\kappa}_d$.}
    \label{fqs_additional_ansatz}
\end{figure}

Finally, we present our results for \verb|FQS|, which are shown in Fig.~\ref{fqs_additional_ansatz}. With the gate freezing threshold value $T=0.01$, the first layer of all ansätze receives the lowest averages. When we set $T=0.005$, the higher averages for ansatz B2 concentrate more on the center of the last layer. On the other hand, the gates with the lowest average values are distributed evenly in the last layer for the C2 ansatz and in the second layer for the D2 ansatz. Finally, when the threshold is set to $T=0.001$, the gate with the highest average in D2 is exactly in the middle of the circuit. The results for Ansatz B2 appear similar to those shown in the Appendix~\ref{appendix_threshold_counts}.

\section{Scaling Matrix Norms} \label{unitary_norms_nqubit_gate_sec}

In this section, we explore the implementation of gate freezing for various gate sizes. Before evaluating whether the optimized gate falls below the freezing threshold, we normalize the computed norm values to a common range. Matrix norms vary in their range of values depending on the size of the unitary matrices. To allow a consistent comparison between different matrix sizes, we normalize all the norms computed to be in the range $[0, 1]$.

We focus on the distance metric derived from the Frobenius norm in Sec.~\ref{matrix_norm_derivation_sec}. Let $U$ and $V$ be unitary matrices of size $2^n \times 2^n$. The Frobenius norm between matrices $U$ and $V$ is defined:

\begin{align}
    \norm{U - V}_F &= \sqrt{\text{Tr}\bigl[(U - V)^\dagger (U - V)\bigr]},
\end{align}

Using the fact that $\Tr(U^\dagger U) = \Tr(V^\dagger V) = \Tr(I) = 2^n$ for $n$-qubit unitaries, we then obtain

\begin{equation}
    \norm{U - V}_F = \sqrt{ 2^{n+1} - 2 \cdot \Re\bigl(\Tr(U^\dagger V)\bigr)}.
\end{equation}

Using a similar argument as in Sec.~\ref{matrix_norm_derivation_sec}, we rotate the Frobenius inner product $\Tr(U^\dagger V)$ onto the positive real axis with the complex phase $e^{i\delta}$. That is, we select $\delta$ such that $\Tr(U^\dagger e^{i\delta} V) = e^{i\delta} \Tr(U^\dagger V) = |\Tr(U^\dagger V)| $. Then, we define the distance for the $n$-qubit unitaries as follows

\begin{eqnarray}
    \D(U, V) \equiv \sqrt{2^{n+1} - 2\cdot |\Tr(U^\dagger V)|}.
\end{eqnarray}\\

Finally, as in Sec.~\ref{matrix_norm_derivation_sec}, we normalize the distance to be in the range $[0,1]$. The expression inside the square root achieves its maximal value when $|\Tr(U^\dagger V)| = 0$ and hence $\D(U, V)$ can have a maximal value of $\sqrt{2^{n+1}}$. To obtain the normalized distance for the $n$-qubit unitaries, we divide the computed distance by its maximal value $\sqrt{2^{n+1}}$.

\bibliography{apssamp}

\end{document}